\documentclass[11pt]{article}
\usepackage[utf8]{inputenc}
\usepackage[T1]{fontenc}
\usepackage{mathtools}
\usepackage{amssymb}
\usepackage{amsthm}
\usepackage{marginnote}

\usepackage[tt=false]{libertine}
\usepackage[libertine]{newtxmath}
\usepackage{microtype}

\usepackage{array}
\usepackage{nicefrac}
\usepackage{xcolor}
\usepackage{colortbl}

\usepackage{xspace}

\usepackage[margin=1.1in]{geometry}
\usepackage[normalem]{ulem}
\usepackage{paralist}
\usepackage{float}
\usepackage{caption}

\usepackage[linesnumbered,boxruled,vlined,noend,nosemicolon]{algorithm2e}

\SetAlFnt{\small}
\SetAlCapFnt{\footnotesize}
\SetAlCapNameFnt{\footnotesize}
\SetAlCapSty{bfseries}
\SetAlCapNameSty{normalfont}
\SetAlgoCaptionSeparator{\enspace}
\SetInd{0.5em}{1em}
\SetAlgoHangIndent{1.5em}
\SetNlSkip{0.75em}
\SetAlgoNlRelativeSize{-2}

\SetNlSty{algolinenumber}{}{}

\SetCommentSty{algocomment}
\SetKwComment{tcc}{\(\triangleright\) }{}
\SetArgSty{textnormal}

\SetKw{Initialize}{Initialize}
\SetKw{Return}{Return}
\SetKwProg{AlgoProc}{Procedure}{:}{}

\usepackage[
    backend=biber,
    style=alphabetic,
    doi=false, maxbibnames=8,
    minbibnames=8,
    maxalphanames=5,
    minalphanames=1,
]{biblatex}
    \uspunctuation
    \DeclareDelimFormat{multicitedelim}{\addcomma\space}
    \DeclareFieldFormat{url}{\href{#1}{\texttt{(Link)}}}
    \AtEveryBibitem{\clearfield{pages}\ifentrytype{inproceedings}{\clearlist{publisher}}{}}
    \renewbibmacro*{in:}{}
    \AtBeginBibliography{\setlength{\emergencystretch}{3em}}

\ExplSyntaxOn
\renewbibmacro*{date}{
    \tl_set:Nx \l_tmpa_tl {\thefield{series}}
    \exp_args:NnV \regex_match:nnTF {\s [0-9]{4} \Z} \l_tmpa_tl
        {}
        {\printdate}
}
\ExplSyntaxOff

\usepackage[bookmarks=false]{hyperref}
\usepackage{cleveref}

\definecolor{DarkGreen}{rgb}{0.2,0.6,0.2}
\definecolor{DarkRed}{rgb}{0.6,0.2,0.2}
\definecolor{DarkBlue}{rgb}{0.2,0.2,0.6}
\definecolor{DarkPurple}{rgb}{0.4,0.2,0.4}

\hypersetup{
    linktocpage=true,
    colorlinks=true,
    linkcolor=DarkBlue,
    citecolor=DarkBlue,
    urlcolor=DarkBlue,
}

\DeclareDelimFormat[textcite]{namelabeldelim}{\addnbspace}
\DeclareDelimFormat[textcite]{multicitedelim}{\addsemicolon\space}
\NewDocumentCommand{\citet}{s O{} m}{\begingroup
    \IfBooleanT{#1}{\AtNextCite{\AtEachCitekey{\defcounter{maxnames}{999}\defcounter{minnames}{999}}}}\textcite[#2]{#3}\endgroup}
\DeclareCiteCommand*{\citeauthor}
    {\boolfalse{citetracker}\boolfalse{pagetracker}\usebibmacro{prenote}}
    {\printnames[][1-999]{labelname}}
    {\multicitedelim}
    {\usebibmacro{postnote}}

\newcolumntype{P}[1]{>{\centering\arraybackslash}p{#1}}

\usepackage{tikz}
    \usetikzlibrary{positioning}

    \let\originalleft\left
\let\originalright\right
\renewcommand{\left}{\mathopen{}\mathclose\bgroup\originalleft}
\renewcommand{\right}{\aftergroup\egroup\originalright}

\newcommand{\abs}[1]{{\left | {#1} \right|}}
\newcommand{\paren}[1]{{\left ( {#1} \right)}}
\newcommand{\bparen}[1]{{\big ( {#1} \big)}}
\newcommand{\Bparen}[1]{{\Big ( {#1} \Big)}}

\newcommand{\set}[1]{{\left \{ {#1} \right\}}}

\newcommand{\from}{:}

\newcommand{\eps}{\varepsilon}
\renewcommand{\epsilon}{\eps}

\newcommand{\N}{\mathbb{N}}
\newcommand{\R}{\mathbb{R}}
\newcommand{\Z}{\mathbb{Z}}

\newcommand{\cA}{\mathcal{A}}
\newcommand{\cB}{\mathcal{B}}

\newcommand{\cG}{\mathcal{G}}

\newcommand{\cI}{\mathcal{I}}

\newcommand{\cL}{\mathcal{L}}
\newcommand{\cM}{\mathcal{M}}

\newcommand{\cP}{\mathcal{P}}

\newcommand{\cW}{\mathcal{W}}
\newcommand{\cX}{\mathcal{X}}
\newcommand{\cY}{\mathcal{Y}}

\newcommand{\ex}[2]{{\ifx&#1& \mathbb{E} \else \underset{#1}{\mathbb{E}} \fi \left[#2\right]}}
\newcommand{\pr}[2]{{\ifx&#1& \mathbb{P} \else \underset{#1}{\mathbb{P}} \fi \left[#2\right]}}
\newcommand{\var}[2]{{\ifx&#1& \mathrm{Var} \else \underset{#1}{\mathrm{Var}} \fi \left[#2\right]}}

\newcommand{\Lap}{\text{\rm Lap}}

\theoremstyle{plain}
\newtheorem{thm}{Theorem}[section]
\newtheorem{theorem}[thm]{Theorem}
\newtheorem{clm}[thm]{Claim}
\newtheorem{lem}[thm]{Lemma}
\newtheorem{cor}[thm]{Corollary}
\newtheorem{prop}[thm]{Proposition}

\newtheorem{obs}[thm]{Observation}
\theoremstyle{definition}

\newtheorem{definition}[thm]{Definition}

\crefname{thm}{Theorem}{Theorems}
\Crefname{thm}{Theorem}{Theorems}
\crefname{lem}{Lemma}{Lemmas}
\Crefname{lem}{Lemma}{Lemmas}
\crefname{clm}{Claim}{Claims}
\Crefname{clm}{Claim}{Claims}
\crefname{prop}{Proposition}{Propositions}
\Crefname{prop}{Proposition}{Propositions}

\theoremstyle{plain}

\newcommand{\mypar}[1]{\smallskip\noindent\textbf{#1}}

\newcommand{\dnspdetail}[3]{\ensuremath{\cP_{{#1},{#2},{#3}}}\xspace}
\newcommand{\dnsp}{\dnspdetail{\dbound}{\dmax}{\teps}}

\newcommand{\sadnspdetail}[3]{\ensuremath{\text{\rm adapt-}\cP_{{#1},{#2},{#3}}}\xspace}
\newcommand{\sadnsp}{\sadnspdetail{\dbound}{\eps}{\delta}}

\newcommand{\svtinit}{\texttt{SVT-initialize}}
\newcommand{\increment}{\texttt{increment}}
\newcommand{\flagproc}{\texttt{flag}}
\newcommand{\counter}{\text{\sf\small count}}
\newcommand{\flag}{\text{\sf \small flag}}

\newcommand{\fsum}{{f_{\text{sum}}}}
\newcommand{\cGvec}[1]{\cG^{{#1}}}
\newcommand{\cGvecbound}[2]{\cG^{#1}_{#2}}
\newcommand{\vsum}{f_{\text{\rm sum}}^d}
\newcommand{\vsumt}[1]{f_{\text{\rm sum}, {#1}}^d}

\newcommand{\hatvsumt}[1]{{\widehat f}_{\text{\rm sum}, {#1}}^d}
\newcommand{\kashintime}{\mathsf{Kashin}}
\newcommand{\flist}{{f_{\text{list}}}}
\newcommand{\ftilde}{{\tilde{f}}}
\newcommand{\tflist}{{\tilde{f}_{\textrm{list}}}}
\newcommand{\tfsum}{{\tilde{f}_{\textrm{sum}}}}

\newcommand{\dbound}{D}
\newcommand{\dmax}{\widehat{\dbound}}
\newcommand{\dmaxN}{\dmax_N}
\newcommand{\teps}{{\tilde \eps}}
\newcommand{\tdelta}{{\tilde \delta}}

\newcommand{\peps}{{\eps_{p}}}
\newcommand{\pdelta}{{\delta_{p}}}
\newcommand{\aeps}{{\eps_{a}}}
\newcommand{\adelta}{{\delta_{a}}}
\newcommand{\seps}{{\eps_{s}}}
\newcommand{\sdelta}{{\delta_{s}}}

\newcommand{\goodevent}{{\text{\sf Good}_{v^*}}}

\let\oldnl\nl\newcommand{\nonl}{\renewcommand{\nl}{\let\nl\oldnl}}

\definecolor{GoodCell}{HTML}{E2F0D9}

\title{Node Differential Privacy via Node-Stable Graph Projections}

\author{
    Ishaq Aden-Ali\thanks{Based on work done while at UC Berkeley.}\hspace{3em}
    Adam Sealfon\thanks{Google Research.}\hspace{3em}
    Adam Smith\thanks{Department of Computer Science, Boston University. Based in part on work done while at Google Deepmind.}\\[0.6em]
    Thomas Steinke\thanks{Based on work done while at Google DeepMind. Now at Anthropic.}\hspace{3em}
    Jonathan Ullman\thanks{Khoury College of Computer Sciences, Northeastern University.  Supported by NSF awards CNS-2232692 and CNS-2247484.}
}
\date{}

\makeatletter
\def\blx@bblfile{\blx@secinit\begingroup\blx@bblstart
\refsection{0}
  \datalist[entry]{anyt/global//global/global/global}
    \entry{BlockiBDS13}{inproceedings}{}{}
      \name{author}{4}{}{{{hash=57802952607bcac8de916b22479f953f}{family={Blocki},
           familyi={B\bibinitperiod},
           given={Jeremiah},
           giveni={J\bibinitperiod}}}{{hash=c348e11a9e7f34b17b0811660fa01649}{family={Blum},
           familyi={B\bibinitperiod},
           given={Avrim},
           giveni={A\bibinitperiod}}}{{hash=c3ec74f0096bebfca003e01940429056}{family={Datta},
           familyi={D\bibinitperiod},
           given={Anupam},
           giveni={A\bibinitperiod}}}{{hash=b6e363cfbfe2baaa9cc167d439779964}{family={Sheffet},
           familyi={S\bibinitperiod},
           given={Or},
           giveni={O\bibinitperiod}}}}
      \strng{namehash}{b1ba0a09263fd83352b72d0a9a9939fb}
      \strng{fullhash}{f789dfd206bf4c2aa027c8a586f83278}
      \strng{fullhashraw}{f789dfd206bf4c2aa027c8a586f83278}
      \strng{bibnamehash}{f789dfd206bf4c2aa027c8a586f83278}
      \strng{authorbibnamehash}{f789dfd206bf4c2aa027c8a586f83278}
      \strng{authornamehash}{b1ba0a09263fd83352b72d0a9a9939fb}
      \strng{authorfullhash}{f789dfd206bf4c2aa027c8a586f83278}
      \strng{authorfullhashraw}{f789dfd206bf4c2aa027c8a586f83278}
      \field{labelalpha}{BBDS13}
      \field{sortinit}{B}
      \field{sortinithash}{d7095fff47cda75ca2589920aae98399}
      \field{labelnamesource}{author}
      \field{labeltitlesource}{title}
      \field{booktitle}{ACM Conference on Innovations in Theoretical Computer Science}
      \field{series}{ITCS 2013}
      \field{title}{Differentially private data analysis of social networks via restricted sensitivity}
      \field{year}{2013}
      \field{pages}{87\bibrangedash 96}
      \range{pages}{10}
      \strng{urlraw}{https://arxiv.org/abs/1208.4586}
      \strng{url}{https://arxiv.org/abs/1208.4586}
    \endentry
    \entry{BalleBG18}{inproceedings}{}{}
      \name{author}{3}{}{{{hash=22e757242574b20b1a043348c734d13f}{family={Balle},
           familyi={B\bibinitperiod},
           given={Borja},
           giveni={B\bibinitperiod}}}{{hash=a591e71debdbd9a86231c933df50ef0b}{family={Barthe},
           familyi={B\bibinitperiod},
           given={Gilles},
           giveni={G\bibinitperiod}}}{{hash=2236e238d6bdc3c54b604fb209e4f6d2}{family={Gaboardi},
           familyi={G\bibinitperiod},
           given={Marco},
           giveni={M\bibinitperiod}}}}
      \strng{namehash}{da4a5c5c69e1b49e9bc5d8f71a2b7b82}
      \strng{fullhash}{af6aeadf9e6a925b2e550f908890cc31}
      \strng{fullhashraw}{af6aeadf9e6a925b2e550f908890cc31}
      \strng{bibnamehash}{af6aeadf9e6a925b2e550f908890cc31}
      \strng{authorbibnamehash}{af6aeadf9e6a925b2e550f908890cc31}
      \strng{authornamehash}{da4a5c5c69e1b49e9bc5d8f71a2b7b82}
      \strng{authorfullhash}{af6aeadf9e6a925b2e550f908890cc31}
      \strng{authorfullhashraw}{af6aeadf9e6a925b2e550f908890cc31}
      \field{labelalpha}{BBG18}
      \field{sortinit}{B}
      \field{sortinithash}{d7095fff47cda75ca2589920aae98399}
      \field{labelnamesource}{author}
      \field{labeltitlesource}{title}
      \field{booktitle}{Advances in Neural Information Processing Systems}
      \field{series}{NeurIPS 2018}
      \field{title}{Privacy Amplification by Subsampling: Tight Analyses via Couplings and Divergences}
      \field{year}{2018}
      \strng{urlraw}{https://arxiv.org/abs/1807.01647}
      \strng{url}{https://arxiv.org/abs/1807.01647}
    \endentry
    \entry{BorgsCSZ18b}{unpublished}{}{}
      \name{author}{4}{}{{{hash=af9e6b19ead4b3f0c62f2ea329b5009b}{family={Borgs},
           familyi={B\bibinitperiod},
           given={Christian},
           giveni={C\bibinitperiod}}}{{hash=85b0bafaa8969b4a9dbe4504a550fde4}{family={Chayes},
           familyi={C\bibinitperiod},
           given={Jennifer},
           giveni={J\bibinitperiod}}}{{hash=5ccecf114bbd51a1a2979a23dd69c78c}{family={Smith},
           familyi={S\bibinitperiod},
           given={Adam},
           giveni={A\bibinitperiod}}}{{hash=a61b01f61a9ef8f4232936460b50b7e5}{family={Zadik},
           familyi={Z\bibinitperiod},
           given={Ilias},
           giveni={I\bibinitperiod}}}}
      \strng{namehash}{5a5a22113f286290b31c3ea82ed95e4e}
      \strng{fullhash}{ddd59e628bd038b6c1bb1ec69351ba0e}
      \strng{fullhashraw}{ddd59e628bd038b6c1bb1ec69351ba0e}
      \strng{bibnamehash}{ddd59e628bd038b6c1bb1ec69351ba0e}
      \strng{authorbibnamehash}{ddd59e628bd038b6c1bb1ec69351ba0e}
      \strng{authornamehash}{5a5a22113f286290b31c3ea82ed95e4e}
      \strng{authorfullhash}{ddd59e628bd038b6c1bb1ec69351ba0e}
      \strng{authorfullhashraw}{ddd59e628bd038b6c1bb1ec69351ba0e}
      \field{labelalpha}{BCSZ18}
      \field{sortinit}{B}
      \field{sortinithash}{d7095fff47cda75ca2589920aae98399}
      \field{labelnamesource}{author}
      \field{labeltitlesource}{title}
      \field{note}{arXiv:1810.12518}
      \field{title}{Private Algorithms Can Always Be Extended}
      \field{year}{2018}
      \strng{urlraw}{https://arxiv.org/abs/1810.12518}
      \strng{url}{https://arxiv.org/abs/1810.12518}
    \endentry
    \entry{BassilyFGT20}{inproceedings}{}{}
      \name{author}{4}{}{{{hash=30743e89fbbe422b35bbc4cf057dcbf9}{family={Bassily},
           familyi={B\bibinitperiod},
           given={Raef},
           giveni={R\bibinitperiod}}}{{hash=2cdf1f22d9f874d0aee7260732af92c3}{family={Feldman},
           familyi={F\bibinitperiod},
           given={Vitaly},
           giveni={V\bibinitperiod}}}{{hash=da0b99fc715a07e97933e280b0d02b2b}{family={Guzmán},
           familyi={G\bibinitperiod},
           given={Cristóbal},
           giveni={C\bibinitperiod}}}{{hash=1754a40f9d600dea756c1dd1047ce170}{family={Talwar},
           familyi={T\bibinitperiod},
           given={Kunal},
           giveni={K\bibinitperiod}}}}
      \strng{namehash}{a06a97b024a9575bba6dc3e588786d72}
      \strng{fullhash}{9800595342011e6605987c6db5969d4f}
      \strng{fullhashraw}{9800595342011e6605987c6db5969d4f}
      \strng{bibnamehash}{9800595342011e6605987c6db5969d4f}
      \strng{authorbibnamehash}{9800595342011e6605987c6db5969d4f}
      \strng{authornamehash}{a06a97b024a9575bba6dc3e588786d72}
      \strng{authorfullhash}{9800595342011e6605987c6db5969d4f}
      \strng{authorfullhashraw}{9800595342011e6605987c6db5969d4f}
      \field{labelalpha}{BFGT20}
      \field{sortinit}{B}
      \field{sortinithash}{d7095fff47cda75ca2589920aae98399}
      \field{labelnamesource}{author}
      \field{labeltitlesource}{title}
      \field{booktitle}{Advances in Neural Information Processing Systems}
      \field{series}{NeurIPS 2020}
      \field{title}{Stability of Stochastic Gradient Descent on Nonsmooth Convex Losses}
      \field{year}{2020}
      \strng{urlraw}{https://arxiv.org/abs/2006.06914}
      \strng{url}{https://arxiv.org/abs/2006.06914}
    \endentry
    \entry{CummingsD18}{inproceedings}{}{}
      \name{author}{2}{}{{{hash=865e8602616e91ebdf605d8cdce55167}{family={Cummings},
           familyi={C\bibinitperiod},
           given={Rachel},
           giveni={R\bibinitperiod}}}{{hash=da09ff987339db08a7d6e004c9a976f0}{family={Durfee},
           familyi={D\bibinitperiod},
           given={David},
           giveni={D\bibinitperiod}}}}
      \strng{namehash}{24754ec759467441b21ca65cb581fbf2}
      \strng{fullhash}{24754ec759467441b21ca65cb581fbf2}
      \strng{fullhashraw}{24754ec759467441b21ca65cb581fbf2}
      \strng{bibnamehash}{24754ec759467441b21ca65cb581fbf2}
      \strng{authorbibnamehash}{24754ec759467441b21ca65cb581fbf2}
      \strng{authornamehash}{24754ec759467441b21ca65cb581fbf2}
      \strng{authorfullhash}{24754ec759467441b21ca65cb581fbf2}
      \strng{authorfullhashraw}{24754ec759467441b21ca65cb581fbf2}
      \field{labelalpha}{CD20}
      \field{sortinit}{C}
      \field{sortinithash}{4d103a86280481745c9c897c925753c0}
      \field{labelnamesource}{author}
      \field{labeltitlesource}{title}
      \field{booktitle}{ACM-SIAM Symposium on Discrete Algorithms}
      \field{series}{SODA 2020}
      \field{title}{Individual Sensitivity Preprocessing for Data Privacy}
      \field{year}{2020}
      \field{pages}{528\bibrangedash 547}
      \range{pages}{20}
      \strng{doi}{10.1137/1.9781611975994.32}
      \strng{urlraw}{https://arxiv.org/abs/1804.08645}
      \strng{url}{https://arxiv.org/abs/1804.08645}
    \endentry
    \entry{ChanSS11}{article}{}{}
      \name{author}{3}{}{{{hash=661f9d525966376a6e8864681a9bed7c}{family={Chan},
           familyi={C\bibinitperiod},
           given={T-H\bibnamedelima Hubert},
           giveni={T\bibinithyphendelim H\bibinitperiod\bibinitdelim H\bibinitperiod}}}{{hash=edb728c71cdde40b9afd81bb8c9e733c}{family={Shi},
           familyi={S\bibinitperiod},
           given={Elaine},
           giveni={E\bibinitperiod}}}{{hash=d458d5f9b64652055bd81d57e4777b22}{family={Song},
           familyi={S\bibinitperiod},
           given={Dawn},
           giveni={D\bibinitperiod}}}}
      \strng{namehash}{c6e02cc6a6e26677c92271342d280163}
      \strng{fullhash}{f6824667d3d9aee2e7b18a1d878e2043}
      \strng{fullhashraw}{f6824667d3d9aee2e7b18a1d878e2043}
      \strng{bibnamehash}{f6824667d3d9aee2e7b18a1d878e2043}
      \strng{authorbibnamehash}{f6824667d3d9aee2e7b18a1d878e2043}
      \strng{authornamehash}{c6e02cc6a6e26677c92271342d280163}
      \strng{authorfullhash}{f6824667d3d9aee2e7b18a1d878e2043}
      \strng{authorfullhashraw}{f6824667d3d9aee2e7b18a1d878e2043}
      \field{labelalpha}{CSS11}
      \field{sortinit}{C}
      \field{sortinithash}{4d103a86280481745c9c897c925753c0}
      \field{labelnamesource}{author}
      \field{labeltitlesource}{title}
      \field{journaltitle}{ACM Transactions on Information and System Security (TISSEC)}
      \field{number}{3}
      \field{title}{Private and continual release of statistics}
      \field{volume}{14}
      \field{year}{2011}
      \field{pages}{26:1\bibrangedash 26:24}
      \range{pages}{-1}
      \strng{doi}{10.1145/2043621.2043626}
      \strng{urlraw}{https://eprint.iacr.org/2010/076}
      \strng{url}{https://eprint.iacr.org/2010/076}
    \endentry
    \entry{ChenZ13}{inproceedings}{}{}
      \name{author}{2}{}{{{hash=6b6b6100a7abbc8bbcd917f29c96e8c8}{family={Chen},
           familyi={C\bibinitperiod},
           given={Shixi},
           giveni={S\bibinitperiod}}}{{hash=d3614e56318171231cbd059b7cc112e7}{family={Zhou},
           familyi={Z\bibinitperiod},
           given={Shuigeng},
           giveni={S\bibinitperiod}}}}
      \strng{namehash}{d3334a078bf975f8a7c5298999e8c355}
      \strng{fullhash}{d3334a078bf975f8a7c5298999e8c355}
      \strng{fullhashraw}{d3334a078bf975f8a7c5298999e8c355}
      \strng{bibnamehash}{d3334a078bf975f8a7c5298999e8c355}
      \strng{authorbibnamehash}{d3334a078bf975f8a7c5298999e8c355}
      \strng{authornamehash}{d3334a078bf975f8a7c5298999e8c355}
      \strng{authorfullhash}{d3334a078bf975f8a7c5298999e8c355}
      \strng{authorfullhashraw}{d3334a078bf975f8a7c5298999e8c355}
      \field{labelalpha}{CZ13}
      \field{sortinit}{C}
      \field{sortinithash}{4d103a86280481745c9c897c925753c0}
      \field{labelnamesource}{author}
      \field{labeltitlesource}{title}
      \field{booktitle}{ACM SIGMOD International Conference on Management of Data}
      \field{series}{SIGMOD 2013}
      \field{title}{Recursive mechanism: towards node differential privacy and unrestricted joins}
      \field{year}{2013}
      \field{pages}{653\bibrangedash 664}
      \range{pages}{12}
      \strng{doi}{10.1145/2463676.2465304}
      \strng{urlraw}{https://arxiv.org/abs/1304.4795}
      \strng{url}{https://arxiv.org/abs/1304.4795}
    \endentry
    \entry{DworkL09}{inproceedings}{}{}
      \name{author}{2}{}{{{hash=c6d8866083f4c03af5e37e032df87089}{family={Dwork},
           familyi={D\bibinitperiod},
           given={Cynthia},
           giveni={C\bibinitperiod}}}{{hash=5716652c08e70b34048d4a87afff54d5}{family={Lei},
           familyi={L\bibinitperiod},
           given={Jing},
           giveni={J\bibinitperiod}}}}
      \list{publisher}{1}{{ACM}}
      \strng{namehash}{534a140e3538e0484ba6c08925f2a841}
      \strng{fullhash}{534a140e3538e0484ba6c08925f2a841}
      \strng{fullhashraw}{534a140e3538e0484ba6c08925f2a841}
      \strng{bibnamehash}{534a140e3538e0484ba6c08925f2a841}
      \strng{authorbibnamehash}{534a140e3538e0484ba6c08925f2a841}
      \strng{authornamehash}{534a140e3538e0484ba6c08925f2a841}
      \strng{authorfullhash}{534a140e3538e0484ba6c08925f2a841}
      \strng{authorfullhashraw}{534a140e3538e0484ba6c08925f2a841}
      \field{labelalpha}{DL09}
      \field{sortinit}{D}
      \field{sortinithash}{6f385f66841fb5e82009dc833c761848}
      \field{labelnamesource}{author}
      \field{labeltitlesource}{title}
      \field{booktitle}{Proceedings of the 41st ACM Symposium on Theory of Computing}
      \field{series}{STOC '09}
      \field{title}{Differential privacy and robust statistics}
      \field{year}{2009}
      \field{pages}{371\bibrangedash 380}
      \range{pages}{10}
    \endentry
    \entry{DayLL16}{inproceedings}{}{}
      \name{author}{3}{}{{{hash=e7f583c13c77d1116ec8564128f42070}{family={Day},
           familyi={D\bibinitperiod},
           given={Wei-Yen},
           giveni={W\bibinithyphendelim Y\bibinitperiod}}}{{hash=3ab7a15935d7d7bbbf292c3407adc036}{family={Li},
           familyi={L\bibinitperiod},
           given={Ninghui},
           giveni={N\bibinitperiod}}}{{hash=6a0a3c4b15cfe39205d66d1c299e952a}{family={Lyu},
           familyi={L\bibinitperiod},
           given={Min},
           giveni={M\bibinitperiod}}}}
      \strng{namehash}{aabd212ac0935e86d4d0040db1427166}
      \strng{fullhash}{61b57de3e39070bb8c3cebfefd6300c1}
      \strng{fullhashraw}{61b57de3e39070bb8c3cebfefd6300c1}
      \strng{bibnamehash}{61b57de3e39070bb8c3cebfefd6300c1}
      \strng{authorbibnamehash}{61b57de3e39070bb8c3cebfefd6300c1}
      \strng{authornamehash}{aabd212ac0935e86d4d0040db1427166}
      \strng{authorfullhash}{61b57de3e39070bb8c3cebfefd6300c1}
      \strng{authorfullhashraw}{61b57de3e39070bb8c3cebfefd6300c1}
      \field{labelalpha}{DLL16}
      \field{sortinit}{D}
      \field{sortinithash}{6f385f66841fb5e82009dc833c761848}
      \field{labelnamesource}{author}
      \field{labeltitlesource}{title}
      \field{booktitle}{ACM SIGMOD International Conference on Management of Data}
      \field{series}{SIGMOD 2016}
      \field{title}{Publishing Graph Degree Distribution with Node Differential Privacy}
      \field{year}{2016}
      \field{pages}{123\bibrangedash 138}
      \range{pages}{16}
      \strng{doi}{10.1145/2882903.2926745}
      \strng{urlraw}{https://doi.org/10.1145/2882903.2926745}
      \strng{url}{https://doi.org/10.1145/2882903.2926745}
    \endentry
    \entry{DworkMNS06}{inproceedings}{}{}
      \name{author}{4}{}{{{hash=c6d8866083f4c03af5e37e032df87089}{family={Dwork},
           familyi={D\bibinitperiod},
           given={Cynthia},
           giveni={C\bibinitperiod}}}{{hash=a0e6c99aabdc2587f248892c8bb23f70}{family={McSherry},
           familyi={M\bibinitperiod},
           given={Frank},
           giveni={F\bibinitperiod}}}{{hash=f420df13b9ad6b7a1a9cbcdba96c8381}{family={Nissim},
           familyi={N\bibinitperiod},
           given={Kobbi},
           giveni={K\bibinitperiod}}}{{hash=5ccecf114bbd51a1a2979a23dd69c78c}{family={Smith},
           familyi={S\bibinitperiod},
           given={Adam},
           giveni={A\bibinitperiod}}}}
      \strng{namehash}{f676b30234623eef69e5ed6be9df75b0}
      \strng{fullhash}{77e096c431cfc8d9fee8a2eef9669db1}
      \strng{fullhashraw}{77e096c431cfc8d9fee8a2eef9669db1}
      \strng{bibnamehash}{77e096c431cfc8d9fee8a2eef9669db1}
      \strng{authorbibnamehash}{77e096c431cfc8d9fee8a2eef9669db1}
      \strng{authornamehash}{f676b30234623eef69e5ed6be9df75b0}
      \strng{authorfullhash}{77e096c431cfc8d9fee8a2eef9669db1}
      \strng{authorfullhashraw}{77e096c431cfc8d9fee8a2eef9669db1}
      \field{extraname}{1}
      \field{labelalpha}{DMNS06}
      \field{sortinit}{D}
      \field{sortinithash}{6f385f66841fb5e82009dc833c761848}
      \field{labelnamesource}{author}
      \field{labeltitlesource}{title}
      \field{booktitle}{Theory of Cryptography Conference}
      \field{series}{TCC 2006}
      \field{title}{Calibrating Noise to Sensitivity in Private Data Analysis}
      \field{year}{2006}
      \field{pages}{265\bibrangedash 284}
      \range{pages}{20}
      \strng{urlraw}{https://link.springer.com/chapter/10.1007/11681878_14}
      \strng{url}{https://link.springer.com/chapter/10.1007/11681878_14}
    \endentry
    \entry{DworkNPR10}{inproceedings}{}{}
      \name{author}{4}{}{{{hash=c6d8866083f4c03af5e37e032df87089}{family={Dwork},
           familyi={D\bibinitperiod},
           given={Cynthia},
           giveni={C\bibinitperiod}}}{{hash=c6af39e79e88306281e6e0b0ccea956c}{family={Naor},
           familyi={N\bibinitperiod},
           given={Moni},
           giveni={M\bibinitperiod}}}{{hash=1813007d96ae80284b22fd39a7996df9}{family={Pitassi},
           familyi={P\bibinitperiod},
           given={Toniann},
           giveni={T\bibinitperiod}}}{{hash=8386b9fcc2a9783ef59c9c2cc1fd2396}{family={Rothblum},
           familyi={R\bibinitperiod},
           given={Guy\bibnamedelima N.},
           giveni={G\bibinitperiod\bibinitdelim N\bibinitperiod}}}}
      \strng{namehash}{f676b30234623eef69e5ed6be9df75b0}
      \strng{fullhash}{91f87a6ad79d250452f8f4125568fdff}
      \strng{fullhashraw}{91f87a6ad79d250452f8f4125568fdff}
      \strng{bibnamehash}{91f87a6ad79d250452f8f4125568fdff}
      \strng{authorbibnamehash}{91f87a6ad79d250452f8f4125568fdff}
      \strng{authornamehash}{f676b30234623eef69e5ed6be9df75b0}
      \strng{authorfullhash}{91f87a6ad79d250452f8f4125568fdff}
      \strng{authorfullhashraw}{91f87a6ad79d250452f8f4125568fdff}
      \field{extraname}{2}
      \field{labelalpha}{DNPR10}
      \field{sortinit}{D}
      \field{sortinithash}{6f385f66841fb5e82009dc833c761848}
      \field{labelnamesource}{author}
      \field{labeltitlesource}{title}
      \field{booktitle}{ACM SIGACT Symposium on Theory of Computing}
      \field{series}{STOC 2010}
      \field{title}{Differential Privacy Under Continual Observation}
      \field{year}{2010}
      \field{pages}{715\bibrangedash 724}
      \range{pages}{10}
      \strng{urlraw}{https://www.cs.toronto.edu/~toni/Papers/dp-observation.pdf}
      \strng{url}{https://www.cs.toronto.edu/~toni/Papers/dp-observation.pdf}
    \endentry
    \entry{DworkNRRV09}{inproceedings}{}{}
      \name{author}{5}{}{{{hash=c6d8866083f4c03af5e37e032df87089}{family={Dwork},
           familyi={D\bibinitperiod},
           given={Cynthia},
           giveni={C\bibinitperiod}}}{{hash=c6af39e79e88306281e6e0b0ccea956c}{family={Naor},
           familyi={N\bibinitperiod},
           given={Moni},
           giveni={M\bibinitperiod}}}{{hash=6ff631f4e02261486e4a7d47b7aa7d0b}{family={Reingold},
           familyi={R\bibinitperiod},
           given={Omer},
           giveni={O\bibinitperiod}}}{{hash=8386b9fcc2a9783ef59c9c2cc1fd2396}{family={Rothblum},
           familyi={R\bibinitperiod},
           given={Guy\bibnamedelima N.},
           giveni={G\bibinitperiod\bibinitdelim N\bibinitperiod}}}{{hash=4a7848d2e829c08c64fd0693e3389940}{family={Vadhan},
           familyi={V\bibinitperiod},
           given={Salil\bibnamedelima P.},
           giveni={S\bibinitperiod\bibinitdelim P\bibinitperiod}}}}
      \strng{namehash}{f676b30234623eef69e5ed6be9df75b0}
      \strng{fullhash}{219f66ef6596958cd19a41bbea522ea0}
      \strng{fullhashraw}{219f66ef6596958cd19a41bbea522ea0}
      \strng{bibnamehash}{219f66ef6596958cd19a41bbea522ea0}
      \strng{authorbibnamehash}{219f66ef6596958cd19a41bbea522ea0}
      \strng{authornamehash}{f676b30234623eef69e5ed6be9df75b0}
      \strng{authorfullhash}{219f66ef6596958cd19a41bbea522ea0}
      \strng{authorfullhashraw}{219f66ef6596958cd19a41bbea522ea0}
      \field{extraname}{3}
      \field{labelalpha}{DNRRV09}
      \field{sortinit}{D}
      \field{sortinithash}{6f385f66841fb5e82009dc833c761848}
      \field{labelnamesource}{author}
      \field{labeltitlesource}{title}
      \field{booktitle}{ACM SIGACT Symposium on Theory of Computing}
      \field{series}{STOC 2009}
      \field{title}{On the complexity of differentially private data release: efficient algorithms and hardness results}
      \field{year}{2009}
      \field{pages}{381\bibrangedash 390}
      \range{pages}{10}
      \strng{urlraw}{https://doi.org/10.1145/1536414.1536467}
      \strng{url}{https://doi.org/10.1145/1536414.1536467}
    \endentry
    \entry{DworkR14}{article}{}{}
      \name{author}{2}{}{{{hash=c6d8866083f4c03af5e37e032df87089}{family={Dwork},
           familyi={D\bibinitperiod},
           given={Cynthia},
           giveni={C\bibinitperiod}}}{{hash=9bb323b3c3d2af68def0dcbde8e6795d}{family={Roth},
           familyi={R\bibinitperiod},
           given={Aaron},
           giveni={A\bibinitperiod}}}}
      \strng{namehash}{4d8bf63261aefa96e8725c0b5785f3b3}
      \strng{fullhash}{4d8bf63261aefa96e8725c0b5785f3b3}
      \strng{fullhashraw}{4d8bf63261aefa96e8725c0b5785f3b3}
      \strng{bibnamehash}{4d8bf63261aefa96e8725c0b5785f3b3}
      \strng{authorbibnamehash}{4d8bf63261aefa96e8725c0b5785f3b3}
      \strng{authornamehash}{4d8bf63261aefa96e8725c0b5785f3b3}
      \strng{authorfullhash}{4d8bf63261aefa96e8725c0b5785f3b3}
      \strng{authorfullhashraw}{4d8bf63261aefa96e8725c0b5785f3b3}
      \field{labelalpha}{DR14}
      \field{sortinit}{D}
      \field{sortinithash}{6f385f66841fb5e82009dc833c761848}
      \field{labelnamesource}{author}
      \field{labeltitlesource}{title}
      \field{journaltitle}{Foundations and Trends in Theoretical Computer Science}
      \field{number}{3-4}
      \field{title}{The algorithmic foundations of differential privacy}
      \field{volume}{9}
      \field{year}{2014}
      \field{pages}{211\bibrangedash 407}
      \range{pages}{197}
      \strng{urlraw}{https://www.microsoft.com/en-us/research/publication/algorithmic-foundations-differential-privacy/}
      \strng{url}{https://www.microsoft.com/en-us/research/publication/algorithmic-foundations-differential-privacy/}
    \endentry
    \entry{DongRS19}{article}{}{}
      \name{author}{3}{}{{{hash=359770e50fa5b6848ef22b3e1c50038f}{family={Dong},
           familyi={D\bibinitperiod},
           given={Jinshuo},
           giveni={J\bibinitperiod}}}{{hash=9bb323b3c3d2af68def0dcbde8e6795d}{family={Roth},
           familyi={R\bibinitperiod},
           given={Aaron},
           giveni={A\bibinitperiod}}}{{hash=87ddac44623b8bb57fc24211977df985}{family={Su},
           familyi={S\bibinitperiod},
           given={Weijie\bibnamedelima J.},
           giveni={W\bibinitperiod\bibinitdelim J\bibinitperiod}}}}
      \strng{namehash}{930434a4e5ef1c26059f3723b203d15c}
      \strng{fullhash}{0d1df7edf815910807e0d8bddb030bdb}
      \strng{fullhashraw}{0d1df7edf815910807e0d8bddb030bdb}
      \strng{bibnamehash}{0d1df7edf815910807e0d8bddb030bdb}
      \strng{authorbibnamehash}{0d1df7edf815910807e0d8bddb030bdb}
      \strng{authornamehash}{930434a4e5ef1c26059f3723b203d15c}
      \strng{authorfullhash}{0d1df7edf815910807e0d8bddb030bdb}
      \strng{authorfullhashraw}{0d1df7edf815910807e0d8bddb030bdb}
      \field{labelalpha}{DRS22}
      \field{sortinit}{D}
      \field{sortinithash}{6f385f66841fb5e82009dc833c761848}
      \field{labelnamesource}{author}
      \field{labeltitlesource}{title}
      \field{journaltitle}{Journal of the Royal Statistical Society Series B: Statistical Methodology}
      \field{number}{1}
      \field{title}{{Gaussian} differential privacy}
      \field{volume}{84}
      \field{year}{2022}
      \field{pages}{3\bibrangedash 37}
      \range{pages}{35}
      \strng{doi}{10.1111/rssb.12454}
      \strng{urlraw}{https://arxiv.org/abs/1905.02383}
      \strng{url}{https://arxiv.org/abs/1905.02383}
    \endentry
    \entry{DworkRV10}{inproceedings}{}{}
      \name{author}{3}{}{{{hash=c6d8866083f4c03af5e37e032df87089}{family={Dwork},
           familyi={D\bibinitperiod},
           given={Cynthia},
           giveni={C\bibinitperiod}}}{{hash=8386b9fcc2a9783ef59c9c2cc1fd2396}{family={Rothblum},
           familyi={R\bibinitperiod},
           given={Guy\bibnamedelima N.},
           giveni={G\bibinitperiod\bibinitdelim N\bibinitperiod}}}{{hash=4a7848d2e829c08c64fd0693e3389940}{family={Vadhan},
           familyi={V\bibinitperiod},
           given={Salil\bibnamedelima P.},
           giveni={S\bibinitperiod\bibinitdelim P\bibinitperiod}}}}
      \strng{namehash}{f676b30234623eef69e5ed6be9df75b0}
      \strng{fullhash}{c4e1c58f3dc7bef0cc711634d94bb080}
      \strng{fullhashraw}{c4e1c58f3dc7bef0cc711634d94bb080}
      \strng{bibnamehash}{c4e1c58f3dc7bef0cc711634d94bb080}
      \strng{authorbibnamehash}{c4e1c58f3dc7bef0cc711634d94bb080}
      \strng{authornamehash}{f676b30234623eef69e5ed6be9df75b0}
      \strng{authorfullhash}{c4e1c58f3dc7bef0cc711634d94bb080}
      \strng{authorfullhashraw}{c4e1c58f3dc7bef0cc711634d94bb080}
      \field{extraname}{4}
      \field{labelalpha}{DRV10}
      \field{sortinit}{D}
      \field{sortinithash}{6f385f66841fb5e82009dc833c761848}
      \field{labelnamesource}{author}
      \field{labeltitlesource}{title}
      \field{booktitle}{IEEE Symposium on Foundations of Computer Science}
      \field{series}{FOCS 2010}
      \field{title}{Boosting and Differential Privacy}
      \field{year}{2010}
      \field{pages}{51\bibrangedash 60}
      \range{pages}{10}
      \strng{doi}{10.1109/FOCS.2010.12}
      \strng{urlraw}{https://privacytools.seas.harvard.edu/publications/boosting-and-differential-privacy}
      \strng{url}{https://privacytools.seas.harvard.edu/publications/boosting-and-differential-privacy}
    \endentry
    \entry{FangDY22}{inproceedings}{}{}
      \name{author}{3}{}{{{hash=ac5f88df285b26636eb3776af0cd2ffc}{family={Fang},
           familyi={F\bibinitperiod},
           given={Juanru},
           giveni={J\bibinitperiod}}}{{hash=7d87c5957b07153c7f18918b92830bf8}{family={Dong},
           familyi={D\bibinitperiod},
           given={Wei},
           giveni={W\bibinitperiod}}}{{hash=91ceec99caada2f2587e7657c721f3f5}{family={Yi},
           familyi={Y\bibinitperiod},
           given={Ke},
           giveni={K\bibinitperiod}}}}
      \strng{namehash}{69b56b6a2f594356748df8fb8a5dbaa1}
      \strng{fullhash}{f2a91620eadc2cb286fd9c8de3967273}
      \strng{fullhashraw}{f2a91620eadc2cb286fd9c8de3967273}
      \strng{bibnamehash}{f2a91620eadc2cb286fd9c8de3967273}
      \strng{authorbibnamehash}{f2a91620eadc2cb286fd9c8de3967273}
      \strng{authornamehash}{69b56b6a2f594356748df8fb8a5dbaa1}
      \strng{authorfullhash}{f2a91620eadc2cb286fd9c8de3967273}
      \strng{authorfullhashraw}{f2a91620eadc2cb286fd9c8de3967273}
      \field{labelalpha}{FDY22}
      \field{sortinit}{F}
      \field{sortinithash}{2638baaa20439f1b5a8f80c6c08a13b4}
      \field{labelnamesource}{author}
      \field{labeltitlesource}{title}
      \field{booktitle}{ACM SIGSAC Conference on Computer and Communications Security}
      \field{series}{CCS 2022}
      \field{title}{Shifted Inverse: {A} General Mechanism for Monotonic Functions under User Differential Privacy}
      \field{year}{2022}
      \field{pages}{1009\bibrangedash 1022}
      \range{pages}{14}
      \strng{doi}{10.1145/3548606.3560567}
      \strng{urlraw}{https://doi.org/10.1145/3548606.3560567}
      \strng{url}{https://doi.org/10.1145/3548606.3560567}
    \endentry
    \entry{GaneshMMWS25}{inproceedings}{}{}
      \name{author}{5}{}{{{hash=0d5c936d18190e3def9f03421604e026}{family={Ganesh},
           familyi={G\bibinitperiod},
           given={Arun},
           giveni={A\bibinitperiod}}}{{hash=f45d63f842a55e478c4519c415500aa0}{family={McKenna},
           familyi={M\bibinitperiod},
           given={Ryan},
           giveni={R\bibinitperiod}}}{{hash=707fa44f08f047ab822f84fe38c69dfb}{family={McMahan},
           familyi={M\bibinitperiod},
           given={H.\bibnamedelimi Brendan},
           giveni={H\bibinitperiod\bibinitdelim B\bibinitperiod}}}{{hash=5ccecf114bbd51a1a2979a23dd69c78c}{family={Smith},
           familyi={S\bibinitperiod},
           given={Adam},
           giveni={A\bibinitperiod}}}{{hash=86756d7b8cb7311613b1c58712b999b3}{family={Wu},
           familyi={W\bibinitperiod},
           given={Fan},
           giveni={F\bibinitperiod}}}}
      \strng{namehash}{6a125f34cdf2c8d18ce244bfc91d8eb4}
      \strng{fullhash}{820406598a22e894e6098b8624e63b0a}
      \strng{fullhashraw}{820406598a22e894e6098b8624e63b0a}
      \strng{bibnamehash}{820406598a22e894e6098b8624e63b0a}
      \strng{authorbibnamehash}{820406598a22e894e6098b8624e63b0a}
      \strng{authornamehash}{6a125f34cdf2c8d18ce244bfc91d8eb4}
      \strng{authorfullhash}{820406598a22e894e6098b8624e63b0a}
      \strng{authorfullhashraw}{820406598a22e894e6098b8624e63b0a}
      \field{labelalpha}{GMMSW25}
      \field{sortinit}{G}
      \field{sortinithash}{32d67eca0634bf53703493fb1090a2e8}
      \field{labelnamesource}{author}
      \field{labeltitlesource}{title}
      \field{booktitle}{International Conference on Machine Learning}
      \field{series}{ICML 2025}
      \field{title}{It's My Data Too: Private {ML} for Datasets with Multi-User Training Examples}
      \field{year}{2025}
      \strng{urlraw}{https://arxiv.org/abs/2503.03622}
      \strng{url}{https://arxiv.org/abs/2503.03622}
    \endentry
    \entry{HayLMJ09}{inproceedings}{}{}
      \name{author}{4}{}{{{hash=a80c8d292c85e81bcde88f81c4b25204}{family={Hay},
           familyi={H\bibinitperiod},
           given={Michael},
           giveni={M\bibinitperiod}}}{{hash=ad699cc0ff44aa8ef2a872dab8e20175}{family={Li},
           familyi={L\bibinitperiod},
           given={Chao},
           giveni={C\bibinitperiod}}}{{hash=53aeb6a90e2fe9e3d137fec53080f31a}{family={Miklau},
           familyi={M\bibinitperiod},
           given={Gerome},
           giveni={G\bibinitperiod}}}{{hash=cd7b617c3b1a4aa8fac0ada3eeb6e6c2}{family={Jensen},
           familyi={J\bibinitperiod},
           given={David\bibnamedelima D.},
           giveni={D\bibinitperiod\bibinitdelim D\bibinitperiod}}}}
      \strng{namehash}{ab9e8370aa23a555e25350a7ca43a7a9}
      \strng{fullhash}{8305d024882e0a97e13cda58c5adcd7f}
      \strng{fullhashraw}{8305d024882e0a97e13cda58c5adcd7f}
      \strng{bibnamehash}{8305d024882e0a97e13cda58c5adcd7f}
      \strng{authorbibnamehash}{8305d024882e0a97e13cda58c5adcd7f}
      \strng{authornamehash}{ab9e8370aa23a555e25350a7ca43a7a9}
      \strng{authorfullhash}{8305d024882e0a97e13cda58c5adcd7f}
      \strng{authorfullhashraw}{8305d024882e0a97e13cda58c5adcd7f}
      \field{labelalpha}{HLMJ09}
      \field{sortinit}{H}
      \field{sortinithash}{23a3aa7c24e56cfa16945d55545109b5}
      \field{labelnamesource}{author}
      \field{labeltitlesource}{title}
      \field{booktitle}{IEEE International Conference on Data Mining}
      \field{series}{ICDM 2009}
      \field{title}{Accurate Estimation of the Degree Distribution of Private Networks}
      \field{year}{2009}
      \field{pages}{169\bibrangedash 178}
      \range{pages}{10}
      \strng{urlraw}{https://www.cs.colgate.edu/~mhay/assets/publications/hay2009accurate.pdf}
      \strng{url}{https://www.cs.colgate.edu/~mhay/assets/publications/hay2009accurate.pdf}
    \endentry
    \entry{HardtR10}{inproceedings}{}{}
      \name{author}{2}{}{{{hash=e771760b6d0d33f8b4dfd907d8d57ac2}{family={Hardt},
           familyi={H\bibinitperiod},
           given={Moritz},
           giveni={M\bibinitperiod}}}{{hash=f106b13a5fc7b67798625e67910f0fc2}{family={Rothblum},
           familyi={R\bibinitperiod},
           given={Guy},
           giveni={G\bibinitperiod}}}}
      \strng{namehash}{b4a07784ed256c62adcc25fb9278f947}
      \strng{fullhash}{b4a07784ed256c62adcc25fb9278f947}
      \strng{fullhashraw}{b4a07784ed256c62adcc25fb9278f947}
      \strng{bibnamehash}{b4a07784ed256c62adcc25fb9278f947}
      \strng{authorbibnamehash}{b4a07784ed256c62adcc25fb9278f947}
      \strng{authornamehash}{b4a07784ed256c62adcc25fb9278f947}
      \strng{authorfullhash}{b4a07784ed256c62adcc25fb9278f947}
      \strng{authorfullhashraw}{b4a07784ed256c62adcc25fb9278f947}
      \field{labelalpha}{HR10}
      \field{sortinit}{H}
      \field{sortinithash}{23a3aa7c24e56cfa16945d55545109b5}
      \field{labelnamesource}{author}
      \field{labeltitlesource}{title}
      \field{booktitle}{IEEE Symposium on Foundations of Computer Science}
      \field{series}{FOCS 2010}
      \field{title}{A Multiplicative Weights Mechanism for Privacy-Preserving Data Analysis}
      \field{year}{2010}
      \field{pages}{61\bibrangedash 70}
      \range{pages}{10}
      \strng{urlraw}{https://www.microsoft.com/en-us/research/publication/a-multiplicative-weights-mechanism-for-privacy-preserving-data-analysis/}
      \strng{url}{https://www.microsoft.com/en-us/research/publication/a-multiplicative-weights-mechanism-for-privacy-preserving-data-analysis/}
    \endentry
    \entry{HenzingerSV26}{article}{}{}
      \name{author}{3}{}{{{hash=1ac7ee7edcf3a1101028c3e28232767c}{family={Henzinger},
           familyi={H\bibinitperiod},
           given={Monika},
           giveni={M\bibinitperiod}}}{{hash=6ef56145c69ac6239f0a0941958e970b}{family={Safavi},
           familyi={S\bibinitperiod},
           given={Roodabeh},
           giveni={R\bibinitperiod}}}{{hash=c915d3380da8c4288e9634bbee5ba983}{family={Vadhan},
           familyi={V\bibinitperiod},
           given={Salil},
           giveni={S\bibinitperiod}}}}
      \strng{namehash}{121a0ee9db1b1637751bff9e845727dd}
      \strng{fullhash}{12ea5fd48d84f01a9c22b1d3d9f9f12c}
      \strng{fullhashraw}{12ea5fd48d84f01a9c22b1d3d9f9f12c}
      \strng{bibnamehash}{12ea5fd48d84f01a9c22b1d3d9f9f12c}
      \strng{authorbibnamehash}{12ea5fd48d84f01a9c22b1d3d9f9f12c}
      \strng{authornamehash}{121a0ee9db1b1637751bff9e845727dd}
      \strng{authorfullhash}{12ea5fd48d84f01a9c22b1d3d9f9f12c}
      \strng{authorfullhashraw}{12ea5fd48d84f01a9c22b1d3d9f9f12c}
      \field{labelalpha}{HSV26}
      \field{sortinit}{H}
      \field{sortinithash}{23a3aa7c24e56cfa16945d55545109b5}
      \field{labelnamesource}{author}
      \field{labeltitlesource}{title}
      \field{eid}{99}
      \field{journaltitle}{Proceedings of the ACM on Management of Data}
      \field{number}{2}
      \field{title}{Concurrent Composition for Differentially Private Continual Mechanisms}
      \field{volume}{4}
      \field{year}{2026}
      \field{pages}{99:1\bibrangedash 99:26}
      \range{pages}{-1}
      \strng{doi}{10.1145/3801895}
      \strng{urlraw}{https://arxiv.org/abs/2411.03299}
      \strng{url}{https://arxiv.org/abs/2411.03299}
    \endentry
    \entry{JerniganM09}{article}{}{}
      \name{author}{2}{}{{{hash=8b57d14d57b8263a5c1783fae0250cb3}{family={Jernigan},
           familyi={J\bibinitperiod},
           given={Carter},
           giveni={C\bibinitperiod}}}{{hash=245ae56416bea8830b6030a28a919fca}{family={Mistree},
           familyi={M\bibinitperiod},
           given={Behram\bibnamedelimb F.\bibnamedelimi T.},
           giveni={B\bibinitperiod\bibinitdelim F\bibinitperiod\bibinitdelim T\bibinitperiod}}}}
      \strng{namehash}{fc32e49018318ff83d4bb491d08115be}
      \strng{fullhash}{fc32e49018318ff83d4bb491d08115be}
      \strng{fullhashraw}{fc32e49018318ff83d4bb491d08115be}
      \strng{bibnamehash}{fc32e49018318ff83d4bb491d08115be}
      \strng{authorbibnamehash}{fc32e49018318ff83d4bb491d08115be}
      \strng{authornamehash}{fc32e49018318ff83d4bb491d08115be}
      \strng{authorfullhash}{fc32e49018318ff83d4bb491d08115be}
      \strng{authorfullhashraw}{fc32e49018318ff83d4bb491d08115be}
      \field{labelalpha}{JM09}
      \field{sortinit}{J}
      \field{sortinithash}{b2f54a9081ace9966a7cb9413811edb4}
      \field{labelnamesource}{author}
      \field{labeltitlesource}{title}
      \field{journaltitle}{First Monday}
      \field{number}{10}
      \field{title}{{Gaydar: Facebook Friendships Expose Sexual Orientation}}
      \field{volume}{14}
      \field{year}{2009}
      \strng{urlraw}{https://doi.org/10.5210/fm.v14i10.2611}
      \strng{url}{https://doi.org/10.5210/fm.v14i10.2611}
    \endentry
    \entry{JainRSS23}{inproceedings}{}{}
      \name{author}{4}{}{{{hash=150cb2d2b437ff25ca436a698053fd17}{family={Jain},
           familyi={J\bibinitperiod},
           given={Palak},
           giveni={P\bibinitperiod}}}{{hash=595cb420436ece2997a0f6b90fa4089f}{family={Raskhodnikova},
           familyi={R\bibinitperiod},
           given={Sofya},
           giveni={S\bibinitperiod}}}{{hash=78e92c8d4de98ab5c0ca334b16e7ba68}{family={Sivakumar},
           familyi={S\bibinitperiod},
           given={Satchit},
           giveni={S\bibinitperiod}}}{{hash=5ccecf114bbd51a1a2979a23dd69c78c}{family={Smith},
           familyi={S\bibinitperiod},
           given={Adam},
           giveni={A\bibinitperiod}}}}
      \strng{namehash}{5e383600db37bd143f7ba51afab43704}
      \strng{fullhash}{dc102c151e1b6a6028acb458cd3ecd9a}
      \strng{fullhashraw}{dc102c151e1b6a6028acb458cd3ecd9a}
      \strng{bibnamehash}{dc102c151e1b6a6028acb458cd3ecd9a}
      \strng{authorbibnamehash}{dc102c151e1b6a6028acb458cd3ecd9a}
      \strng{authornamehash}{5e383600db37bd143f7ba51afab43704}
      \strng{authorfullhash}{dc102c151e1b6a6028acb458cd3ecd9a}
      \strng{authorfullhashraw}{dc102c151e1b6a6028acb458cd3ecd9a}
      \field{extraname}{1}
      \field{labelalpha}{JRSS23}
      \field{sortinit}{J}
      \field{sortinithash}{b2f54a9081ace9966a7cb9413811edb4}
      \field{labelnamesource}{author}
      \field{labeltitlesource}{title}
      \field{booktitle}{International Conference on Machine Learning}
      \field{series}{ICML 2023}
      \field{title}{The Price of Differential Privacy under Continual Observation}
      \field{year}{2023}
      \field{pages}{14654\bibrangedash 14678}
      \range{pages}{25}
      \strng{urlraw}{https://arxiv.org/abs/2112.00828}
      \strng{url}{https://arxiv.org/abs/2112.00828}
    \endentry
    \entry{JainSW24}{inproceedings}{}{}
      \name{author}{3}{}{{{hash=150cb2d2b437ff25ca436a698053fd17}{family={Jain},
           familyi={J\bibinitperiod},
           given={Palak},
           giveni={P\bibinitperiod}}}{{hash=5ccecf114bbd51a1a2979a23dd69c78c}{family={Smith},
           familyi={S\bibinitperiod},
           given={Adam},
           giveni={A\bibinitperiod}}}{{hash=f53d4350b43224b6d37785e4d10bc981}{family={Wagaman},
           familyi={W\bibinitperiod},
           given={Connor},
           giveni={C\bibinitperiod}}}}
      \strng{namehash}{5e383600db37bd143f7ba51afab43704}
      \strng{fullhash}{3f2a14b757a47c43d2b4531b9425b646}
      \strng{fullhashraw}{3f2a14b757a47c43d2b4531b9425b646}
      \strng{bibnamehash}{3f2a14b757a47c43d2b4531b9425b646}
      \strng{authorbibnamehash}{3f2a14b757a47c43d2b4531b9425b646}
      \strng{authornamehash}{5e383600db37bd143f7ba51afab43704}
      \strng{authorfullhash}{3f2a14b757a47c43d2b4531b9425b646}
      \strng{authorfullhashraw}{3f2a14b757a47c43d2b4531b9425b646}
      \field{extraname}{2}
      \field{labelalpha}{JSW24}
      \field{sortinit}{J}
      \field{sortinithash}{b2f54a9081ace9966a7cb9413811edb4}
      \field{labelnamesource}{author}
      \field{labeltitlesource}{title}
      \field{booktitle}{IEEE Symposium on Security and Privacy}
      \field{series}{IEEE S\&P 2024}
      \field{title}{Time-Aware Projections: Truly Node-Private Graph Statistics under Continual Observation}
      \field{year}{2024}
      \strng{doi}{10.1109/SP54263.2024.00196}
      \strng{urlraw}{https://arxiv.org/abs/2403.04630}
      \strng{url}{https://arxiv.org/abs/2403.04630}
    \endentry
    \entry{KasiviswanathanNRS13}{inproceedings}{}{}
      \name{author}{4}{}{{{hash=07b9a8a032f65e512da346e8b71786b4}{family={Kasiviswanathan},
           familyi={K\bibinitperiod},
           given={Shiva\bibnamedelima P.},
           giveni={S\bibinitperiod\bibinitdelim P\bibinitperiod}}}{{hash=f420df13b9ad6b7a1a9cbcdba96c8381}{family={Nissim},
           familyi={N\bibinitperiod},
           given={Kobbi},
           giveni={K\bibinitperiod}}}{{hash=595cb420436ece2997a0f6b90fa4089f}{family={Raskhodnikova},
           familyi={R\bibinitperiod},
           given={Sofya},
           giveni={S\bibinitperiod}}}{{hash=5ccecf114bbd51a1a2979a23dd69c78c}{family={Smith},
           familyi={S\bibinitperiod},
           given={Adam},
           giveni={A\bibinitperiod}}}}
      \strng{namehash}{910521e44fb75252a4fa203499886f74}
      \strng{fullhash}{c5de6ef5d543f36fad05b00defe1d7fb}
      \strng{fullhashraw}{c5de6ef5d543f36fad05b00defe1d7fb}
      \strng{bibnamehash}{c5de6ef5d543f36fad05b00defe1d7fb}
      \strng{authorbibnamehash}{c5de6ef5d543f36fad05b00defe1d7fb}
      \strng{authornamehash}{910521e44fb75252a4fa203499886f74}
      \strng{authorfullhash}{c5de6ef5d543f36fad05b00defe1d7fb}
      \strng{authorfullhashraw}{c5de6ef5d543f36fad05b00defe1d7fb}
      \field{labelalpha}{KNRS13}
      \field{sortinit}{K}
      \field{sortinithash}{c02bf6bff1c488450c352b40f5d853ab}
      \field{labelnamesource}{author}
      \field{labeltitlesource}{title}
      \field{booktitle}{Theory of Cryptography Conference}
      \field{series}{TCC 2013}
      \field{title}{Analyzing Graphs with Node Differential Privacy}
      \field{year}{2013}
      \field{pages}{457\bibrangedash 476}
      \range{pages}{20}
      \strng{urlraw}{https://www.iacr.org/archive/tcc2013/77850455/77850455.pdf}
      \strng{url}{https://www.iacr.org/archive/tcc2013/77850455/77850455.pdf}
    \endentry
    \entry{KalemajRST23}{inproceedings}{}{}
      \name{author}{4}{}{{{hash=b05eca50dcc31e4183471fe8b187df97}{family={Kalemaj},
           familyi={K\bibinitperiod},
           given={Iden},
           giveni={I\bibinitperiod}}}{{hash=595cb420436ece2997a0f6b90fa4089f}{family={Raskhodnikova},
           familyi={R\bibinitperiod},
           given={Sofya},
           giveni={S\bibinitperiod}}}{{hash=de399363c3a749f51200c9a48984656f}{family={Smith},
           familyi={S\bibinitperiod},
           given={Adam\bibnamedelima D.},
           giveni={A\bibinitperiod\bibinitdelim D\bibinitperiod}}}{{hash=ad596aaa3fe4ccdb67388ccb833a8666}{family={Tsourakakis},
           familyi={T\bibinitperiod},
           given={Charalampos\bibnamedelima E.},
           giveni={C\bibinitperiod\bibinitdelim E\bibinitperiod}}}}
      \strng{namehash}{fde4060760b1f385cc98128789793be5}
      \strng{fullhash}{9a98336cb421b697990ff11bf50dc3ec}
      \strng{fullhashraw}{9a98336cb421b697990ff11bf50dc3ec}
      \strng{bibnamehash}{9a98336cb421b697990ff11bf50dc3ec}
      \strng{authorbibnamehash}{9a98336cb421b697990ff11bf50dc3ec}
      \strng{authornamehash}{fde4060760b1f385cc98128789793be5}
      \strng{authorfullhash}{9a98336cb421b697990ff11bf50dc3ec}
      \strng{authorfullhashraw}{9a98336cb421b697990ff11bf50dc3ec}
      \field{labelalpha}{KRST23}
      \field{sortinit}{K}
      \field{sortinithash}{c02bf6bff1c488450c352b40f5d853ab}
      \field{labelnamesource}{author}
      \field{labeltitlesource}{title}
      \field{booktitle}{ACM SIGMOD-SIGACT-SIGAI Symposium on Principles of Database Systems}
      \field{series}{PODS 2023}
      \field{title}{Node-Differentially Private Estimation of the Number of Connected Components}
      \field{year}{2023}
      \field{pages}{183\bibrangedash 194}
      \range{pages}{12}
      \strng{doi}{10.1145/3584372.3588671}
      \strng{urlraw}{https://arxiv.org/abs/2304.05890}
      \strng{url}{https://arxiv.org/abs/2304.05890}
    \endentry
    \entry{LinderRSS25}{inproceedings}{}{}
      \name{author}{4}{}{{{hash=7e7c3633d5e014986bbfdab263eb4a58}{family={Linder},
           familyi={L\bibinitperiod},
           given={Ephraim},
           giveni={E\bibinitperiod}}}{{hash=595cb420436ece2997a0f6b90fa4089f}{family={Raskhodnikova},
           familyi={R\bibinitperiod},
           given={Sofya},
           giveni={S\bibinitperiod}}}{{hash=5ccecf114bbd51a1a2979a23dd69c78c}{family={Smith},
           familyi={S\bibinitperiod},
           given={Adam},
           giveni={A\bibinitperiod}}}{{hash=a8ee507f496e43705d0c170feeb81861}{family={Steinke},
           familyi={S\bibinitperiod},
           given={Thomas},
           giveni={T\bibinitperiod}}}}
      \strng{namehash}{b78e1a89eb17d12d6fff5026cca06f6f}
      \strng{fullhash}{a558cadf60db480167eeb8cdf501f09c}
      \strng{fullhashraw}{a558cadf60db480167eeb8cdf501f09c}
      \strng{bibnamehash}{a558cadf60db480167eeb8cdf501f09c}
      \strng{authorbibnamehash}{a558cadf60db480167eeb8cdf501f09c}
      \strng{authornamehash}{b78e1a89eb17d12d6fff5026cca06f6f}
      \strng{authorfullhash}{a558cadf60db480167eeb8cdf501f09c}
      \strng{authorfullhashraw}{a558cadf60db480167eeb8cdf501f09c}
      \field{labelalpha}{LRSS25}
      \field{sortinit}{L}
      \field{sortinithash}{7c47d417cecb1f4bd38d1825c427a61a}
      \field{labelnamesource}{author}
      \field{labeltitlesource}{title}
      \field{booktitle}{ACM Symposium on Theory of Computing}
      \field{series}{STOC 2025}
      \field{title}{Privately Evaluating Untrusted Black-Box Functions}
      \field{year}{2025}
      \field{pages}{2350\bibrangedash 2361}
      \range{pages}{12}
      \strng{doi}{10.1145/3717823.3718247}
      \strng{urlraw}{https://arxiv.org/abs/2503.19268}
      \strng{url}{https://arxiv.org/abs/2503.19268}
    \endentry
    \entry{LyuSL17}{article}{}{}
      \name{author}{3}{}{{{hash=6a0a3c4b15cfe39205d66d1c299e952a}{family={Lyu},
           familyi={L\bibinitperiod},
           given={Min},
           giveni={M\bibinitperiod}}}{{hash=b5b2027bcfd3d18f5afb62ce44957714}{family={Su},
           familyi={S\bibinitperiod},
           given={Dong},
           giveni={D\bibinitperiod}}}{{hash=3ab7a15935d7d7bbbf292c3407adc036}{family={Li},
           familyi={L\bibinitperiod},
           given={Ninghui},
           giveni={N\bibinitperiod}}}}
      \strng{namehash}{a9ed3ccf82d800509e57cd6272f9c1fe}
      \strng{fullhash}{32b0d0dffc5725356fc5504712fdeab3}
      \strng{fullhashraw}{32b0d0dffc5725356fc5504712fdeab3}
      \strng{bibnamehash}{32b0d0dffc5725356fc5504712fdeab3}
      \strng{authorbibnamehash}{32b0d0dffc5725356fc5504712fdeab3}
      \strng{authornamehash}{a9ed3ccf82d800509e57cd6272f9c1fe}
      \strng{authorfullhash}{32b0d0dffc5725356fc5504712fdeab3}
      \strng{authorfullhashraw}{32b0d0dffc5725356fc5504712fdeab3}
      \field{labelalpha}{LSL17}
      \field{sortinit}{L}
      \field{sortinithash}{7c47d417cecb1f4bd38d1825c427a61a}
      \field{labelnamesource}{author}
      \field{labeltitlesource}{title}
      \field{journaltitle}{Proceedings of the {VLDB} Endowment}
      \field{number}{6}
      \field{title}{Understanding the Sparse Vector Technique for Differential Privacy}
      \field{volume}{10}
      \field{year}{2017}
      \field{pages}{637\bibrangedash 648}
      \range{pages}{12}
      \strng{doi}{10.14778/3055330.3055331}
      \strng{urlraw}{http://www.vldb.org/pvldb/vol10/p637-lyu.pdf}
      \strng{url}{http://www.vldb.org/pvldb/vol10/p637-lyu.pdf}
    \endentry
    \entry{LyubarskiiV10}{article}{}{}
      \name{author}{2}{}{{{hash=1a1ccff9637bfc36ff93dde486d583d2}{family={Lyubarskii},
           familyi={L\bibinitperiod},
           given={Yurii},
           giveni={Y\bibinitperiod}}}{{hash=711c5aa2fe76a4ff674a2c6e2c7a2d91}{family={Vershynin},
           familyi={V\bibinitperiod},
           given={Roman},
           giveni={R\bibinitperiod}}}}
      \strng{namehash}{eb7c88a9f1f9f05747dd9de6ed8c22a6}
      \strng{fullhash}{eb7c88a9f1f9f05747dd9de6ed8c22a6}
      \strng{fullhashraw}{eb7c88a9f1f9f05747dd9de6ed8c22a6}
      \strng{bibnamehash}{eb7c88a9f1f9f05747dd9de6ed8c22a6}
      \strng{authorbibnamehash}{eb7c88a9f1f9f05747dd9de6ed8c22a6}
      \strng{authornamehash}{eb7c88a9f1f9f05747dd9de6ed8c22a6}
      \strng{authorfullhash}{eb7c88a9f1f9f05747dd9de6ed8c22a6}
      \strng{authorfullhashraw}{eb7c88a9f1f9f05747dd9de6ed8c22a6}
      \field{labelalpha}{LV10}
      \field{sortinit}{L}
      \field{sortinithash}{7c47d417cecb1f4bd38d1825c427a61a}
      \field{labelnamesource}{author}
      \field{labeltitlesource}{title}
      \field{journaltitle}{IEEE Transactions on Information Theory}
      \field{number}{7}
      \field{title}{Uncertainty principles and vector quantization}
      \field{volume}{56}
      \field{year}{2010}
      \field{pages}{3491\bibrangedash 3501}
      \range{pages}{11}
      \strng{urlraw}{https://arxiv.org/abs/math/0611343}
      \strng{url}{https://arxiv.org/abs/math/0611343}
    \endentry
    \entry{McShane34}{article}{}{}
      \name{author}{1}{}{{{hash=b67f867fdcbeb4351fda3979bc4b709d}{family={McShane},
           familyi={M\bibinitperiod},
           given={E.\bibnamedelimi J.},
           giveni={E\bibinitperiod\bibinitdelim J\bibinitperiod}}}}
      \strng{namehash}{b67f867fdcbeb4351fda3979bc4b709d}
      \strng{fullhash}{b67f867fdcbeb4351fda3979bc4b709d}
      \strng{fullhashraw}{b67f867fdcbeb4351fda3979bc4b709d}
      \strng{bibnamehash}{b67f867fdcbeb4351fda3979bc4b709d}
      \strng{authorbibnamehash}{b67f867fdcbeb4351fda3979bc4b709d}
      \strng{authornamehash}{b67f867fdcbeb4351fda3979bc4b709d}
      \strng{authorfullhash}{b67f867fdcbeb4351fda3979bc4b709d}
      \strng{authorfullhashraw}{b67f867fdcbeb4351fda3979bc4b709d}
      \field{labelalpha}{McS34}
      \field{sortinit}{M}
      \field{sortinithash}{4625c616857f13d17ce56f7d4f97d451}
      \field{labelnamesource}{author}
      \field{labeltitlesource}{title}
      \field{journaltitle}{Bulletin of the American Mathematical Society}
      \field{number}{12}
      \field{title}{Extension of range of functions}
      \field{volume}{40}
      \field{year}{1934}
      \field{pages}{837\bibrangedash 842}
      \range{pages}{6}
      \strng{doi}{10.1090/S0002-9904-1934-05978-0}
      \strng{urlraw}{https://doi.org/10.1090/S0002-9904-1934-05978-0}
      \strng{url}{https://doi.org/10.1090/S0002-9904-1934-05978-0}
    \endentry
    \entry{NissimRS07}{inproceedings}{}{}
      \name{author}{3}{}{{{hash=f420df13b9ad6b7a1a9cbcdba96c8381}{family={Nissim},
           familyi={N\bibinitperiod},
           given={Kobbi},
           giveni={K\bibinitperiod}}}{{hash=595cb420436ece2997a0f6b90fa4089f}{family={Raskhodnikova},
           familyi={R\bibinitperiod},
           given={Sofya},
           giveni={S\bibinitperiod}}}{{hash=5ccecf114bbd51a1a2979a23dd69c78c}{family={Smith},
           familyi={S\bibinitperiod},
           given={Adam},
           giveni={A\bibinitperiod}}}}
      \strng{namehash}{0d03418b9d3d3553c60775874cd4ab22}
      \strng{fullhash}{f421067e62fbcac3e0219c59e1f88613}
      \strng{fullhashraw}{f421067e62fbcac3e0219c59e1f88613}
      \strng{bibnamehash}{f421067e62fbcac3e0219c59e1f88613}
      \strng{authorbibnamehash}{f421067e62fbcac3e0219c59e1f88613}
      \strng{authornamehash}{0d03418b9d3d3553c60775874cd4ab22}
      \strng{authorfullhash}{f421067e62fbcac3e0219c59e1f88613}
      \strng{authorfullhashraw}{f421067e62fbcac3e0219c59e1f88613}
      \field{labelalpha}{NRS07}
      \field{sortinit}{N}
      \field{sortinithash}{22369a73d5f88983a108b63f07f37084}
      \field{labelnamesource}{author}
      \field{labeltitlesource}{title}
      \field{booktitle}{ACM SIGACT Symposium on Theory of Computing}
      \field{series}{STOC 2007}
      \field{title}{Smooth sensitivity and sampling in private data analysis}
      \field{year}{2007}
      \field{pages}{75\bibrangedash 84}
      \range{pages}{10}
      \strng{doi}{10.1145/1250790.1250803}
      \strng{urlraw}{https://people.csail.mit.edu/asmith/PS/stoc321-nissim.pdf}
      \strng{url}{https://people.csail.mit.edu/asmith/PS/stoc321-nissim.pdf}
    \endentry
    \entry{RothR10}{inproceedings}{}{}
      \name{author}{2}{}{{{hash=9bb323b3c3d2af68def0dcbde8e6795d}{family={Roth},
           familyi={R\bibinitperiod},
           given={Aaron},
           giveni={A\bibinitperiod}}}{{hash=efce6f6d496cdcc1bb07fb47dfecb3aa}{family={Roughgarden},
           familyi={R\bibinitperiod},
           given={Tim},
           giveni={T\bibinitperiod}}}}
      \strng{namehash}{2f8c0849aabef8b7a6c0b33476c5e575}
      \strng{fullhash}{2f8c0849aabef8b7a6c0b33476c5e575}
      \strng{fullhashraw}{2f8c0849aabef8b7a6c0b33476c5e575}
      \strng{bibnamehash}{2f8c0849aabef8b7a6c0b33476c5e575}
      \strng{authorbibnamehash}{2f8c0849aabef8b7a6c0b33476c5e575}
      \strng{authornamehash}{2f8c0849aabef8b7a6c0b33476c5e575}
      \strng{authorfullhash}{2f8c0849aabef8b7a6c0b33476c5e575}
      \strng{authorfullhashraw}{2f8c0849aabef8b7a6c0b33476c5e575}
      \field{labelalpha}{RR10}
      \field{sortinit}{R}
      \field{sortinithash}{5e1c39a9d46ffb6bebd8f801023a9486}
      \field{labelnamesource}{author}
      \field{labeltitlesource}{title}
      \field{booktitle}{ACM SIGACT Symposium on Theory of Computing}
      \field{series}{STOC 2010}
      \field{title}{Interactive privacy via the median mechanism}
      \field{year}{2010}
      \field{pages}{765\bibrangedash 774}
      \range{pages}{10}
      \strng{urlraw}{https://arxiv.org/abs/0911.1813}
      \strng{url}{https://arxiv.org/abs/0911.1813}
    \endentry
    \entry{RaskhodnikovaS16}{inproceedings}{}{}
      \name{author}{2}{}{{{hash=595cb420436ece2997a0f6b90fa4089f}{family={Raskhodnikova},
           familyi={R\bibinitperiod},
           given={Sofya},
           giveni={S\bibinitperiod}}}{{hash=5ccecf114bbd51a1a2979a23dd69c78c}{family={Smith},
           familyi={S\bibinitperiod},
           given={Adam},
           giveni={A\bibinitperiod}}}}
      \strng{namehash}{94c988edbbf64d23d9f4e60ee7f113e7}
      \strng{fullhash}{94c988edbbf64d23d9f4e60ee7f113e7}
      \strng{fullhashraw}{94c988edbbf64d23d9f4e60ee7f113e7}
      \strng{bibnamehash}{94c988edbbf64d23d9f4e60ee7f113e7}
      \strng{authorbibnamehash}{94c988edbbf64d23d9f4e60ee7f113e7}
      \strng{authornamehash}{94c988edbbf64d23d9f4e60ee7f113e7}
      \strng{authorfullhash}{94c988edbbf64d23d9f4e60ee7f113e7}
      \strng{authorfullhashraw}{94c988edbbf64d23d9f4e60ee7f113e7}
      \field{labelalpha}{RS16}
      \field{sortinit}{R}
      \field{sortinithash}{5e1c39a9d46ffb6bebd8f801023a9486}
      \field{labelnamesource}{author}
      \field{labeltitlesource}{title}
      \field{booktitle}{IEEE Symposium on Foundations of Computer Science}
      \field{series}{FOCS 2016}
      \field{title}{Lipschitz Extensions for Node-Private Graph Statistics and the Generalized Exponential Mechanism}
      \field{year}{2016}
      \field{pages}{495\bibrangedash 504}
      \range{pages}{10}
      \strng{urlraw}{https://cs-people.bu.edu/sofya/pubs/Lipschitz-deg-ext-FOCS16.pdf}
      \strng{url}{https://cs-people.bu.edu/sofya/pubs/Lipschitz-deg-ext-FOCS16.pdf}
    \endentry
    \entry{SealfonU19}{inproceedings}{}{}
      \name{author}{2}{}{{{hash=e61602764483f76e0c720118163123e9}{family={Sealfon},
           familyi={S\bibinitperiod},
           given={Adam},
           giveni={A\bibinitperiod}}}{{hash=9e9335807e70a8b4ca12eda1833b293d}{family={Ullman},
           familyi={U\bibinitperiod},
           given={Jonathan},
           giveni={J\bibinitperiod}}}}
      \strng{namehash}{3d6dbe7c064d46b72af5dda42ba2348e}
      \strng{fullhash}{3d6dbe7c064d46b72af5dda42ba2348e}
      \strng{fullhashraw}{3d6dbe7c064d46b72af5dda42ba2348e}
      \strng{bibnamehash}{3d6dbe7c064d46b72af5dda42ba2348e}
      \strng{authorbibnamehash}{3d6dbe7c064d46b72af5dda42ba2348e}
      \strng{authornamehash}{3d6dbe7c064d46b72af5dda42ba2348e}
      \strng{authorfullhash}{3d6dbe7c064d46b72af5dda42ba2348e}
      \strng{authorfullhashraw}{3d6dbe7c064d46b72af5dda42ba2348e}
      \field{labelalpha}{SU19}
      \field{sortinit}{S}
      \field{sortinithash}{b164b07b29984b41daf1e85279fbc5ab}
      \field{labelnamesource}{author}
      \field{labeltitlesource}{title}
      \field{booktitle}{Advances in Neural Information Processing Systems}
      \field{series}{NeurIPS 2019}
      \field{title}{{Efficiently Estimating Erdős-Rényi Graphs with Node Differential Privacy}}
      \field{year}{2019}
      \strng{urlraw}{https://arxiv.org/abs/1905.10477}
      \strng{url}{https://arxiv.org/abs/1905.10477}
    \endentry
  \enddatalist
\endrefsection
\global\toggletrue{blx@bbldone}\blx@generate@bbl@mdfivesum@found{\jobname.tex}\blx@bblend\endgroup
\csnumgdef{blx@labelnumber@\the\c@refsection}{0}}
\makeatother

\begin{document}
\pagenumbering{gobble}
\maketitle

\begin{abstract}
    Node-differentially private algorithms enable learning on network-structured datasets in which records reflect interactions among individuals. They protect the privacy of each individual's information, including all their associated records. A key challenge in the design of such algorithms is to exploit the structure of realistic graphs---where most nodes have low degree---to achieve high accuracy, without compromising on the worst-case privacy guarantee. 
    
    In this work we give two new tools for designing accurate, computationally efficient node-differentially private algorithms.  First, we give a nearly linear time, randomized, node-stable graph projection: the projection leaves graphs of maximum degree $D$ unchanged, maps arbitrary graphs to subgraphs of maximum degree $\hat D \approx D$, and ensures that changing one node of the input graph has only a small effect on the distribution of the edges selected among the remaining vertices. Unlike prior approaches, which abstain when the input violates the degree bound, our projection acts as degree clipping: even when $D$ is set well below the true maximum degree (trading truncation bias at high-degree outliers for lower privacy noise), it retains at least as many edges as a maximal $D$-bounded subgraph.  
    
    Second, for graphs with bounded edge weights, we give a deterministic nearly linear time Lipschitz extension for the total edge weight that agrees with the original sum on graphs of maximum degree $D$ yet is $D$-Lipschitz on all graphs.  
    
    Finally, we show how to apply these methods to design node-private algorithms for high-dimensional vector summation and gradient descent on datasets where records are associated with graph edges, such as email threads.
\end{abstract}
\vfill\newpage

    \setcounter{tocdepth}{2}
    \tableofcontents
    \vfill\newpage

\pagenumbering{arabic}
\section{Introduction}

Many important data analysis and machine learning tasks operate on network-structured data, where records reflect interactions among individuals---from training language models on email or text messages, where each message corresponds to an edge between sender and recipient, to estimating clustering coefficients or disease spread in contact-tracing or social networks. Such data are often highly sensitive. Differential privacy (DP)~\cite{DworkMNS06} is a standard framework for analyzing privacy leakage rigorously. For network data, there are two well-studied versions of DP \cite{HayLMJ09}: \emph{edge-level} DP (edge-DP)~\cite{NissimRS07} conceals the presence or absence of a single edge (e.g., email thread), whereas \emph{node-level} DP (node-DP)~\cite{HayLMJ09,BlockiBDS13,KasiviswanathanNRS13,ChenZ13} hides the presence of a node together with all of its incident edges and their associated records. When nodes correspond to individuals, node-DP aligns directly with protecting ``one individual's data''; for instance, edge-DP allows revealing aggregate information about one person's contacts, which might be very sensitive~\cite{JerniganM09}. 

Node differential privacy is challenging to achieve. It requires that the algorithm's outputs be similarly distributed on any pair of graphs that differ by the insertion of a node and its adjacent edges. However, most graph computations are very sensitive to adding or removing a node: for example, in a sparse graph (say, of constant average degree), a single node can contribute a constant fraction of the entire dataset. Nevertheless, a rich set of algorithmic techniques has been developed for node-DP (see \Cref{sec:intro-background}). At a high level, these techniques identify a family of ``nice'' inputs---most commonly, the set $\cG_\dbound$ of graphs that satisfy a given maximum degree bound $\dbound$---for which they guarantee good accuracy, while requiring \emph{privacy to hold unconditionally on every input graph}.

Bridging bounded-degree utility and unconditional privacy is delicate. In particular, simply preprocessing the input by selecting a maximal or maximum-size subgraph of maximum degree $\dbound$ (often called ``contribution bounding'') is not sufficient: adding or removing one node can trigger cascading changes along alternating paths, altering $\Omega(n)$ edges in the selected subgraph; \Cref{fig:greedy-instability} illustrates instability on two neighboring chain graphs when $\dbound=1$. Notably, recent work on practical methods for training models on edge- and hyperedge-associated records (\citet{GaneshMMWS25}) set aside this degree-truncation step in the privacy analysis, assuming instead that the graph topology is fixed. An important goal for privacy research is to develop methods with end-to-end privacy guarantees when the graph topology---which is often sensitive---must also be protected.

\mypar{This paper.}
We present two new, efficient algorithmic primitives for controlling sensitivity in node-level differentially private algorithms. Both are based on careful modifications of the greedy algorithm for maximal $\dbound$-bounded subgraphs and computable in near-linear time $O(m\log m)$ on graphs with $n$ vertices and $m$ edges.

\begin{itemize}
    \item We give a general-purpose, \emph{randomized} graph projection (\Cref{sec:stable-projection}) for contribution bounding. Given parameter $\dbound$, it maps an arbitrary input graph to a subgraph of maximum degree $\dmax \approx \dbound$ based on the graph structure alone, while leaving graphs of maximum degree $\dbound$ unchanged. 
    To formalize its guarantees, we introduce the notion of  \emph{distributionally node-stable projection}: adding or removing a node does not significantly change the distribution of selected edges in the rest of the graph. This allows black-box composition with any downstream node-private algorithm designed for bounded-degree graphs. Ours is the first projection with a tight node-level stability guarantee. 
    \item For graphs with real-valued, bounded edge weights, we give a deterministic \emph{Lipschitz extension} (\Cref{sec:extension-for-sums}) of the total edge weight $\fsum(G) = \sum_{e \in E} x(e)$ that agrees with $\fsum$ on $\dbound$-bounded graphs and is $\dbound$-Lipschitz on all graphs. This expands the set of computations addressed by efficient Lipschitz extensions to include gradient estimation in first-order optimization methods. 
    \item To illustrate both tools, we show how each yields node-private algorithms for vector summation and noisy gradient descent over edge-associated records (\Cref{sec:applications}); this setting captures training models on email threads, SMS messages and other data associated with pairs of individuals. Our methods match the accuracy achievable under a \emph{promised} degree bound~\cite{GaneshMMWS25}.
\end{itemize}

\subsection{Background: graph projections and Lipschitz extensions}
\label{sec:intro-background}

Let $\cG$ denote the set of undirected simple graphs\footnote{This paper focuses on relational data where each record involves two individuals. The few works that have addressed the more general \textit{hypergraph} setting either obtained inefficient algorithms~\cite{FangDY22,LinderRSS25} or assumed a promised degree bound~\cite{GaneshMMWS25}. Extending our results to hypergraphs is an open problem.
}
$G = (V,E,x)$, where $x \colon E \to \cX$ assigns an attribute or weight to each edge (when omitted, $x(e) = 1$). A graph $G$ is \emph{$\dbound$-bounded} if it has maximum degree at most $\dbound$, and $\cG_\dbound \subseteq \cG$ denotes the set of $\dbound$-bounded graphs.

Two graphs $G, G' \in \cG$ are \emph{node-neighbors} if one is obtained from the other by removing a vertex and all of its incident edges; the \emph{node distance} $d(G,G')$ is the shortest-path metric induced by this relation. Two random variables $X, X'$ over the same domain $\cY$ are \emph{$(\eps,\delta)$-indistinguishable}, written $X \approx_{\eps,\delta} X'$, if $\pr{}{X \in Y} \leq e^{\eps}\pr{}{X' \in Y} + \delta$ and $\pr{}{X' \in Y} \leq e^{\eps}\pr{}{X \in Y} + \delta$ for all events $Y \subseteq \cY$. An algorithm $\cA \colon \cG \to \cY$ satisfies \emph{$(\eps,\delta)$-node differential privacy} (node-DP)~\cite{HayLMJ09,BlockiBDS13,KasiviswanathanNRS13,ChenZ13} if $\cA(G) \approx_{\eps,\delta} \cA(G')$ for all node-neighboring $G, G' \in \cG$.

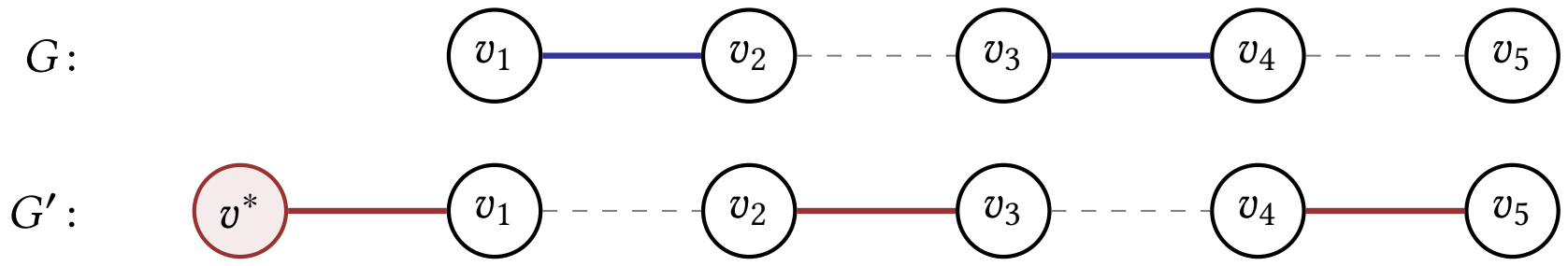
\begin{figure}[t]
\centering
\begin{tikzpicture}[
    vertex/.style={circle, draw, thick, minimum size=6.5mm, inner sep=0pt, font=\small},
    newvertex/.style={circle, draw=DarkRed, fill=DarkRed!10, thick, minimum size=6.5mm, inner sep=0pt, font=\small},
    selectedG/.style={very thick, DarkBlue},
    selectedGprime/.style={very thick, DarkRed},
    unselected/.style={thin, gray, dashed}
]
    \node[font=\small\bfseries, anchor=east] at (-0.9, 0) {$G\colon$};
    \node[vertex] (g1) at (1.8, 0) {$v_1$};
    \node[vertex] (g2) at (3.6, 0) {$v_2$};
    \node[vertex] (g3) at (5.4, 0) {$v_3$};
    \node[vertex] (g4) at (7.2, 0) {$v_4$};
    \node[vertex] (g5) at (9.0, 0) {$v_5$};

    \draw[selectedG]  (g1) -- (g2);
    \draw[unselected] (g2) -- (g3);
    \draw[selectedG]  (g3) -- (g4);
    \draw[unselected] (g4) -- (g5);

    \node[font=\small\bfseries, anchor=east] at (-0.9, -1.1) {$G'\colon$};
    \node[newvertex] (u)  at (0.0, -1.1) {$v^*$};
    \node[vertex]    (h1) at (1.8, -1.1) {$v_1$};
    \node[vertex]    (h2) at (3.6, -1.1) {$v_2$};
    \node[vertex]    (h3) at (5.4, -1.1) {$v_3$};
    \node[vertex]    (h4) at (7.2, -1.1) {$v_4$};
    \node[vertex]    (h5) at (9.0, -1.1) {$v_5$};

    \draw[selectedGprime] (u)  -- (h1);
    \draw[unselected]     (h1) -- (h2);
    \draw[selectedGprime] (h2) -- (h3);
    \draw[unselected]     (h3) -- (h4);
    \draw[selectedGprime] (h4) -- (h5);
\end{tikzpicture}
\caption{Instability of naive degree truncation for $\dbound = 1$ (where a $1$-bounded subgraph is a matching) on node-neighboring chain graphs $G$ and $G' = G \cup \{v^*\}$. Left-to-right greedy selection (or maximum matching) chooses the solid blue edges in $G$, but the disjoint set of solid red edges in $G'$.}
\label{fig:greedy-instability}
\end{figure}

Two main approaches are used to design algorithms that are accurate on $\cG_\dbound$ while guaranteeing node-DP on all of $\cG$:

\mypar{Graph projections.}
The most modular approach seeks a \emph{graph projection}~\cite{BlockiBDS13,KasiviswanathanNRS13,JainSW24} that maps an arbitrary graph $G \in \cG$ to a bounded-degree subgraph $H = \cP(G)$ while acting as the identity on $\cG_\dbound$, so that any downstream algorithm for bounded-degree graphs can be run on $H$. This is challenging because any deterministic projection that enforces a hard degree cutoff $\dbound$ can suffer cascading edge replacements along alternating paths (\Cref{fig:greedy-instability}), altering $\Omega(n)$ edges and vertices when a single node is added~\cite{BlockiBDS13,RaskhodnikovaS16}. The most relevant prior construction, by \citet*{JainSW24} (building on~\cite{BlockiBDS13,DayLL16}), shows that a deterministic greedy projection has bounded node and edge sensitivity on the restricted class of graphs with at most $\ell$ vertices of degree above $\dbound$, and combines this projection with a private safety test that abstains outside this class. As we discuss in \Cref{sec:intro-proj-results}, however, this test-and-project paradigm still incurs an $\Omega(\log(1/\delta)/\eps)$ blowup in node distance even on $\dbound$-bounded graphs, and abstains on graphs with many high-degree vertices. 

\mypar{Lipschitz extensions.}
A function $h \colon \cG \to \R^d$ is \emph{$\Delta$-Lipschitz} (in a norm $\|\cdot\|$) on $S \subseteq \cG$ if $\|h(G) - h(G')\| \leq \Delta \cdot d(G,G')$ for all $G, G' \in S$; when $S = \cG$, we simply say that $h$ is $\Delta$-Lipschitz. When a target query $f$ is $\Delta$-Lipschitz on $\cG_\dbound$, a large body of work~\cite{BlockiBDS13,KasiviswanathanNRS13,ChenZ13,RaskhodnikovaS16,DayLL16,CummingsD18,BorgsCSZ18b,SealfonU19,KalemajRST23} constructs a \emph{Lipschitz extension} $\ftilde \colon \cG \to \R^d$ that agrees with $f$ on $\cG_\dbound$ and is $\Delta'$-Lipschitz on all of $\cG$ (where $\Delta'/\Delta$ is the \emph{stretch}). Adding Laplace or Gaussian noise scaled to $\Delta'$ then yields node-DP on all graphs without spending privacy budget on preprocessing. However, Lipschitz extensions must be tailored to a specific function class; general LP-based constructions~\cite{KasiviswanathanNRS13,RaskhodnikovaS16} require high-degree polynomial time, while prior near-linear-time combinatorial extensions~\cite{DayLL16} were restricted to unweighted edge counts and degree histograms.

\subsection{Our results}
\label{sec:intro-results}

\subsubsection{Distributionally node-stable graph projections (\Cref{sec:stable-projection})}
\label{sec:intro-proj-results}

\suppressfloats[t]
\begin{table}[t]
    \centering
    \small
    \renewcommand{\arraystretch}{1.25}
    \begin{tabular}{>{\raggedright\arraybackslash}m{1.2in}||>{\centering\arraybackslash}m{2.80in}|>{\centering\arraybackslash}m{1.30in}}
        \hline
        Approach
            & Requirement on downstream $\cA$ to ensure $(\eps,\delta)$-node-DP
            & When $k$ nodes have $\deg > \dbound$
            \\
        \hline
        Greedy truncation (unstable) \cite{GaneshMMWS25}
            & Cannot achieve $(\eps,\delta)$-node-DP and any nontrivial accuracy guarantee on $\cG_\dbound$
            & Clips degrees to $\dbound$
            \\
        \hline
        Test-and-project (node stable version) \cite{JainSW24}
            & $\cA$ is $(\eps/\ell,\delta/e^{\ell})$-node-DP on $\cG_{D}$ 
                \mbox{for $\ell = \Omega(\log(1/\delta)/\eps)$}
            & Outputs $\bot$ unless $\dbound+\ell = \Omega(k)$
            \\
        \hline
        Test-and-project (edge stable version) \cite{JainSW24}
            & $\cA$ is $(\eps/(D+\ell),\delta/e^{(D+\ell)})$-edge-DP on $\cG_{D}$ 
            \mbox{for $\ell = \Omega(\log(1/\delta)/\eps)$}
            & Outputs $\bot$ unless $\dbound+\ell = \Omega(k)$
            \\
        \hline
        \textbf{Randomized projection} (\Cref{thm:informal-proj})
            & $\cA$ is $(\eps/2,\delta/2)$-node-DP on $\cG_{\dmax}$ 
            \mbox{for $\dmax \approx \dbound + \log(1/\delta)/\eps^2$}
            & Clips degrees to $\dmax$
            \\
        \hline
        \end{tabular}
    \caption{Comparison of general projection methods for node-DP with target degree bound $\dbound$ and target privacy parameters $\eps$ and $\delta$.
        These methods preprocess the graph for a downstream algorithm on bounded-degree graphs.
        The second column describes the requirements on the downstream algorithm to achieve the desired privacy guarantee.
        The third column describes the behavior of the algorithm on input graphs that do not have maximum degree $\dbound$.
        All methods have running time $O(m \log m)$.
    }
    \label{tab:projection-comparison}
\end{table}

We give a new, randomized projection algorithm for node-private degree bounding. 
Given $0 < \dbound < \dmax$, a randomized algorithm $\cP$ is a \emph{$(\dbound,\dmax)$-projection} (\Cref{def:projection}) if, for every $G \in \cG$, the output $\cP(G)$ is a $\dmax$-bounded subgraph of $G$ on the same vertex set, and $\cP(G) = G$ whenever $G \in \cG_\dbound$. 

Rather than requiring that the projections of neighboring graphs be close in the space of graphs, we introduce a weaker but sufficient requirement: the output \emph{distributions} on the shared vertices should be nearly indistinguishable. Specifically, we say $\cP$ is \emph{$(\eps,\delta)$-distributionally node-stable} (\Cref{def:distrib-stability}) if, for every pair of node-neighboring graphs $G, G'$ where $G'$ is obtained from $G$ by adding a vertex $v^*$ with arbitrary incident edges,
\[
    \cP(G) \approx_{\eps,\delta} \cP(G')_{-v^*} \, .
\]
Here, $\cP(G')_{-v^*}$ denotes the graph obtained by deleting $v^*$ and its incident edges from $\cP(G')$. This constraint is similar to closeness in a Wasserstein metric on distributions in $\Delta(\cG_{\dmax})$. Note that $\cP(G)$ is not itself differentially private (indeed, $\cP(G) = G$ on $\cG_\dbound$) and is kept internal as the input to a downstream node-DP algorithm.

Our first main result is a near-linear-time projection that achieves this guarantee. Our construction circumvents the cascading instability of \Cref{fig:greedy-instability} via two relaxations: \emph{bicriteria degree slack} (preserving a maximal $\dbound$-bounded subgraph while capping degrees at $\dmax \approx \dbound$) and \emph{randomized per-vertex cutoffs} in $[\dbound,\dmax]$, which absorb the effect of adding $v^*$ into an $(\eps,\delta)$ distributional shift at its direct neighbors while keeping $\cP(G)$ and $\cP(G')$ at node distance $1$. When the number of vertices $n$ is public, $\dmax$ is a deterministic function of $(n,\dbound,\eps,\delta)$; when $n$ is itself sensitive, our algorithm privately estimates an upper bound on $n$ and releases the resulting degree bound $\dmax(G)$ alongside $\cP(G)$:

\begin{theorem}[Informal version of \Cref{lem:stable-proj} and \Cref{thm:private-size-stable-proj}]\label{thm:informal-proj}
    For every target degree bound $\dbound \in \N$ and stability parameters $\eps, \delta \in (0,1)$, there is a randomized algorithm $\sadnsp$ that, given a graph $G=(V,E)$ with $n$ vertices and $m$ edges, runs in time $\widetilde{O}(m)$ and outputs a subgraph $\cP(G)$ and a released degree bound $\dmax(G)$ such that:
    \begin{enumerate}
        \item \emph{(Projection and edge preservation)} Conditioned on any released value $\dmax = \dmax(G)$, the map $\cP$ is a $(\dbound,\dmax)$-projection. Moreover, on every input graph $G$, $\cP(G)$ retains at least as many edges as some maximal $\dbound$-bounded subgraph of $G$.
        \item \emph{(Distributional node stability)} For every pair of node-neighboring graphs $G, G'$ where $G'$ is obtained from $G$ by adding a node $v^*$ with arbitrary incident edges,
        \[
            \bparen{\cP(G),\ \dmax(G)} \approx_{\eps,\delta} \bparen{\cP(G')_{-v^*},\ \dmax(G')}.
        \]
        \item \emph{(Released degree bound)} When $\delta = n^{-\Theta(1)}$, with probability $1$,
        \[
            \dmax(G) \leq (1+o(1))\dbound + \frac{\operatorname{polylog}(n)}{\eps^2}.
        \]
    \end{enumerate}
\end{theorem}

\mypar{Black-box composition with downstream algorithms.}
Distributional node stability allows $\sadnsp$ to serve as a black-box preprocessing step: given any family of algorithms $\{\cA_\ell \colon \cG_\ell \to \cY\}_{\ell \in \N}$ that are $(\aeps,\adelta)$-node-DP on $\ell$-bounded graphs, running $\sadnspdetail{\dbound}{\peps}{\pdelta}(G)$ to get $(H,\dmax)$ and releasing $(\dmax, \cA_{\dmax}(H))$ is $(\peps+\aeps, \pdelta+\adelta)$-node-DP on all graphs. On every $\dbound$-bounded graph $G \in \cG_\dbound$ its output is distributed identically to $\cA_{\dmax}(G)$ (see \Cref{lem:projection-plus-dp}). For graphs outside of $\cG_{\dbound}$, it runs $\cA_{\dmax}$ on a large $\dmax$-bounded subgraph. 

\mypar{Comparison to \cite{JainSW24}: tight node stability and clipping versus testing.}
\citet{JainSW24}, henceforth ``JSW,'' study deterministic greedy projections\footnote{The results of \cite{JainSW24} are stated in a \emph{dynamic} (or \emph{continual release}) setting where the graph evolves over $T$ time steps.  To make a direct comparison we focus on just the static setting, which is the special case of their results where $T=1$.  Our projection can also be adapted to the dynamic setting.
}
onto $\cG_\dbound$ that are stable only on \emph{$(\dbound,\ell)$-bounded graphs}, which are graphs in which at most $\ell$ vertices have degree exceeding $\dbound$.  On this class, the projections of node-neighboring graphs lie at node distance at most $2\ell+1$ and edge distance at most $O(\dbound+\ell)$. Because these projections can change $\Omega(n)$ vertices and edges outside this class, JSW use the propose-test-release method \cite{DworkL09} to design an $(\eps,\delta)$-node-DP \emph{safety test} that abstains and outputs $\bot$ whenever the input graph is not $(\dbound,\ell)$-bounded.  Relative to JSW, our randomized projection has two main advantages:
\begin{itemize}
    \item \emph{Tight node stability:} Whenever $\dbound < n-1$, any $(\eps,\delta)$-node-DP safety test must tolerate a margin of at least $\ell = \Omega(\log(1/\delta)/\eps)$ high-degree vertices in order to accept any graph at all, even graphs in $\cG_\dbound$. Consequently, on the safe class accepted by the test, JSW's projection only guarantees node-to-node distance $2\ell+1 = \Omega(\log(1/\delta)/\eps)$. Composing JSW's projection with a downstream node-DP algorithm on $\cG_\dbound$ via group privacy therefore degrades the downstream privacy budget by a multiplicative factor of $2\ell+1 = \Omega(\log(1/\delta)/\eps)$, which in nearly all applications would blow up the final error by at least this factor.  Note that this blowup occurs, even when the input graph actually lies in $\cG_{\dbound}$.  By contrast, under our projection, $\cP(G)$ and $\cP(G')$ can be approximately coupled to have node distance $1$ on \emph{all} graphs, so the downstream node-DP algorithm experiences no group-privacy blowup.  The cost of our randomized projection appears only as an additive slack in the released degree bound $\dmax$.
    
    \item \emph{Graceful degradation:} When the input graph is not in $\cG_{\dbound}$, our projection acts as a clipping of the degree of each node, and thus returns some meaningful output graph even when the parameter $\dbound$ is much smaller than the actual maximum degree $\max_v \deg(v)$.  In contrast, JSW would abstain and output $\bot$ on graphs that are not $(\dbound,\ell)$-bounded.  Real-world networks almost always contain high-degree outlier nodes, such as hubs or bots.  For better utility, we would like to be able to tune $\dbound$ to something closer to the average degree or some quantile of the degree distribution, which would trade some small truncation bias on the outlier nodes for a large reduction in the privacy noise.  In JSW, even a small dense cluster forces the algorithm to either abstain or calibrate the noise to the outliers. For example, on a graph consisting of a $k$-clique together with a matching on the remaining $n-k$ nodes, JSW will obstain unless $\dbound + \ell = \Omega(k)$. So using JSW requires blowing up the sensitivity by $\Omega(k)$, regardless of whether we use the node- or edge- stability bound.  In contrast, setting $\dbound=1$ in our projection retains the entire matching while capping degrees in the clique at $\dmax = \operatorname{polylog}(n/\delta)/\eps^2$ and maintaining node distance $1$. We work through this example for estimating the mean of i.i.d.\ $d$-dimensional edge attributes in \Cref{app:jsw-brittleness}.
\end{itemize}

\noindent
\Cref{tab:projection-comparison} summarizes how our projection compares to existing approaches to node-DP degree truncation.

\subsubsection{Lipschitz extensions for weighted edge sums (\Cref{sec:extension-for-sums})}
\label{sec:intro-lip-results}

\suppressfloats[t]
\begin{table}[t]
    \centering
    \small
    \renewcommand{\arraystretch}{1.25}
    \begin{tabular}{>{\raggedright\arraybackslash}m{1.55in}||>{\centering\arraybackslash}m{1.25in}|>{\centering\arraybackslash}m{1.50in}|>{\centering\arraybackslash}m{1.00in}}
        \hline
        Approach
            & Applicable Functions
            & When some nodes have degree $> \dbound$
            & Time
            \\
        \hline
        Edge-count extensions 
            \cite{DayLL16} \newline (builds on \cite{KasiviswanathanNRS13,RaskhodnikovaS16})
            & Edge count $|E(G)|$
            & Size of some maximal $\dbound$-bounded subgraph
            & $O(m\log m)$
            \\
        \hline
        LP-based extensions \newline
            \cite{KasiviswanathanNRS13,ChenZ13}
            & Weighted edge sum $\fsum(G)=\sum_{e} x(e)$ (also weighted subgraph counts)
            & Maximum weight $\dbound$-bounded subgraph
            & LP solve with $m$ vars and $n$ constraints
            \\
        \hline
        \textbf{Sorted-greedy extension}\newline
            (\Cref{thm:informal-lip})
            & Weighted edge sums $\fsum(G)=\sum_{e} x(e)$
            & Weight of some maximal $\dbound$-bounded subgraph
            & $O(m\log m)$
            \\
        \hline
        \end{tabular}
    \caption{Comparison of Lipschitz extensions for counting edges or summing edge weights for node-DP with target degree bound $\dbound$.  The second column describes the applicable function.  The third column describes the behavior of the algorithm on input graphs that do not have maximum degree $\dbound$.  The fourth column describes the running time of the algorithm.  All algorithms compute a $\dbound$-Lipschitz extension of the target function from $\cG_\dbound$ to all of $\cG$.
    }
    \label{tab:lipschitz-comparison}
\end{table}

When the goal is specifically to sum bounded edge attributes, we can avoid both the degree slack $\dmax - \dbound$ and the preprocessing privacy cost by constructing a direct Lipschitz extension. For graphs $G = (V,E,x)$ with edge weights $x \colon E \to [0,1]$, the total edge weight $\fsum(G) = \sum_{e \in E} x(e)$ is $\dbound$-Lipschitz on $\cG_\dbound$. 
Efficient Lipschitz extensions for the \emph{number of edges} (the unweighted case $x \equiv 1$) were known from prior work~\cite{KasiviswanathanNRS13,ChenZ13,DayLL16}; however, those algorithms either do not have low sensitivity when used to sum general edge weights (for example, in the greedy construction of \citet{DayLL16}, adding a node can trigger a long alternating path of edge replacements whose weights differ significantly), or require solving weighted matching problems or linear programs, making them hard to use. 

We adapt the combinatorial extension of \citet{DayLL16}---running the greedy $\dbound$-bounded subgraph algorithm on edges sorted in \emph{decreasing order of weight} (\Cref{alg:sum-proj})---to obtain an optimal, stretch-$1$ Lipschitz extension of $\fsum$ in near-linear time:

\begin{theorem}[Informal version of Corollary~\ref{cor:Lip-ext-sum}]\label{thm:informal-lip}
    For every $\dbound \in \N$, there is a deterministic algorithm (\Cref{alg:sum-proj}) running in time $O(m\log m)$ on graphs $G = (V,E,x)$ with $n$ vertices, $m$ edges, and edge weights $x \colon E \to [0,1]$ that computes a $\dbound$-Lipschitz extension $\tfsum \colon \cG \to \R$ of $\fsum$ from $\cG_\dbound$ to $\cG$:
    \begin{enumerate}
        \item \emph{(Extension)} For every $\dbound$-bounded graph $G \in \cG_\dbound$, $\tfsum(G) = \fsum(G)$.
        \item \emph{(Stretch-$1$ Lipschitz property)} For every pair of node-neighboring graphs $G, G' \in \cG$,
        \[
            \abs{\tfsum(G) - \tfsum(G')} \leq \dbound.
        \]
    \end{enumerate}
\end{theorem}

\noindent
\Cref{tab:lipschitz-comparison} summarizes how our Lipschitz extension compares with previous extensions.

\subsubsection{Applications to vector summation and SCO (\Cref{sec:applications})}
\label{sec:intro-apps}

\suppressfloats[t]
\begin{table}[t]
    \centering
    \small
    \renewcommand{\arraystretch}{1.25}
    \begin{tabular}{>{\raggedright\arraybackslash}m{1.80in}||>{\centering\arraybackslash}m{2.65in}|>{\centering\arraybackslash}m{0.95in}}
        \hline
        Approach
            & Error bound
            & Running time
            \\
        \hline
        Subsampled SGD with na\"ive contribution bounding \newline (does not satisfy DP)
            & $\tilde{O}\paren{\frac{1}{\sqrt{m}} + \frac{\dbound\sqrt{d \log n}}{m\eps}}$
            & $O(dm^2)$
            \\
        \hline
        \textbf{Full-batch gradient descent via Lipschitz extensions}\newline
            (\Cref{prop:sco-extension})
            & $O\paren{\frac{1}{\sqrt{m}} + \frac{\dbound\sqrt{d \log n}}{m\eps}}$
            & $\tilde{O}(d^2m^3)$
            \\
        \hline
        \textbf{Subsampled SGD via one-time stable projection}\newline
            (\Cref{prop:sco-projection})
            & $\tilde{O}\paren{\frac{1}{\sqrt{m}} + \frac{\dmax\sqrt{d \log n}}{m\eps}}$\par
                for $\dmax \approx \dbound + \log(1/\delta)/\eps^2$
            & $O(dm^2)$
            \\
        \hline
        \end{tabular}
    \caption{Comparison of accuracy and running time for stochastic convex optimization on graphs with maximum degree $\dbound$.
        The nonprivate contribution-bounding baseline is compared with our Lipschitz-extension and stable-projection methods, using $T=m^2$ gradient steps.
    }
    \label{tab:sco-comparison}
\end{table}

In \Cref{sec:applications}, we illustrate both primitives by applying them to high-dimensional vector summation---computing $\vsum(G) = \sum_{e \in E(G)} x(e)$ for edge vectors $x(e) \in \R^d$ with $\|x(e)\|_2 \leq 1$---and to first-order optimization over edge-associated records, where each $x(e)$ defines a loss function on a parameter space $\cW \subseteq \R^d$. The methods we develop satisfy node-level privacy (for all graphs and, with gradient clipping, for all loss functions). We prove accuracy bounds for the setting of convex, 1-Lipschitz losses over a unit-diameter domain $\cW$, assuming each edge's record is drawn i.i.d.\ from an unknown population (a setting often referred to as \emph{stochastic convex optimization}, or SCO) and the maximum degree of the graph is actually $\dbound$.  Our two primitives yield two complementary algorithms:
\begin{itemize}
    \item \emph{Full-batch gradient descent via Lipschitz extensions (\Cref{thm:vector-sum-gdp,prop:sco-extension}):} Combining our scalar extension $\tfsum$ with Kashin representations~\cite{LyubarskiiV10}, a family of nonlinear transformations that, roughly, embed $\ell_2$-bounded vectors into $\ell_\infty$,  gives a constant-stretch Lipschitz extension for vector sums in the $\ell_2$ norm (\Cref{lem:vector-sum-ext}). Using this extension to compute noisy full-batch subgradients matches the asymptotic error of a \emph{promised} degree bound $\dbound$---enforcing end-to-end node-DP without changing the asymptotic accuracy cost---at the expense of increased running time.

    \item \emph{Subsampled SGD via one-time stable projection (\Cref{thm:vector-sum-oracle,prop:sco-projection}):} Preprocessing the graph once with $\sadnsp$ builds a reusable subsampling oracle that can answer adaptive vector-sum queries in sublinear time, while paying the projection privacy cost only once. Plugging this oracle into stochastic subgradient descent matches the running time of standard subsampled SGD (up to the $O(m\log m)$ preprocessing), with the released bound $\dmax$ in place of $\dbound$.
\end{itemize}

We summarize the two methods' properties in \Cref{tab:sco-comparison}, alongside a baseline that applies \emph{nonprivate} contribution bounding (as in \cite{GaneshMMWS25}). We assume $\eps=\Omega(1/n)$ and $\delta = n^{-\Theta(1)}$, and set the number of gradient steps to $T = m^2$ to optimize accuracy. This choice of $T$ accounts for the $m^2$ factor in the running times below: each full-batch step takes $O(dm(d + \log m))$ time, while each subsampled SGD step takes $O(d)$ time after $O(m\log m)$ one-time preprocessing. We bound the running time of the Kashin representation as $O(d^2)$, though faster implementations may exist.

These applications are intended as concrete illustrations of our techniques, highlighting their tradeoff between running time and accuracy, and not an exhaustive treatment of SCO on network data. 

\subsection{Technical overview}
\label{sec:intro-tech-overview}

Both of our main constructions modify the classical greedy algorithm for finding a maximal $\dbound$-bounded subgraph: maintain a degree counter $c(u)$ for each vertex $u$, scan the edges in a fixed order, and include $\{u,v\}$ in $H$ if and only if $\max\{c(u), c(v)\} < \dbound$, incrementing both endpoints' counters when $\{u,v\}$ is added. \citet*{DayLL16} showed that the \emph{degree distribution} of the projected graph is stable; unfortunately the actual set of edges included in the projections of two neighboring graphs can differ dramatically.

\subsubsection{Constructing and analyzing the randomized projection}

The basic idea behind our randomized projection $\cP$ is to break the long chains of dependencies in greedy subgraph selection by randomizing the degree cutoff at each vertex. As in the greedy algorithm, we scan through the edges in a fixed order, but each vertex $u$ maintains a randomized counter $t(u)$ that raises a flag to stop $u$ from accepting new edges once its degree in the selected subgraph roughly exceeds $\dbound$.

Consider two node-neighboring graphs $G$ and $G'$ that differ by the insertion of a vertex $v^*$ with an arbitrary set of incident edges, so that $G = G'_{-v^*}$. At a high level, we can ``couple'' two executions of the projection---one on $G$ and one on $G'$---by slightly shifting the internal randomness in the counters of the (at most $\dmax \approx \dbound$) neighbors of $v^*$ that receive an edge from $v^*$, so that all choices made on the edges of $G$ are \emph{identical} in both executions. This induces an approximate coupling between $\cP(G)$ and $\cP(G')$ under which the two projected graphs always lie at node distance $1$, and choosing the randomized counters carefully lets us bound the distributional shift. For simplicity, we assume here that $G$ and $G'$ share the same public vertex set $V'$ of size $n$, with $v^*$ isolated in $G$.

\begin{figure}[t]
\centering
\begin{tikzpicture}[
    font=\small,
    vertex/.style={circle, draw=black!65, fill=white, thick,
        minimum size=5.5mm, inner sep=0pt},
    affected/.style={vertex, draw=DarkBlue, fill=DarkBlue!12},
    retained/.style={very thick, DarkBlue},
    rejected/.style={thick, dashed, black!60},
    common/.style={thin, black!25}
]
    \draw[rounded corners=8pt, draw=black!45, fill=black!2]
        (2.5,-3.7) rectangle (11.7,3.15);
    \node at (7.1,-3.3) {$G=G'_{-v^*}$};

    \draw[rounded corners=6pt, draw=DarkBlue!50, fill=DarkBlue!5]
        (3.1,0.95) rectangle (7.05,2.05);
    \node[text=DarkBlue] at (5.075,2.45)
        {$S=N_H(v^*),\quad |S|\leq\dmax$};

    \draw[rounded corners=6pt, draw=black!30, fill=white]
        (3.1,-2.05) rectangle (7.05,-0.95);
    \node at (5.075,-2.4) {$N_{G'}(v^*)\setminus S$};

    \node[align=center] at (9.35,2.15)
        {Other vertices\\$V(G)\setminus N_{G'}(v^*)$};
    \node[align=center] at (9.35,-2.6)
        {No edge to $v^*$ in $G'$};

    \node[vertex, draw=DarkRed, fill=DarkRed!10,
        minimum size=8mm] (vstar) at (0,0) {$v^*$};
    \node[align=center] at (0,-0.85) {added to $G$\\to obtain $G'$};

    \node[affected] (s1) at (3.7,1.5) {};
    \node[affected] (s2) at (5.075,1.5) {};
    \node[affected] (s3) at (6.45,1.5) {};
    \node[vertex] (r1) at (3.9,-1.5) {};
    \node[vertex] (r2) at (6.25,-1.5) {};
    \node[vertex] (w1) at (8.5,1.05) {};
    \node[vertex] (w2) at (10.45,0.25) {};
    \node[vertex] (w3) at (8.5,-1.3) {};
    \node[vertex] (w4) at (10.45,-1.3) {};

    \draw[common] (s3) -- (w1) -- (w2) -- (w4) -- (w3) -- (r2);
    \draw[common] (w1) -- (w3);
    \draw[common] (w2) -- (w3);

    \foreach \u in {s1,s2,s3}
        \draw[retained] (vstar) -- (\u);
    \foreach \u in {r1,r2}
        \draw[rejected] (vstar) -- (\u);

    \draw[retained] (0.25,-4.3) -- (1.05,-4.3);
    \node[anchor=west] at (1.2,-4.3) {edge retained in $H$};
    \draw[rejected] (5.05,-4.3) -- (5.85,-4.3);
    \node[anchor=west] at (6,-4.3) {edge in $G'$ rejected from $H$};
    \draw[common] (0.25,-4.75) -- (1.05,-4.75);
    \node[anchor=west] at (1.2,-4.75) {edge of the common input graph $G$};
\end{tikzpicture}
\caption{The stability analysis of our randomized projection.
Here $H$ is the subgraph constructed by the main loop of
\Cref{alg:stable-proj} on $G'$. 
The shaded region contains all nodes of $G=G'_{-v^*}$;
adding $v^*$ and its solid and dashed incident edges gives $G'$.
Only the counters at nodes in $S=N_H(v^*)$ receive a differing input
at their pair with $v^*$, once each, so at most $\dmax$ counters are
affected by the presence of $v^*$'s edges. We couple the two executions by altering the randomness of only those nodes' counters.
}
\label{fig:stable-projection-counters}
\end{figure}
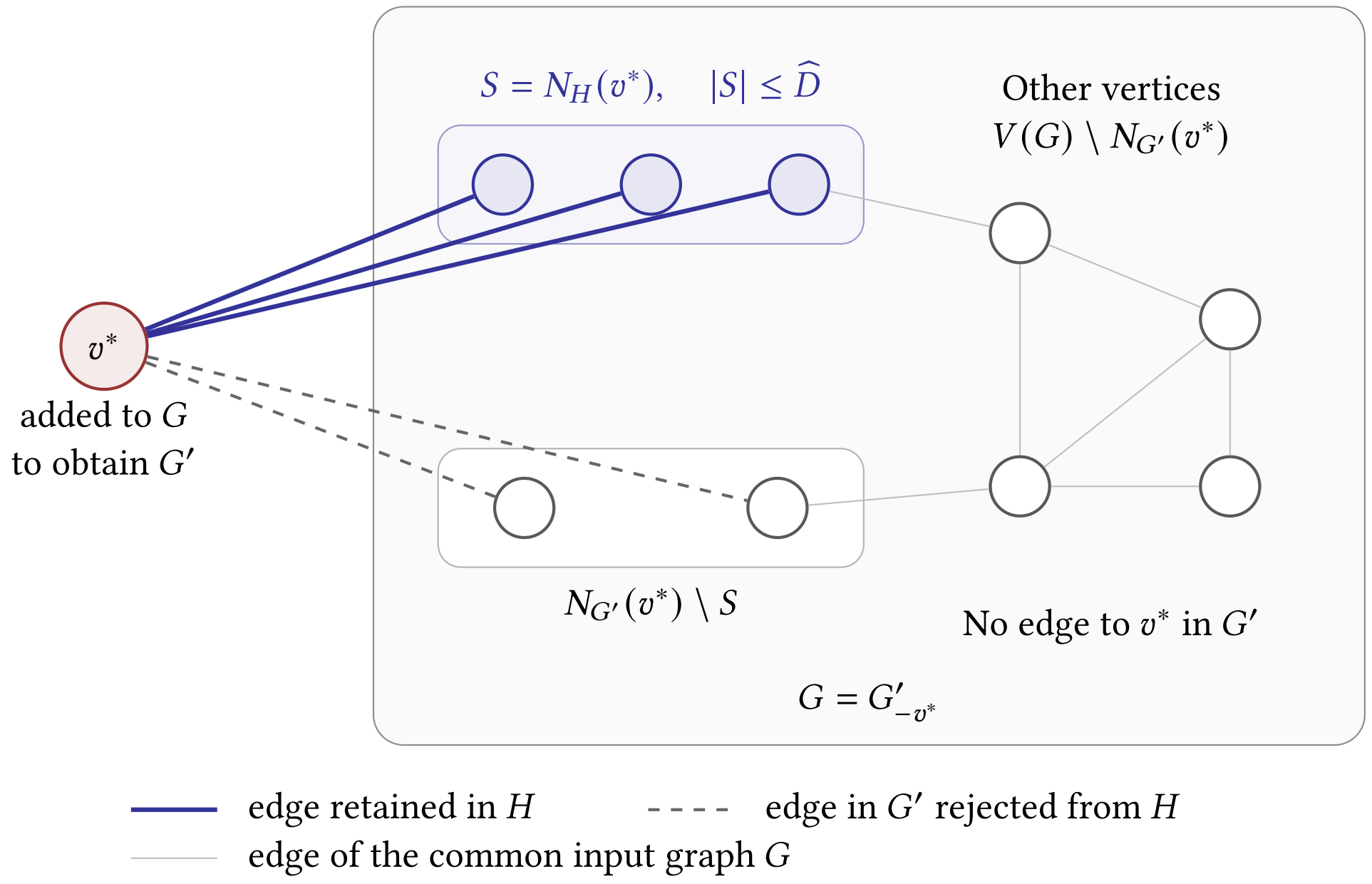

\mypar{Randomized counters and greedy execution.}
We instantiate each counter $t(u)$ using a continual-release threshold mechanism based on the Sparse Vector Technique (SVT)~\cite{DworkNRRV09,RothR10,HardtR10,DworkNPR10,ChanSS11} (\Cref{alg:svt-detector}), which continually announces whether $u$'s degree cutoff has been met while satisfying $(\teps,0)$-DP under a single-unit change to its stream of increments. Given $\dbound$ and the per-counter privacy parameter $\teps$, we set $\dmax = \dbound + O\bparen{\frac{\log(n/\delta)}{\teps}}$ so that, with probability at least $1-\delta$, every vertex $u$'s counter raises its flag when its number of accepted edges lies in $[\dbound, \dmax]$: no flag is raised before $u$'s degree reaches $\dbound$, and every flag is raised once $u$'s degree reaches $\dmax$ (\Cref{lem:svt-detector}).

Initially, the subgraph $H$ is empty. We scan through all pairs $\{u,v\}$ in a fixed ordering of $\binom{V'}{2}$: if $\{u,v\}$ is in the graph and neither $t(u)$ nor $t(v)$ has raised its flag, $\{u,v\}$ is added to $H$ and both counters receive an increment of $1$; otherwise, both counters receive an increment of $0$. Regardless, both counters update their output flags. 

Subject to all flags being raised within $[\dbound,\dmax]$, the ``projection and edge preservation'' guarantees of \Cref{thm:informal-proj} (\Cref{prop:edge-preservation}) hold immediately.\footnote{Checking at the end whether any counter flagged outside $[\dbound,\dmax]$ and falling back to any maximal $\dmax$-bounded subgraph on that $\delta$-probability event ensures these guarantees hold with probability $1$.} While scanning all pairs naively takes $\Theta(n^2)$ time, the algorithm can be implemented in $O(m\log m)$ time by ``fast-forwarding'' counters past runs of non-edges (\Cref{app:fast-stable-proj}).

\mypar{Analyzing distributional stability.}
The stability analysis is much more delicate. Since $v^*$ itself is deleted when comparing $\cP(G)$ and $\cP(G')_{-v^*}$, we can condition on a fixed ``good'' draw of $t(v^*)$'s randomness under which $v^*$ accepts at most $\dmax$ incident edges in $H' = \cP(G')$. Let $S \subseteq V'\setminus\{v^*\}$ be the (random) set of neighbors of $v^*$ in $H'$, so $|S| \leq \dmax$. Notice that for any neighbor of $v^*$ in $G'$ that is \emph{not} in $S$ (because $t(v^*)$ or $t(u)$ had already flagged), feeding $0$-increments on both skipped edges and non-edges ensures that $t(u)$ sees the exact same update at $\{u,v^*\}$ in $G$ and $G'$. Meanwhile, for the vertices in $S$, because their counters satisfy $(\teps,0)$-DP, we can shift their internal randomness to hide the single extra edge (increment of $1$ vs.\ $0$) that each receives from $v^*$ relative to the execution on $G$. After this shift, the executions on $G$ and $G'$ are \emph{identical} on all remaining pairs, and the randomness of counters outside $S$ need not be changed at all.

By the ``strong composition'' property of differential privacy~\cite{DworkRV10} across the $|S| \leq \dmax$ counters in $S$, the total distributional shift scales as $\teps\sqrt{\dmax \log(1/\delta)}$. Requiring this shift to be at most $\eps$ alongside the counter accuracy condition imposes the following constraints (suppressing constants):
\[
    \teps\sqrt{\dmax \log(1/\delta)} \leq \eps
    \qquad\text{and}\qquad
    \dmax \geq \dbound + \frac{\log(n/\delta)}{\teps}
    \, .
\]
Choosing the largest $\teps$ and smallest $\dmax$ satisfying both constraints (with $\delta = n^{-\Theta(1)}$) resolves to the bound on $\dmax$ in \Cref{thm:informal-proj}.

\mypar{Handling adaptive dependencies via concurrent composition.}
A rigorous proof is subtle because the set $S$ of $v^*$'s neighbors in $H'$ is random and potentially depends on the choices made by all other counters in the graph. Fortunately, this dependency is mediated entirely through the public outputs (flags) of the counters. The recently introduced \emph{concurrent composition} framework of \citet*{HenzingerSV26} (HSV) allows for exactly this type of adaptive dependency (\Cref{prop:HSV}), and a substantial part of our proof in \Cref{sec:proof-of-stable-proj} (\Cref{clm:stability-simulation}) consists of showing that our algorithm's structure fits their framework with sparsity $\dmax$. It is surprising (to us) that the HSV framework---which was designed to allow composition in unknown external environments---appears to be needed to analyze our algorithm, which is an entirely closed system; nevertheless, we are not aware of a proof of our projection's stability that does not either use their framework or reprove something of similar generality. 

Finally, for the setting where the graph size $n$ is private, we give a wrapper (\Cref{thm:private-size-stable-proj}) that privately upper-bounds $n$ and pads the graph with isolated dummy vertices before invoking the fixed-$n$ projection.

\subsubsection{Constructing and analyzing the Lipschitz extension for sums}

We now turn to graphs $G=(V,E)$ with edge weights given by a function $x:E\to [0,1]$. Recall that we wish to design a function $\tfsum$ that agrees with the sum of weights, $\fsum(G)$, on $\dbound$-bounded graphs, but is $\dbound$-Lipschitz on all of $\cG$. We define $\tfsum(G)$ as the result of running the deterministic greedy projection algorithm, scanning edges in decreasing order of weight (breaking ties, say, lexicographically), and returning the sum of the weights of the retained edges. When the input is $\dbound$-bounded, we retain all the edges and hence return $\fsum(G)$. 

To bound the Lipschitz constant, consider graphs $G$ and $G'$ that differ by the insertion of the node $v^*$ and its adjacent edges, so that $G = G'_{-v^*}$. Observe that the order of the shared edges (those in $G$) is the same in both executions. Let $\tilde{E}$ and $\tilde E'$ be the sets of edges retained by the projections on $G$ and $G'$, respectively. Our starting point is the analysis of \citet*{DayLL16}, who considered the unweighted setting. They argued that the symmetric difference $\tilde{E}\triangle \tilde E'$ can be written as a union of at most $\dbound$ edge-disjoint, simple paths, each of which \emph{alternates} between edges of $\tilde{E}$ and $\tilde E'$, originating at $v^*$ (\Cref{fig:alternating-path}). If we start from $\tilde E$ and make the changes represented by one path at a time, the internal nodes' degrees do not change; only the endpoints' degrees can change by 1. As a result, the sum of the degrees changes by at most $2\dbound$ and hence the number of edges, by at most $\dbound$.  In the weighted case, this analysis is insufficient, since the set of edges that change between the two graphs could have size $\Omega(n)$ even for $\dbound =1$. If the weights alternate between 0 and 1 along the paths, then the sum of weights of retained edges could vary dramatically between the two graphs. 

The key claim for our analysis is that, when we sort the edges by weight, \emph{the weights of the edges along each alternating path are monotonically nonincreasing} (\Cref{clm:path-diff}), roughly because the differences along the path happen at later and later steps of the algorithm (more precisely, we construct the paths by tracing through the two executions, extending the paths as the executions progress). If the edges on the path are $\hat e_1,\ldots,\hat e_r$, we have $1 \geq x(\hat{e}_1) \geq x(\hat{e}_2) \geq \cdots \geq x(\hat{e}_r) \geq 0$. Since the edges alternate between $\tilde{E}$ and $\tilde E'$, the net difference in $\tfsum(G)$ due to this path of changes is a telescoping sum,
$$x(\hat{e}_1) \underbrace{- x(\hat e_2) + x(\hat e_3)}_{\leq 0} - \cdots \pm  x(\hat{e}_r) \leq  x(\hat{e}_1) \leq 1 \, ,
$$
where the first inequality follows from the observation that the last term, $\pm x(\hat e_r)$, is either negative (when $r$ is even) or paired with the previous term (when $r$ is odd). A lower bound on the net change follows by a symmetric argument.
Since there are at most $\dbound$ paths, the total change in the sum is at most $\dbound$.

In \Cref{thm:Lip-ext-sum}, we actually prove a stronger property, namely that  the $\ell_1$ change in the list of sums of edge weights by node---the weighted analogue of the list of degrees---is bounded by $2\dbound$; see \Cref{sec:extension-for-sums}. 

\subsection{AI disclosure}

The main results and proof ideas in this paper were generated by the authors, without artificial assistance. LLMs were used to help with writing and editing parts of the paper, working out calculations, parameters and more standard proofs in the appendices, and to debug proofs. 

\section{Preliminaries}
\label{sec:preliminaries}

When stating asymptotic bounds, we assume $n=|V|\geq2$ and
$m=|E|\geq2$. When simplifying running times, we additionally assume
$n=O(m)$. 

Recall that $\cG$ is the set of all graphs, and $\cG_\dbound$ the set of graphs with maximum degree at most $\dbound$ (called $\dbound$-bounded graphs). Similarly, 
let $\cG^n$ be the set of $n$-node graphs, where the vertex set $V$ is identified with $[n]$, and let  $\cG^n_\dbound$ be the set of $D$-bounded graphs on $n$ nodes.

\begin{definition}[Node-neighboring graphs and node distance]
    Two graphs $G, G' \in \cG$ are {\em node-neighbors} if one can be obtained from the other by removing a vertex and all of its adjacent edges.

    The {\em node distance} $d(G, G')$ between two graphs $G$ and $G'$ is the smallest number of modifications needed to obtain $G'$ from $G$, where each modification is either the removal of one node and all of its adjacent edges, or the insertion of one node with arbitrary edges incident on it. This distance is the shortest-path metric defined by the node-neighboring relation.\footnote{The vertices $V$ are labeled, and thus it is easy to check if two graphs are neighbors, so there is no issue of isomorphism. However, determining the distance between two graphs is, in general, NP-hard.}
\end{definition}

\begin{obs}
    Two graphs $G = (V,E,x)$ and $G' = (V',E',x')$ are node-neighbors if $|V \triangle V'| = 1$ and either $G$ is an \emph{induced} subgraph of $G'$ or vice versa.  The node distance between two graphs $G=(V,E,x),G'=(V',E',x')$ is equal to $|V \setminus V_H| + |V' \setminus V_H| $, where $H=(V_H,E_H,x_H)$ is a largest graph that is an induced subgraph of both $G$ and $G'$.
\end{obs}

\begin{definition}[Fixed-vertex node-neighboring graphs]\label{def:fixed-node-neighbor}
    Two graphs $G = (V,E,x)$ and $G' = (V',E',x')$ are \emph{fixed-vertex node-neighbors} if there is a node $v^*\in[n]$ such that $G_{-v^*}=G'_{-v^*}$ and $v^*$ is isolated in one of $G,G'$ and $x$ and $x'$ agree on the common edges. Equivalently, one graph can be obtained from the other by deleting all edges incident to $v^*$, while keeping the vertex set $[n]$ fixed. We call $v^*$ the differing node.
\end{definition}

Node-neighboring graphs on a fixed vertex set are sometimes defined in terms of an arbitrary rewiring of all edges incident to a given node $v^*$. Such graphs are at node distance at most two according to our definition (since they are both adjacent to the graph in which $v^*$ is isolated).

\begin{definition}[Indistinguishability of distributions]
    Two distributions $P,P'$ over the same domain $\cY$ are $(\eps,\delta)$-indistinguishable if for every event $Y \subseteq \cY$
    \[
        P(Y) \leq e^{\eps} P'(Y) + \delta \quad \text{and} \quad
        P'(Y) \leq e^{\eps} P(Y) + \delta \, .
    \]
    We denote this relationship $P \approx_{\eps,\delta} P'$.
    For random variables $X$ and $X'$ taking values in the same set $\cY$, we say $X$ and $X'$ are  $(\eps,\delta)$-indistinguishable (and write $X\approx_{\eps,\delta} X'$) when their distributions are $(\eps,\delta)$-indistinguishable.
\end{definition}

\begin{definition}[Node differential privacy~\cite{HayLMJ09,BlockiBDS13,KasiviswanathanNRS13,ChenZ13}]
    An algorithm $A \from \cG \to \cY$ satisfies $(\eps,\delta)$-node differential privacy if for every pair of node-neighboring graphs $G,G' \in \cG$, we have 
    $$A(G) \approx_{\eps,\delta} A(G').$$
    \end{definition}

\begin{definition}[$\Delta$-Lipschitz functions]
    A function $h \colon \cG \to \R$  is {\em $\Delta$-Lipschitz} on a set $S$ of graphs if $|h(G) - h(G')| \leq \Delta \cdot d(G, G')$ for all $G,G'\in S$.  When $S=\cG,$ we simply say $f$ is $\Delta$-Lipschitz.
\end{definition}

\section{A Distributionally Node-Stable Projection}
\label{sec:stable-projection}

In this section we describe our distributionally node-stable projection onto the set of bounded degree graphs.

\begin{definition}[Graph projections]\label{def:projection}
Given  $\dbound,\dmax \in \R$ such that $0<\dbound<\dmax$, an algorithm $\cP$ is a \emph{$(\dbound,\dmax)$-projection} if, for every $G\in \cG$,
\begin{enumerate}
    \item $\cP(G)$ is a $\dmax$-bounded subgraph of $G$  with the same vertex set as $G$, and 
    \item $\cP$ acts as the identity on $\cG_\dbound$.
\end{enumerate}
If $\cP$ is randomized, we require these to hold with probability 1.
\end{definition}

Our construction has the additional property that, on input a graph of maximum degree larger than $\dbound$, it will return a $\dbound$-bounded subgraph with as many edges as possible, up to a factor of 2 (\Cref{prop:edge-preservation}). 

\begin{definition}[Distributional Node Stability]\label{def:distrib-stability}
    A randomized graph projection algorithm $\cP$ is 
    $(\eps,\delta)$-\emph{distributionally node-stable} 
    if, for every pair of node-neighboring graphs $G,G'$ such that $G'$ is obtained from $G$ by adding a node $v^*$ together with arbitrary incident edges,
    \[
        \cP(G) \approx_{\eps,\delta} \cP(G')_{-v^*}.
    \]
    Here $\cP(G')_{-v^*}$ denotes the random variable obtained by deleting $v^*$ and its adjacent edges from the random output $\cP(G')$; both random variables are graphs on the common vertex set $V(G)$.

    We also use a fixed-vertex set variant: We say $\cP$ is $(\eps,\delta)$-distributionally node-stable \emph{on $n$-node graphs} if, for every pair of fixed-vertex node-neighboring graphs $G,G'\in\cG^n$ with differing node $v^*$ (\Cref{def:fixed-node-neighbor}),
    \[
        \cP(G)_{-v^*} \approx_{\eps,\delta} \cP(G')_{-v^*}.
    \]

\end{definition}

The algorithm $\cP_\dbound$ is described in \Cref{alg:stable-proj}.
A quasilinear-time implementation, obtained by fast-forwarding the zero updates in the threshold detectors, is given in \Cref{app:fast-stable-proj}. 
The algorithm uses the sparse-vector mechanism of Hardt and Rothblum~\cite{HardtR10}, stated in the required form as the ``SVT detector'' \Cref{alg:svt-detector}.

The projection's accuracy and stability guarantees are captured in the following theorem.

\begin{algorithm}[t]
    \caption{Distributionally Node-stable Projection (\dnsp)}
    \label{alg:stable-proj}
    \KwIn{$G=(V,E)$, integer parameters $0<\dbound<\dmax$ and $\teps>0$}
    \KwOut{A subgraph of $G$}
    \BlankLine
    \Initialize $\tau \gets  (\dbound+\dmax)/2$\;
    \tcc{In the analysis, we also use parameter $k := \tfrac 1 2(\dmax-\dbound) = \dmax - \tau = \tau- \dbound$.}
    \For{$u \in V$}{
         $t(u) \gets $ \svtinit($\tau,\teps$)\;
         \tcc{If $k$ is large enough, \Cref{lem:svt-detector} ensures that all nodes raise their flag for some counter value in $[\dbound,\dmax]$.}
         }
    \Initialize $H \gets (V, \emptyset)$\;
    \BlankLine
    \For{unordered pairs $u,v \in V$ in lexicographic order}{
        \If{$\{u,v\}\in E$ and $t(u)\text{\rm .}\flagproc() =
            t(v)\text{\rm .}\flagproc() = \perp$}
            {
                Add $(u,v)$ to $H$\;
                $a=1$ \tcc{This will increment the endpoints' counters}
            }
        \Else{$a=0$ \tcc{No increment (due to a non-edge or to an edge that is skipped)}}
        $t (u)\text{.}\increment(a)$ \;
        $t (v)\text{.}\increment(a)$ \;
    }
    \BlankLine
    \If{\label{line:fallback} any counter raised its flag when the true value was below $\dbound$, or failed to raise its flag when the true value reached $\dmax$}
    {
        \Return any maximal $\dmax$-bounded subgraph of $G$\;
    }
    \Else
    {
        \Return $H$\;
    }
\end{algorithm}

\begin{lem}[Stability of Randomized Projection]\label{lem:stable-proj}
    For every $\dbound,\dmax\in\N$ with $0<\dbound<\dmax$ and every $\teps>0$, the algorithm $\cP_{\dbound,\dmax,\teps}$ satisfies the following.
    \begin{enumerate}
        \item (Projection) \label{item:proj-projection}\label{item:proj-low-degree}\label{item:proj-high-degree} 
        $\cP_{\dbound,\dmax,\teps}$ is a $(\dbound,\dmax)$-projection in the sense of \Cref{def:projection}.
        Moreover, for every input $G$, with probability $1$, its output $H=\cP_{\dbound,\dmax,\teps}(G)$ satisfies
        \begin{equation}\label{eq:proj-edge-preservation}
            \{u,v\}\in E(G)\setminus E(H)
            \quad\Longrightarrow\quad
            \max\{\deg_H(u),\deg_H(v)\}\geq \dbound.
        \end{equation}

        \item (Efficiency) \label{item:proj-runtime} $\cP_{\dbound,\dmax,\teps}$ can be implemented in time $O(m\log m)$ on inputs with $n$ vertices and $m$ edges.

        \item (Stability) \label{item:proj-stability} Let $c_1>0$ be the constant from \Cref{lem:svt-detector}.
        For all $\delta\in (0,1)$ and $n \in \N$,
        if
        \[
            \dmax - \dbound \geq \frac{2c_1\log(9n/\delta)}{\teps},
        \]
        then $\cP_{\dbound,\dmax,\teps}$ is $(\eps,\delta)$-distributionally node-stable on $n$-node graphs (\Cref{def:distrib-stability}), where
        \[
            \eps = \teps\sqrt{2\dmax \ln(3/\delta)} + \dmax \teps \frac{e^\teps -1 }{e^\teps + 1},
        \]
        provided $\eps\leq 1$.
        Assuming $\teps = O\Bparen{1/\sqrt{\dmax}}$, this simplifies to $\eps = O\Bparen{\teps\sqrt{\dmax \ln(1/\delta)}}$.
    \end{enumerate}
\end{lem}

For input graphs $G$ that are $\dbound$-bounded, the projection condition implies that the retained graph $H$ is $G$ itself. For inputs whose maximum degree exceeds $\dbound$, condition~\eqref{eq:proj-edge-preservation} ensures that not too many edges are dropped:

\begin{prop}[Edge preservation on general graphs]\label{prop:edge-preservation}
    Let $G=(V,E)$ be any graph and let $H$ be a subgraph of $G$ on $V$ satisfying \Cref{eq:proj-edge-preservation}. Then $H$ retains at least as many edges as some maximal $\dbound$-bounded subgraph of $G$. In particular, $H$ retains at least half as many edges as a maximum-size $\dbound$-bounded subgraph of $G$.
\end{prop}
\begin{proof}
    Start with $H$ and repeatedly delete an edge $\{v,u\}$ incident to a vertex $v$ of degree greater than $\dbound$. The only possible new violations of \Cref{eq:proj-edge-preservation} are omitted edges incident to $u$, and these can arise only if $u$ drops from degree $\dbound$ to $\dbound-1$. If a violation arises, adding one such edge restores $u$'s degree to $\dbound$ and repairs all violations. Each step reduces the total excess degree above $\dbound$ without increasing the number of edges. The process therefore ends with a $\dbound$-bounded subgraph $F$ that is maximal in $G$ and has at most $|E(H)|$ edges. Since any maximal $\dbound$-bounded subgraph $F$ is a $\frac{1}{2}$-approximation to a maximum-size $\dbound$-bounded subgraph of $G$, the second claim follows.
\end{proof}

In most cases, it makes sense to think of the final $\eps$ and $\delta$ for the stability guarantee as the true inputs, rather than $\teps$ and $\tdelta$. 
The following theorem packages \Cref{lem:stable-proj} in this form, without assuming that the number of vertices is public.  The wrapper uses part of the stability budget to compute a private upper estimate of the input size, and then invokes the projection with parameters calibrated to that estimate. We defer the proof to \Cref{app:private-size-projection}.

\begin{theorem}[Stable projection with a private input size]\label{thm:private-size-stable-proj}
    There is a randomized algorithm {\sadnsp} that, given a graph $G$, a degree
    bound $\dbound\in\N$, and stability parameters $\eps,\delta\in(0,1)$,
    outputs a pair $(\cP(G),\dmax(G))$ consisting of a graph and a released
    degree bound.

    \begin{enumerate}
        \item Conditioned on any possible value $\dmax$ of the second output, 
        the first component $\cP$ is a $(\dbound,\dmax)$-projection in the sense of \Cref{def:projection}.
        Moreover, with probability $1$, every edge of $G$ omitted from $\cP(G)$ has an endpoint of degree at least $\dbound$ in $\cP(G)$.
        \item For every choice of the parameters and every pair of node-neighboring graphs $G,G'$ such that $G'$ is obtained from $G$ by adding a node $v^*$ together with arbitrary incident edges,
        \[
            \bparen{\cP(G),\ \dmax(G)}
            \approx_{\eps,\delta}
            \paren{\cP(G')_{-v^*},\ \dmax(G')}.
        \]
        \item If $G$ has $n$ nodes, $\delta\leq1/2$, and $\eps=\Omega(1/n)$,
        then with probability $1$,
        \[
            \dmax(G)
            \leq
            \dbound+O\paren{\lambda_n\sqrt{\dbound}+\lambda_n^2},
            \qquad\text{where}\qquad
            \lambda_n
            =\frac{\log(n/\delta)\sqrt{\ln(1/\delta)}}{\eps}.
        \]
        \item On inputs with $n$ nodes and $m$ edges, 
        assuming $\delta\leq1/2$ and $\eps=\Omega(1/n)$,
        the algorithm can be implemented in time
        $O(m\log(n/\delta))$.
    \end{enumerate}

\end{theorem}

\begin{algorithm}[t]
    \caption{Sparse-vector-based Threshold Detector \cite{HardtR10,LyuSL17}}
    \label{alg:svt-detector}
    \KwIn{Threshold $\tau$, privacy parameter $\teps>0$}
    \BlankLine
    \AlgoProc{\svtinit($\tau,\teps$)}{
        \Initialize $\counter\gets 0$, $\flag\gets\bot$\;
        $\hat\tau \gets \tau + \eta$, where $\eta\sim\Lap(2/\teps)$\;
    }
    \BlankLine
    \AlgoProc{\increment($b$), where $b\in[0,1]$}{
        \If{$\flag=\bot$}{
            $\counter\gets\counter+b$\;
            $\hat c\gets\counter+\xi$, where $\xi\sim\Lap(4/\teps)$\;
            \lIf{$\hat c>\hat\tau$}{$\flag\gets\top$}
        }
    }
    \BlankLine
    \AlgoProc{\flagproc()}{
        \Return $\flag$\;
    }
\end{algorithm}

\subsection{Analyzing the randomized projection: proof of \Cref{lem:stable-proj}}
\label{sec:proof-of-stable-proj}

We use two standard properties of the SVT detector: it is $(\teps,0)$-differentially private as a continual-release mechanism under single-update changes to its input stream, even with adaptively chosen inputs \cite{JainRSS23}, and its threshold behavior is accurate when the Laplace noises are bounded. We state our required accuracy claim here, for completeness:

\begin{lem}[Accuracy of SVT Counter]\label{lem:svt-detector}
    For some $c_1>0$ and for all $n\in\N$, all graphs $G=(V,E)$ with $|V|\leq n$, and all $\beta < \frac 1 2$, with probability at least $1-\beta$,
    the threshold detectors $t(u)$ instantiated with SVT (\Cref{alg:svt-detector}) satisfy the following conditions, simultaneously for all $u\in V$, with $k = k_{n,\eps,\beta} = c_1\log(n/\beta)/\eps$:
    \begin{compactenum}
        \item The noise random variables satisfy $|\eta_u|, |\xi_{u,\ell}|<\frac 1 2 k$, for every call to $\increment$ on $t(u)$.
        \item $\flag$ is $\bot$ at all time steps for which $\counter < \tau - k$;
        \item $\flag$ is $\top$ at all time steps for which $\counter \geq \tau + k$.
    \end{compactenum}
\end{lem}

\begin{proof}
    Let $\eta_u\sim \Lap(2/\eps)$ be the threshold noise and let $\xi_{u,\ell}\sim \Lap(4/\eps)$ be the query noise for the $\ell$th call to $\increment$ on $t(u)$. There are at most $n$ variables of the first type and at most $n(n-1)$ of the second, since each detector receives one call to $\increment$ for each other vertex. 
    By the Laplace tail bound $\Pr{|\Lap(b)|>x}=e^{-x/b}$ and a union bound, for some absolute constant $c_1$, with probability at least $1-\beta$ we have $|\eta_u|,|\xi_{u,\ell}| < \frac{1}{2} c_1\log(n/\beta)/\eps = \frac 1 2 k$ simultaneously for all $u,\ell$. Conditioning on this event, if $\counter<\tau-k$ then $\hat c=\counter+\xi_{u,t}<\tau+\eta_u=\hat\tau$, so $\flag$ stays $\bot$; similarly, if $\counter \geq \tau+k$ then $\hat c  > \hat\tau$, so $\flag=\top$.
\end{proof}

We can now prove \Cref{lem:stable-proj}.  Fix $\dbound,\dmax\in\N$ with $0<\dbound<\dmax$ and $\teps>0$, and let
\[
    k=\frac{\dmax-\dbound}{2},\qquad \tau=\dbound+k.
\]
These are the values computed by \Cref{alg:stable-proj}.

\subsection{\Cref{item:proj-projection,item:proj-runtime}: analyzing correctness and running time}

\paragraph{\Cref{item:proj-projection}:}

The output is always a subgraph of $G$.  If the final check in \Cref{alg:stable-proj} fails, then the algorithm returns a maximal $\dmax$-bounded subgraph of $G$, so the output is $\dmax$-bounded.  If the final check does not fail, then every detector has raised its flag by the time its true counter reaches $\dmax$.  Thus, after a node has received $\dmax$ edges in $H$, all later incident edges are rejected, so $H$ is $\dmax$-bounded and the algorithm returns $H$.

For \Cref{eq:proj-edge-preservation}, if the final check fails, maximality of the returned $\dmax$-bounded subgraph ensures that every omitted edge has an endpoint of output degree $\dmax>\dbound$.  Otherwise, each omitted edge was rejected because an endpoint had already raised its flag.  Since no flag was raised below counter value $\dbound$, that endpoint has degree at least $\dbound$ in the output.

It remains to check that the algorithm acts as the identity on $\cG_\dbound$.  Fix $G\in\cG_\dbound$.  If the final check fails, then the algorithm returns a maximal $\dmax$-bounded subgraph of $G$.  Since $G$ is itself $\dbound$-bounded and $\dbound<\dmax$, the only maximal $\dmax$-bounded subgraph of $G$ is $G$ itself.  
If the final check does not fail, then no detector raised its flag at a time when its true counter was less than $\dbound$. Since $G$ is $\dbound$-bounded, immediately before any edge of $G$ is processed, each endpoint's true counter is less than $\dbound$. Thus both endpoint flags are still $\bot$, so every edge of $G$ is inserted into $H$.

\paragraph{\Cref{item:proj-runtime}:}
As written, \Cref{alg:stable-proj} takes time $\Theta(n^2)$ regardless of the actual number of edges in the graph. However, it can be implemented to run in time $O(m\log m)$ by ``fast-forwarding'' the counters in between actual edges. We describe this (reasonably standard) variation in
\Cref{app:fast-stable-proj}; \Cref{lem:stable-proj-fast} gives the exact statement.

\subsection{\Cref{item:proj-stability}: analyzing stability}

We now prove \Cref{item:proj-stability}.  Let $n\in\N$ and $\delta\in(0,1)$ satisfy
\(
    \dmax-\dbound\geq \frac{2c_1\log(9n/\delta)}{\teps},
\)
and assume $\eps\leq 1$.
Let $\tdelta=\delta/3$ and $\beta=\delta/9=\tdelta/3$, and observe that $k \geq \frac{c_1\log(n/\beta)}{\teps}$, where $k=(\dmax-\dbound)/2$ is the value computed in \Cref{alg:stable-proj}.

Let $G,G'\in\cG^n$ be fixed-vertex node-neighboring simple graphs with differing node $v^*$. Without loss of generality, assume that $v^*$ is isolated in $G$ and has an arbitrary set of incident edges in $G'$. We compare the induced outputs on $[n]\setminus\{v^*\}$.

The fallback step (see \Cref{line:fallback}) in \Cref{alg:stable-proj} is taken only if the accuracy guarantee of \Cref{lem:svt-detector} fails for at least one detector.  Since $k \geq \frac{c_1\log(n/\beta)}{\teps}$, \Cref{lem:svt-detector} implies that this happens with probability at most $\beta$.

Let $\widetilde{\cP}_{\dbound,\dmax,\teps}$ denote a modified version of \Cref{alg:stable-proj} which always returns the graph $H$ constructed in the main loop, without applying the final fallback step. It suffices to prove stability for $\widetilde{\cP}_{\dbound,\dmax,\teps}$ and then account for the fallback event on both inputs.

To analyze $\widetilde{\cP}_{\dbound,\dmax,\teps}$, we separate out the randomness of $v^*$'s detector. View the internal randomness of $t(v^*)$ as a presampled vector $\omega=(\eta_{v^*},\xi_{v^*,1},\ldots,\xi_{v^*,n-1})$ of independent Laplace random variables, where $\eta_{v^*}\sim\Lap(2/\teps)$ is the threshold noise and $\xi_{v^*,\ell}\sim\Lap(4/\teps)$ is the query noise used if $\flag=\bot$ at the $\ell$-th call to $\increment$ on $t(v^*)$. Consider the event
\begin{equation}\label{eq:goodevent}
    \goodevent=\set{\text{the presampled noise variables of $t(v^*)$ satisfy $|\eta_{v^*}|<\tfrac{k}{2}$ and $|\xi_{v^*,\ell}|<\tfrac{k}{2}$ for all $\ell\in[n-1]$}}.
\end{equation}

\Cref{lem:svt-detector} implies, in particular, that $\goodevent$ occurs with probability at least $1-\beta$.  
Observe that, conditioned on $\goodevent$, for every possible stream of increments given to $t(v^*)$, the detector has flag $\bot$ whenever $\counter<\tau-k = \dbound$ and has flag $\top$ whenever $\counter\geq\tau+k = \dmax$.

Fix any realization of the randomness $\omega$ of the detector $t(v^*)$ in $\goodevent$. 
Consider the distributions of the induced projections $\widetilde{\cP}_{\dbound,\dmax,\teps}(G)_{-v^*} $ and $\widetilde{\cP}_{\dbound,\dmax,\teps}(G')_{-v^*}$, in both cases with the same fixed randomness $\omega$ for $t(v^*)$.

The underlying induced graph on $V\setminus\{v^*\}$ is the same in both executions (namely, the edges of $G$).  The only ``input'' differences between the two executions arise from increments corresponding to edges incident on $v^*$.
On input $G$, none of these increments occur because $v^*$ is isolated. On input $G'$, the fact that  $\omega\in \goodevent$ 
ensures that at most $\dmax$ of these input differences occur, though the actual set of increments that differ may depend on the randomness and outputs of the other nodes' detectors. In our analysis, we want to get leakage (that is, divergence between the two distributions) that scales only with $\sqrt{\dmax}$. \Cref{fig:stable-projection-counters} illustrates the set of counters that receive a differing input at their pair with $v^*$.

Let $P^0_{\omega}$ and $P^1_{\omega}$ denote the distributions of $\widetilde{\cP}_{\dbound,\dmax,\teps}(G)_{-v^*} $ and $\widetilde{\cP}_{\dbound,\dmax,\teps}(G')_{-v^*}$, respectively.
To show that $P^0_{\omega}\approx_{\eps,\tdelta} P^1_{\omega}$, we show below that these two distributions can be realized as the two possible output distributions of a composition of the $n-1$ detectors, each of which is $(\teps,0)$-differentially private in the \emph{continual release model} \cite{DworkNPR10,ChanSS11}. This composition requires care to analyze, since it is \emph{concurrent}---the mechanisms are running simultaneously, and an output of one mechanism may affect later inputs to other mechanisms. We first give some background on concurrent composition (\Cref{sec:conc-comp-background}), and then analyze our algorithm using that framework (\Cref{sec:stab-conc-comp}).

\subsubsection{Background on concurrent composition}
\label{sec:conc-comp-background}

To analyze the composition, we use the  framework of \cite{HenzingerSV26}, of which we present a special case that is sufficient for our needs\footnote{The remarkably general framework of \cite{HenzingerSV26} allows for a wider range of privacy definitions in the component mechanisms, dynamic creation of mechanisms, dynamic setting of privacy parameters, and unbounded execution lengths; we do not need these since \Cref{alg:stable-proj} is highly structured.}. Following their framework, we define a \emph{concurrent parallel composition} of randomized mechanisms as an interaction between several parties: an adversary $\cA$, an identifier $\cI$ that receives as input a single bit $b$ (unknown to the adversary), and a collection of randomized interactive mechanisms $\{\cM^j\}_{j\in N}$, each with their own private randomness and state. In an execution of this interaction, the  adversary adaptively specifies a sequence of tuples $$(m_1[0],m_1[1],j_1), (m_2[0],m_2[1],j_2),...$$ Specifically, at step $i$, the adversary decides on and sends a tuple $(m_i[0],m_i[1],j_i)$ to the identifier, who forwards $m_i[b]$ to mechanism $j_i$ (thus, the adversary does not know which of its two messages was actually given as input to $\cM_{j_i}$). The mechanism $\cM_{j_i}$ then produces an output that is sent to the adversary, and the interaction moves on to step $i+1$. After a predetermined number of steps, the interaction terminates and the adversary produces its output. Let $\cA*\cI[b] *\{\cM^j\}_{j\in N}$ denote the random variable corresponding to an execution of this composition with input $b$ for the identifier.

We need a refinement to this concept, also due to \cite{HenzingerSV26}: For a given $k\in \N$,  a \emph{concurrent $s$-sparse parallel composition} is one for which two conditions hold: 
\begin{enumerate} 
\item in every execution, there are at most $s$ steps $i$ where the message pairs differ (that is, where $m_i[0]\neq m_i[1]$), and 
\item each mechanism is the recipient $j_i$ for at most one of these differing pairs. 
\end{enumerate}
They show that the information revealed about $b$ by such an interaction depends only on $s$, and not the total number of mechanisms. Specifically, we use the following result:

\begin{prop}[Corollary 5.5 in \cite{HenzingerSV26}]\label{prop:HSV}
    If $\cA, \cI$ define a $s$-sparse composition of mechanisms $\set{\cM_j}_{j\in \N}$ that are each $(\teps,0)$-differentially private (in the appropriate interactive model, called \emph{continual release with adaptive inputs}), we have
\[
\Bparen{\cA*\cI[0] *\{\cM^j\}}
\approx_{\eps,\delta'}
\Bparen{\cA*\cI[1] *\{\cM^j\}}
\]
where $\eps,\delta'$ are privacy parameters that apply to the standard, noninteractive composition of $s$ mechanisms that are $(\teps,0)$-differentially private. In particular, we may take $\eps = \teps\sqrt{2s\ln(1/\delta')}+s\teps\frac{e^{\teps}-1}{e^{\teps}+1}$ for any $\delta'>0$.
\end{prop}

\subsubsection{Proving distributional stability via concurrent composition}
\label{sec:stab-conc-comp}

\begin{clm}\label{clm:stability-simulation}
    For every $\omega \in \goodevent$ (see \Cref{eq:goodevent}) and every pair of node-neighboring graphs $G,G'$ that differ on node $v^*$, there is an adversary $\cA$ such that the distributions $P^b_{\omega}$ can be realized by the concurrent $\dmax$-sparse composition of $n-1$ copies of the SVT detector mechanism, using adversary $\cA$, in which the identifier's input is $b$. That is, for both $b\in \set{0,1}$,
    \[ \Bparen{ \cA*\cI[b]*\set{t(u)}_{u \neq v^*}}\sim P^b_{\omega} \, .
    \]
\end{clm}

\begin{proof}
    Fix $\omega\in \goodevent$ and neighboring graphs $G,G'$ as above, with $v^*$ isolated in $G$.  We describe an adversary $\cA$
    witnessing that $P^b_{\omega}$ is the output distribution of a concurrent parallel composition $\cA* \cI[b] * \{t(u):u\in V\setminus\{v^*\}\}$. Namely, when the identifier gets input $b=0$, the concurrent composition should simulate an execution of $\widetilde{\cP}_{\dbound,\dmax,\teps}$ on  $G$; when $b=1$, the composition should simulate an execution on $G'$.

    The adversary $\cA$ simulates \Cref{alg:stable-proj} on the hardwired induced graph on $V\setminus\{v^*\}$, using the fixed randomness $\omega$ for $t(v^*)$, and interacting with the detector $t(u)$ through mechanism $u$ whenever the algorithm needs to query $\flagproc()$ or invoke $\increment(\cdot)$ on that detector.

    For every unordered pair $\{x,y\}$, the adversary generates one or two message pairs (of the form tuple $(m[0],m[1],j)$, where $j$ is the recipient), and possibly updates the state of $t(v^*)$. 
    The message pairs should be understood branchwise: for each fixed identifier bit $b$, the messages actually delivered by $\cI[b]$ reproduce the execution on one graph: for $b=0$ (left messages), we reproduce an execution on $G$; for $b=1$, on $G'$.  The adversary maintains a single simulated copy of $t(v^*)$, using the fixed randomness $\omega$, that tracks the state it would have in the $G'$ execution.
    We show below that this does not distort the induced output on $V\setminus\{v^*\}$ when $b=0$.  
    
    Specifically, for each pair $\set{x,y}$:
    \begin{itemize}
    \item If $\{x,y\}$ does not involve $v^*$, the graph edge is the same in $G$ and $G'$, so $\cA$ sends identical messages to the corresponding mechanisms (that is, it sends either $(0,0,x)$ and $(0,0,y)$ or $(1,1,x)$ and $(1,1,y)$ to the identifier $\cI$). The state of $t(v^*)$ is unchanged.

    \item If $\{u,v^*\}$ is \emph{not} an edge in $G'$, then again the two graphs are identical; thus, $\cA$ sends the tuple $(0,0,u)$ and runs $t(v^*).\increment(0)$.

    \item For a pair $\{u,v^*\}$  that \emph{is} an edge of $G'$:
    \begin{compactitem}
        \item If, at the moment this pair is processed, either the fixed detector $t(v^*)$ or the detector $t(u)$ has already raised its flag, then this pair cannot affect the execution in either branch, so again $\cA$ sends $(0,0,u)$ and runs $t(v^*).\increment(0)$.
        \item Otherwise, $\cA$ sends a differing message pair  to mechanism $u$---namely, it sends the tuple $(0,1,u)$---and runs $t(v^*).\increment(1)$.
    \end{compactitem}
    \end{itemize}
    Finally, the adversary outputs only the subgraph of the projection $H$ induced by $V\setminus\set{v^*}$.

    When the identifier's input is $b=1$, the execution of this adversary with $\cI[1]$ reproduces exactly the execution of the projection on input $G'$.
    The more subtle argument concerns the setting $b=0$. In the concurrent composition with $\cA$ and $\cI[0]$, the detector $t(v^*)$ may fire (since it is being incremented as if the graph is $G'$).
    Such an event could never occur in a real execution on $G$, since $\omega\in \goodevent$ and the true count for $v^*$ is always 0. Nevertheless, the \emph{inputs to the other detectors will be the same} in the concurrent composition with $b=0$ as they are in the real execution with $G$---either way, $t(u)$ will be incremented with the bit 0.
    Therefore, the behavior of the projection restricted to the induced graph on $V\setminus \set{v^*}$  in $\cA*\cI[0]*\set{t(u)}_{u \neq v^*}$ is exactly given by $P^0_{\omega}$.

    It remains to argue that the concurrent composition is $\dmax$-sparse.
    Since $\omega\in \goodevent$, the detector $t(v^*)$ raises its flag once its true counter reaches $\dmax$. Every differing pair of inputs corresponds to an increment of 1 to $t(v^*)$ 
    in the $G'$ execution; there may thus be at most $\dmax$ differing inputs. Furthermore, each differing input goes to a different endpoint $u$. Thus, for each $b\in\{0,1\}$, the output distribution of
    $\cA * \cI[b] * \{t(u)\}_{u\neq v^*}$
    is exactly $P^b_{\omega}$, as claimed. 
\end{proof}

We can  apply \Cref{prop:HSV} (with $\delta'=\tdelta$) to the concurrent composition $\cA * \cI[b] * \{t(u)\}_{u\neq v^*}$ defined in the previous claim (\Cref{clm:stability-simulation}). 
Using the fact that the SVT detector mechanism is $(\teps,0)$-differentially private in the continual release model with adaptively chosen inputs, under single-update changes to the increment stream (see, for example, \cite[Theorem 2]{LyuSL17} for a proof in the nonadaptive setting, and \cite{JainRSS23} for discussion of the adaptive setting),
we conclude that

\begin{equation}\label{eq:eps-stab-bound}
P^0_{\omega} \approx_{\eps,\tdelta}P^1_{\omega} \quad \text{for}\quad
    \eps=\teps\sqrt{2\dmax\ln(1/\tdelta)}+\dmax\teps\frac{e^{\teps}-1}{e^{\teps}+1}.
\end{equation}

Since this holds for every realization $\omega\in \goodevent$, we can average over $\omega$, and translate back to the notation of projections, to see that the distributions are close conditioned on $\goodevent$. Letting $X\big|_A$ denote the (distribution of) the random variable $X$ conditioned on an event $A$, we have:
\[
    \widetilde{\cP}_{\dbound,\dmax,\teps}(G)_{-v^*}\Big|_{\goodevent}
    \quad
    \approx_{\eps,\tdelta}
    \quad 
    \widetilde{\cP}_{\dbound,\dmax,\teps}(G')_{-v^*}\Big|_{\goodevent}.
\]
Removing the conditioning on $\goodevent$ (which occurs with probability at least $1-\beta$ on the common probability space of $\omega$) incurs an additional failure probability of at most $\beta$. Therefore
\(
    \widetilde{\cP}_{\dbound,\dmax,\teps}(G)_{-v^*}\approx_{\eps,\tdelta+\beta}\widetilde{\cP}_{\dbound,\dmax,\teps}(G')_{-v^*}.
\)
As observed above, replacing $\widetilde{\cP}_{\dbound,\dmax,\teps}$ by the actual algorithm $\cP_{\dbound,\dmax,\teps}$ only changes the output on the fallback event, which has probability at most $\beta$ in each execution. Since modifying both sides of an $(\eps,\tdelta+\beta)$-indistinguishable pair on events of probability at most $\beta$ increases the additive slack by at most $(1+e^\eps)\beta$, we obtain
\[
    \cP_{\dbound,\dmax,\teps}(G)_{-v^*}
    \ 
    \approx_{\eps,\,\tdelta+(2+e^\eps)\beta}
    \  
    \cP_{\dbound,\dmax,\teps}(G')_{-v^*}.
\]
Since $\eps\leq 1$, $\tdelta=\delta/3$, and $\beta=\delta/9$, we have $\tdelta+(2+e^\eps)\beta \leq \frac{\delta}{3}+\frac{2+e}{9}\delta < \delta$, giving the desired $(\eps,\delta)$-distributional node stability.  Finally, substituting $\tdelta=\delta/3$ into \Cref{eq:eps-stab-bound} gives the desired bound on $\eps$.
This completes the proof of \Cref{item:proj-stability}, and thus of \Cref{lem:stable-proj}.

\subsection{Using the projection as preprocessing}
\label{sec:gen-trans-composition}

Our projection can be combined with \emph{any} algorithm that is differentially private on the domain of degree-$\dbound$ graphs to obtain an algorithm that is accurate on the

\begin{lem}[Composing a stable projection with a private algorithm]\label{lem:projection-plus-dp}
    Fix $\dbound\in\N$ and stability parameters $\peps,\pdelta\in(0,1)$.  On input $G$, let $(H,\dmax)$ be the output of \sadnspdetail{\dbound}{\peps}{\pdelta} (\Cref{thm:private-size-stable-proj}).  For each possible released bound $\ell$, let $\cA_\ell\colon\cG_\ell\to\cY$ be an $(\aeps,\adelta)$-node-differentially private algorithm. Then:
    \begin{enumerate}
        \item The algorithm
    \(
        \cB(G)=\bparen{\dmax,\cA_{\dmax}(H)},
    \)
    is $(\peps+\aeps,\pdelta+\adelta)$-node-differentially private on $\cG$. 
    \item For every $G\in\cG_\dbound$, conditioned on any possible released value $\dmax=\widehat d$, its second output has exactly the distribution of $\cA_{\widehat d}(G)$.
    \end{enumerate}  
\end{lem}

We would like to prove the Lemma by applying the composition theorem for approximate differential privacy. The usual adaptive composition framework does not fit directly, since it assumes that the second algorithm receives only the input and the (public) output from the first algorithm. In our setting, however, what is passed along is the projected graph, which is sensitive. Nevertheless, once we fix two adjacent graphs $G$ and $G'$, we can make a version of the projected graph public without harming privacy. The proof shows how to modify the process appropriately so that we can use the general result. 
We defer the detailed proof to \Cref{app:projection-plus-dp}.

\section{A Combinatorial Lipschitz Extension for Sums}
\label{sec:extension-for-sums}

Consider the function $\fsum$ that takes a graph with bounded edge weights and computes the sum of edge weights.  More precisely, $\fsum: \cG\to \R$ is defined by $\fsum(G) =\sum_{e \in E} x(e)$, where the weight function $x(e)$ takes values in $[0,1]$.  This function is $\dbound$-Lipschitz on the set $\cG_\dbound$ with respect to node distance, since adding a node with degree $\dbound$ adds $\dbound$ terms to the sum.
McShane's extension theorem~\cite{McShane34} states that every real-valued
$c$-Lipschitz function defined on a subset of a metric space extends to the
entire space with the same Lipschitz constant.
Therefore, there exists a $\dbound$-Lipschitz extension to the entire set $\cG$.  

In this section we give a nearly linear time algorithm that computes such a Lipschitz extension.  Our extension is based on the technique of Day, Li, and Lyu \cite{DayLL16} for unweighted graphs, but we order edges by descending $x(e)$, followed by lexicographic ordering to break ties.  Our analysis also borrows some ideas from Jain, Smith and Wagaman \cite{JainSW24}.

Our Lipschitz extension actually computes the more general \emph{weighted degree list}, which is a vector in $\R^V$ that contains the sum of edge weights incident to each vertex.  

\begin{definition}\label{def:f-list}
    The function $\flist:\cG\to\R^*$ takes a weighted graph $G=(V,E,x)$, where $x:E\to[0,1]$, and returns the \emph{weighted degree list} $\flist(G) \in \R^V$, that is the vector $s$ with entries $s(u) = \sum_{v:(u,v) \in E} x(u,v)$.
    \end{definition}

When the input is restricted to the set $\cG_\dbound$, this function is $2\dbound$-Lipschitz in the $\ell_1$ norm, since adding single edge changes two entries of $\flist(G)$ by at most 1 each, and neighboring graphs in $\cG_\dbound$ differ in at most $\dbound$ edges. A technical issue arises from the fact that two neighboring graphs $G$ and $G'$ may have different numbers of vertices, and so the vectors $\flist(G)$ and $\flist(G')$ may have different dimensions.  We resolve this by padding the smaller vector with zeros to make the two vectors the same dimension when computing their $\ell_1$ distance. 

Because $\fsum(G) = \frac{1}{2}\|\flist(G)\|_1$, a $2\dbound$-Lipschitz extension of $\flist$ from $\cG_\dbound$ to all of $\cG$ immediately gives a $\dbound$-Lipschitz extension of $\fsum$. \Cref{alg:sum-proj} presents an extension, $\tflist$, which can be computed in nearly linear time.

\begin{algorithm}[t]
    \caption{Computing the Lipschitz extension $\tflist$}
    \label{alg:sum-proj}
    \KwIn{$G=(V,E, x)$, parameter $\dbound > 0$}
    \KwOut{$\tflist(G) \in \R^V$}
    \BlankLine
    \Initialize $n = |V|$\;
    \For{$u \in V$}{
         \Initialize counter $t(u) = 0$  \;
         \Initialize sum $\tilde{s}(u) = 0$ \;
    }
    Sort $E$ in decreasing-lexicographical order to get ordering $S$ \;
    $\tilde{E} \gets \emptyset$ \;
    \For{edges $\{u,v\} \in S$ (in \emph{decreasing-lexicographic} order)}{
        \If{$\max \{ t(u) , t(v) \} < \dbound$}
        {
            Add $\{u,v\}$ to $\tilde{E}$ \;
            $\tilde{s}(u) \gets \tilde s(u) + x(\{u,v\})$ and $t(u)\text{++}$ \;
            $\tilde{s}(v) \gets \tilde s(v) + x(\{u,v\})$ and $t(v)\text{++}$ \;
        }
        }
    \Return $\tilde{s}$\;
\end{algorithm}

\begin{theorem} \label{thm:Lip-ext-sum}
    For every $\dbound \in \N$, the function $\tflist$ (\Cref{alg:sum-proj}) is a $2\dbound$-Lipschitz extension of $\flist$ from $\cG_\dbound$ to $\cG$, where outputs are compared in $\ell_1$ after padding missing vertices with zeros, that is computable in time $O(m\log m)$ on graphs with $n$ nodes and $m$ edges.
\end{theorem}

Observe that this Lipschitz guarantee is as tight as possible, since $2\dbound$ is the Lipschitz constant of $\flist$ on $\cG_D$. That is, this Lipschitz extension has \emph{stretch 1}.

This theorem immediately gives a nearly linear time Lipschitz extension for edge sums.

\begin{cor} \label{cor:Lip-ext-sum}
    For every $\dbound \in \N$, the function $\tfsum(G) = \frac12 \|\tflist(G)\|_1$ is a $\dbound$-Lipschitz extension of $\fsum$ from $\cG_\dbound$ to $\cG$ that is computable in time $O(m\log m)$ on graphs with $n$ nodes and $m$ edges.
\end{cor}

The rest of this section is devoted to the proof of \Cref{thm:Lip-ext-sum}.  The running time is immediate from the algorithm description, so we focus on proving the Lipschitz property.

\begin{proof}[Proof of \Cref{thm:Lip-ext-sum}]
Fix two neighboring graphs $G = (V,E,x)$ and $G' = (V',E',x')$ and assume without loss of generality that $G$ is an induced subgraph of $G'$.  That is, $V \subseteq V'$ and $E \subseteq E'$, and $x(e) = x'(e)$ for all $e \in E$.  Since the edge attributes $x$ and $x'$ agree on all edges in $E\subseteq E'$, we use $x$ to denote edge weights in both graphs, with the convention that $x(e)=x'(e)$ for $e\in E'$.  Let $v^*$ be the unique vertex in $V' \setminus V$ and let $E_{v^*} = \{e \in E': v^* \in e\}$ be the set of edges adjacent to $v^*$.  Let $\tilde{G} = (V,\tilde{E}, x)$ and $\tilde{G}' = (V',\tilde{E}',x)$ be the projections of $G$ and $G'$ computed by \Cref{alg:sum-proj}.  Since the projection does not modify the weights, we use $x$ to denote edge attributes in both projections.  Finally, let $\tilde{s}$ and $\tilde{s}'$ denote the weighted degree lists for the graphs $\tilde{G}$ and $\tilde{G}'$, respectively.

We would like to show that for any two neighboring graphs $G$ and $G'$, \Cref{alg:sum-proj} produces projections $\tilde{G} = (V,\tilde{E}, x)$ and $\tilde{G}' = (V',\tilde{E}',x)$ with weighted degree lists $\tilde{s} \in \R^V$ and $\tilde{s}' \in \R^{V'}$ such that
\begin{equation}\label{eq:sum_stability_bound}
\left\| \tilde{s} - \tilde{s}' \right\|_1 = \sum_{e \in \tilde{E'} \atop \text{incident on $v^*$}} x(e) + \sum_{v \in V}\Bigg| \sum_{e \in \tilde{E} \atop \text{incident on $v$}} x(e) - \sum_{e \in \tilde{E}' \atop \text{incident on $v$}} x(e) \Bigg|  \le 2D.
\end{equation}
That is, when we compute the projection of the two neighboring graphs $G$ and $G'$, the sum of the edge weights of the two projections should be close.  Note that the way we compute the $\ell_1$ norm in \Cref{eq:sum_stability_bound} implicitly pads $\tilde{s}$ so that the weighted degree of $v^*$ in $\tilde G$ is $0$.

Let $k = |E_{v^*}|$ be the degree of the node $v^*$, and let $(e_1, \dots, e_k)$ be the edges of $E_{v^*}$ listed in decreasing-lexicographic order so that $x(e_1) \geq x(e_2) \geq \dots \geq x(e_k)$. We consider the sequence of graphs $G_0, G_1, \dots, G_k$ formed by adding the edges of $E_{v^*}$ one at a time in this ordering.  Specifically, $G_0 = G$ and for every $i \in [k]$, $G_i = (V', E_i, x)$ where $E_i = E \cup \{e_1, \dots,  e_i\}.$ We show \Cref{eq:sum_stability_bound} by bounding the difference between the projections on consecutive graphs $G_{i-1}$ and $G_{i}$, and then applying the triangle inequality.

For every $i \in \set{0,...,k},$ let $\tilde{G}_{i} = (V', \tilde{E}_{i},x)$ denote the projection of $G_i$, and $\tilde{s}_{i}$ denote its weighted degree list. We will show that for every $i \in [k],$ 
\begin{equation}\label{eq:sum_increment}
\left\| \tilde{s}_{i} - \tilde{s}_{i-1} \right\|_1 \le 2.
\end{equation}
Furthermore, at most $\dbound$ of these differences can have nonzero norm. Our bound uses the following claim about the way the extension differs on two adjacent graphs.

\begin{clm}\label{clm:path-diff}
    For each $i \in [k]$, the symmetric difference $\tilde{E}_{i-1} \triangle\tilde{E}_{i}$ is a path starting at $v^*$ and along which the edge weights $x(\cdot)$ are nonincreasing. Furthermore, the path alternates between edges of $\tilde{E}_{i}$ and $\tilde{E}_{i-1}$.
\end{clm}

\begin{figure}[t]
\centering
\begin{tikzpicture}[
    vertex/.style={circle, draw, thick, minimum size=8mm, inner sep=0pt},
    unmatched/.style={thick},
    matched/.style={thick, dashed},
    weight/.style={font=\scriptsize, fill=white, inner sep=1pt},
    every node/.style={font=\small}
]

\node[vertex] (u)      at (0,0) {$v^*$};
\node[vertex] (v1)     at (2.0,0) {$v_1$};
\node[vertex] (v2)     at (4.0,0) {$v_2$};
\node[vertex, draw=none, minimum size=13mm] (dots) at (6.25,0) {$\cdots$};
\node[vertex] (vrm2)   at (8.5,0) {$v_{r-2}$};
\node[vertex] (vrm1)   at (10.5,0) {$v_{r-1}$};
\node[vertex] (vr)     at (12.5,0) {$v_r$};

\draw[unmatched] (u)    -- node[weight, above] {$\hat e_1$}     (v1);
\draw[matched]   (v1)   -- node[weight, above] {$\hat e_2$}     (v2);
\draw[unmatched] (v2)   -- node[weight, above] {$\hat e_3$}     (dots);
\draw[unmatched] (dots) -- node[weight, above] {$\hat e_{r-2}$} (vrm2);
\draw[matched]   (vrm2) -- node[weight, above] {$\hat e_{r-1}$} (vrm1);
\draw[unmatched] (vrm1) -- node[weight, above] {$\hat e_r$}     (vr);

\end{tikzpicture}
\caption{An alternating path
$v^*,v_1,\dots,v_r$ for odd $r$.
Solid edges represent additions to the graph, and dashed edges represent removals from the graph.}
\label{fig:alternating-path}
\end{figure}
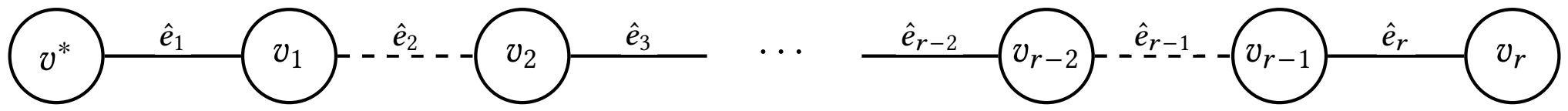

The content of the claim is illustrated in \Cref{fig:alternating-path}.  The proof of \Cref{clm:path-diff} follows ideas of \cite{DayLL16}, which showed that the difference is an alternating path, and \cite{JainSW24}, which observed that the path is monotone in the insertion order. We give a self-contained argument. 

\begin{proof}[Proof of \Cref{clm:path-diff}:]
Fix any $i \in [k]$.  
Consider the point at which \Cref{alg:sum-proj} reaches the unique differing edge $e_i = \{v^*, v_1\}$. If either of its endpoints is already saturated (that is, if its counter has reached $\dbound$), then $e_i$ will have no effect on the projected graph. The remaining edges will be processed identically in the two executions and the symmetric difference 
$\tilde{E}_{i-1} \triangle\tilde{E}_{i}$ will be empty (i.e. a path of length 0). So the Claim, in this case, is true. 

If neither endpoint is saturated, then 
the edge $e_i$ is added to $\tilde{E}_i$ and the counters of its endpoints are increased. Let $\hat e_1 = e_i$ (the first edge on the path.) 
The counters now differ only at $v^*$ and $v_1$.  
The list of remaining edges is the same in both $G_{i-1}$ and $G_i$, the order of the edges is the same. We can thus advance the two executions in lock step. The next time they differ (if they differ at all), it must be at $v_1$, and it must be that an edge, say $\hat e_2=(v_1,v_2)$  is added to $\tilde E_{i-1}$ but not to $\tilde E_i$. We have thus extended the alternating path by one edge. 
Because $\hat e_2$ occurred later in the sequence, $x(\hat e_1)\geq x(\hat e_2)$. After that edge is processed the two executions' counters will differ at $v^*$ and $v_2$. 

The same observations apply as the path grows.  Only $v^*$ and the current endpoint have different counters, so the next disagreement must involve that endpoint.  When this happens, its counters are $\dbound$ and $\dbound-1$.  The execution with the smaller counter accepts the edge, making the endpoint saturated in both executions and transferring the discrepancy to a new vertex.  Every endpoint left behind this way stays saturated, so it cannot occur in a later differing edge; nor can the path return to $v^*$. The execution getting the extra switches at each step.  Thus the differing edges form a simple alternating path $P=(\hat e_1,\ldots,\hat e_r)$ starting at $v^*$.

Since the algorithm processes edges by decreasing weight, the weight of the edges in $P$ are non-increasing, $1\geq x(\hat e_1) \geq \cdots \geq x(\hat e_r) \geq 0$, as desired.
\end{proof}

We can now justify \Cref{eq:sum_increment}. If the difference $\tilde{E}_{i-1} \triangle\tilde{E}_{i}$ is empty, the inequality is trivially satisfied. Otherwise, let $v^*,v_1,...,v_r$ be the nodes on the path $P$ along which $\tilde G_{i-1}$ and $\tilde G_i$ differ. We can expand the $\ell_1$ difference between the two weighted degree lists into the difference in the weighted degree of each node in $P$, where we group the nodes into the two endpoints $v^*$ and $v_{r}$ and the interior nodes $v_{j}$ for $j \in [r-1]$.
\begin{equation} \label{eq:sum_l1_expansion}
    \left\| \tilde{s}_{i} - \tilde{s}_{i-1} \right\|_1 = |\tilde{s}_{i}(v^*) - \tilde{s}_{i-1}(v^*)| + \sum_{j=1}^{r-1} |\tilde{s}_{i}(v_j) - \tilde{s}_{i-1}(v_j)| + |\tilde{s}_{i}(v_r) - \tilde{s}_{i-1}(v_r)|
\end{equation}
To bound \Cref{eq:sum_l1_expansion}, we can compute the absolute change in weight on different types of nodes:
\begin{itemize}
    \item For the node $v^*$, the absolute change in weight is $|\tilde{s}_{i}(v^*) - \tilde{s}_{i-1}(v^*)| = x(\hat{e}_1)$.
    
    \item For interior nodes $v_{j}$, the difference in the weighted degree is either $x(\hat{e}_j) - x(\hat{e}_{j+1})$ or $x(\hat{e}_{j+1}) - x(\hat{e}_{j})$, depending on the parity of $j$. Because the weights are nonincreasing along the path, the absolute value is always $x(\hat{e}_{j}) - x(\hat{e}_{j+1})$. 
    $$|\tilde{s}_{i}(v_j) - \tilde{s}_{i-1}(v_j)|  = x(\hat{e}_j) - x(\hat{e}_{j+1}).$$

    \item For the final node $v_{r}$, the absolute value of the difference is $|\tilde{s}_{i}(v_r) - \tilde{s}_{i-1}(v_r)| = x(\hat e_r)$, where the sign depends on whether $r$ is even or odd.
    \end{itemize}

Thus, we can rewrite \Cref{eq:sum_l1_expansion} to obtain \Cref{eq:sum_increment}:
\begin{align*}
    \left\| \tilde{s}_{i} - \tilde{s}_{i-1} \right\|_1 
    ={} & x(\hat{e}_1) + \sum_{j=1}^{r-1} \Bparen{x(\hat{e}_{j}) - x(\hat{e}_{j+1})} + x(\hat{e}_{r})
    ={} 2x(\hat{e}_1) \leq 2
\end{align*}

We now show \Cref{eq:sum_stability_bound}.  Notice that for any $i \in [k]$, $\tilde{E}_{i-1} \not= \tilde{E}_i$ if and only if the differing edge $e_i \in E_i \setminus E_{i-1}$ is included in $\tilde{E}_{i}$.  By construction of \cref{alg:sum-proj}, any vertex $v \in V_i$ can have at most $\dbound$ of its edges added to the projection, so it follows that $\tilde{E}_{i-1} \not= \tilde{E}_i$ at most $\dbound$ times.  Using the triangle inequality and \cref{eq:sum_increment} together with this observation gives us \Cref{eq:sum_stability_bound}. 
\end{proof}

\section{Applications to Vector Sums and SCO}
\label{sec:applications}

In this section, we demonstrate how both of our primitives---the combinatorial Lipschitz extension (\Cref{sec:extension-for-sums}) and the distributionally node-stable projection (\Cref{sec:stable-projection})---extend beyond scalar graph queries to high-dimensional machine learning and optimization over edge-associated records. In \Cref{sec:summing-vectors}, we lift both tools to the task of summing $\ell_2$-bounded edge vectors in $\R^d$, obtaining a constant-stretch Lipschitz extension for one-shot vector summation and a reusable sublinear-time query oracle for adaptive sequences of vector sums. In \Cref{sec:SCO-application}, we apply both vector-sum primitives to stochastic convex optimization (SCO) over edge records via noisy projected subgradient descent. Proofs and proof sketches for the results in this section appear in \Cref{app:applications-proofs}.

\subsection{Summing $\ell_2$-bounded edge attributes}
\label{sec:summing-vectors}

Let $B_2^d=\{z\in\R^d:\|z\|_2\leq1\}$, and let
$\cGvec{d}$ denote graphs $G=(V,E,x)$ with edge attributes
$x:E\to B_2^d$.
Write $\cGvecbound{d}{\dbound}$ for the $\dbound$-bounded graphs in this domain, and define
\[
    \vsum(G)=\sum_{e\in E(G)}x(e).
\]

The degree bound $\dbound$ and dimension $d$ are public parameters.
All privacy guarantees below hold on arbitrary graphs with the specified
norm-bounded attributes; accuracy is asserted on $\dbound$-bounded inputs.
Dividing the estimates by a public normalization gives corresponding
averaging guarantees.

\subsubsection{A constant-stretch extension and Gaussian privacy}

The vector sum $\vsum$ has $\ell_2$ sensitivity $\dbound$ on $\dbound$-bounded graphs.
Naively applying our scalar $\dbound$-Lipschitz extension (Corollary~\ref{cor:Lip-ext-sum}) to each of the $d$ coordinates separately would yield $\ell_2$ sensitivity $\dbound\sqrt{d}$, since each coordinate of a unit $\ell_2$ vector can be as large as $1$. To avoid losing a $\sqrt{d}$ factor, we first pass each edge vector through a Kashin representation, which spreads its norm nearly uniformly across $O(d)$ coordinates of magnitude $O(1/\sqrt{d})$:

Fix a public Kashin frame~\cite{LyubarskiiV10}: a matrix
$U\in\R^{d\times k}$ with $k=O(d)$ and $UU^\top=I_d$, together with
a deterministic encoder $a$ satisfying
\[
    Ua(z)=z,\qquad \|a(z)\|_\infty\leq C/\sqrt{k}
    \quad\text{for every }z\in B_2^d.
\]
The encoder need not be linear. Let $\kashintime(d)$ denote the time to
compute $a(z)$ for this fixed frame; a standard exact encoder has
$\kashintime(d)=\widetilde O(d^2)$. Decoding with $U$ takes $O(d^2)$ time.

Given an input graph $G=(V,E,x)$, we encode each edge attribute as $a(x(e))\in [-C/\sqrt{k},C/\sqrt{k}]^k$, apply our scalar extension $\tfsum$ (Corollary~\ref{cor:Lip-ext-sum}) coordinate-wise to the positive and negative parts of the encoded vectors (rescaled to $[0,1]$), and apply $U$ to decode the resulting $k$-dimensional vector back to $\R^d$. Because each of the $k=O(d)$ encoded coordinates has sensitivity $O(\dbound/\sqrt{k})$ and $UU^\top=I_d$, this preserves the $\ell_2$ sensitivity $\dbound$ up to a universal constant, independently of the dimension:

\begin{lem}[Lipschitz extension for vector sums]\label[lem]{lem:vector-sum-ext}
    For every $\dbound,d\in\N$, there is a map
    $F_{\dbound}:\cGvec{d}\to\R^d$ such that
    \begin{enumerate}
        \item $F_{\dbound}(G)=\vsum(G)$ for every $G\in\cGvecbound{d}{\dbound}$;
        \item $\|F_{\dbound}(G)-F_{\dbound}(G')\|_2\leq C\dbound$
        for every pair of node-neighboring inputs in $\cGvec{d}$,
        where $C$ is a universal constant.
    \end{enumerate}
    The map uses one Kashin representation per edge attribute and $O(d)$
    calls to the scalar extension in Corollary~\ref{cor:Lip-ext-sum}.
    On graphs with $n$ nodes and $m$ edges, it is computable in time
    \[
        O\left(m\,\kashintime(d)+dm\log m+d^2\right).
    \]
\end{lem}

We use Gaussian differential privacy (GDP)~\cite{DongRS19} with node
adjacency. Equivalently, an algorithm is $\mu$-node-GDP if, for every
$\eps\geq0$, it is $(\eps,\delta_\mu(\eps))$-node-DP, where
\[
    \delta_\mu(\eps)
    =\Phi\left(\frac\mu2-\frac\eps\mu\right)
      -e^\eps\Phi\left(-\frac\mu2-\frac\eps\mu\right),
\]
and $\Phi$ is the standard normal cumulative distribution function.

\begin{theorem}[Vector summation with constant-factor Gaussian noise]
\label{thm:vector-sum-gdp}
    For every $\dbound,d\in\N$ and $\mu>0$, there is a
    $\mu$-node-GDP algorithm on $\cGvec{d}$ of the form
    \[
        A(G)=F_{\dbound}(G)+Z,
        \qquad Z\sim\mathcal N(0,\sigma^2I_d),
        \qquad \sigma=C\dbound/\mu,
    \]
    for a universal constant $C$. On every $\dbound$-bounded input, its output
    is exactly $\vsum(G)+Z$. In particular,
    \[
        \left(\ex{}{\|A(G)-\vsum(G)\|_2^2}\right)^{1/2}
        =O\left(\frac{\dbound\sqrt d}{\mu}\right).
    \]
\end{theorem}

Under a promise that the graph is $\dbound$-bounded, the ordinary Gaussian
mechanism would use standard deviation $\dbound/\mu$ in each coordinate.
Thus \Cref{thm:vector-sum-gdp} achieves privacy on arbitrary graphs with
only a universal constant-factor increase in noise, and no additional
error on $\dbound$-bounded inputs.

\subsubsection{Sublinear-time summation with one-time preprocessing}

Now fix a public upper bound $T$ on the number of queries, and let each
edge of $G=(V,E,x)$ have an attribute $x(e)$ in a set $\cX$.
In round $t\in[T]$, a query specifies a function
$q_t:\cX\to B_2^d$, chosen using the previous answers, whose edge sum is
\[
    \vsumt{t}(G)=\sum_{e\in E(G)}q_t(x(e)).
\]
Assume random access to edge attributes: given an edge $e$ and a query
$q_t$, evaluating $q_t(x(e))$ takes $O(d)$ time.

To answer $T$ adaptive queries in sublinear time per query, we preprocess the graph topology once using $\sadnsp$ (\Cref{thm:private-size-stable-proj}) to obtain a retained subgraph $H$ and a released degree bound $\dmax$, storing references to the retained edges in an array. In each round $t\in[T]$, the oracle forms a mini-batch $B_t\subseteq E(H)$ by including each edge of $H$ independently with probability $p$, and returns the inverse-probability-weighted estimator $\frac{1}{p}\sum_{e\in B_t} q_t(x(e))$ perturbed with Gaussian noise calibrated via privacy amplification by Poisson subsampling~\cite{BalleBG18} and group privacy of size $\dmax$. The preprocessing reads only the graph topology and does not evaluate the queries on any edge attributes; both $H$ and the sampled edge identities $B_t$ remain internal.

\begin{theorem}[A reusable oracle for vector sums]
\label{thm:vector-sum-oracle}
    Fix $\dbound,d,T\in\N$, projection parameters
    $\peps,\pdelta\in(0,1/2)$, and per-query parameters
    $\seps,\sdelta\in(0,1/2)$.
    There is a preprocessing algorithm that outputs an internal data
    structure and a released degree bound $\dmax$. If $\peps=\Omega(1/n)$,
    where $n$ is the number of input vertices, then on graphs with $n$ nodes and $m$ edges preprocessing takes $O(m\log(n/\pdelta))$ time (which is $O(m\log m)$ when $\pdelta=n^{-\Theta(1)}$), independently of $d,T$, and
    \(
        \dmax\leq\dbound+O\left(\lambda_n\sqrt{\dbound}+\lambda_n^2\right),
    \)
    where $\lambda_n$ is as in \Cref{thm:private-size-stable-proj}, evaluated
    at $(\peps,\pdelta)$.
    In particular, if $\pdelta=n^{-\Theta(1)}$, then
    \[
        \dmax\leq\dbound+o(\dbound)
            +\frac{\operatorname{polylog}(n)}{\peps^2}.
    \]
    
    For any sampling rate $p\in[\seps/\dmax,1]$ chosen as a function of the released $\dmax$,
    the data structure supports at most $T$ adaptively chosen functions
    $q_t:\cX\to B_2^d$, with the following guarantees.
    \begin{enumerate}
        \item \emph{Privacy on arbitrary graphs.}
        For every $\delta_c\in(0,1)$, the joint transcript of $\dmax$ and all
        answers is
        \[
            \left(
                \peps+O\left(\seps\sqrt{T\log(1/\delta_c)}+T\seps^2\right),
                \pdelta+T\sdelta+\delta_c
            \right)\text{-node-DP}.
        \]
        In particular, the projection's privacy cost is paid only once.

        \item \emph{Accuracy on $\dbound$-bounded graphs.}
        Each answer $\hatvsumt{t}$ is unbiased and satisfies
        \[
            \ex{}{\|\hatvsumt{t}-\vsumt{t}(G)\|_2^2}
            \leq
            m\frac{1-p}{p}
            +O\left(\frac{\dmax^2d\log(\dmax/\sdelta)}{\seps^2}\right).
        \]
        These statements hold conditionally on $\dmax$, the preceding
        transcript, and the query $q_t$.

        \item \emph{Query time.}
        If the retained graph is $H$, each query takes expected time
        $O(d(1+p|E(H)|))$. On $\dbound$-bounded inputs this is $O(d(1+pm))$.
    \end{enumerate}
\end{theorem}

The oracle pays for the stable projection once and then answers each query
using a fresh sample of the retained edges. Its error combines sampling
variance and Gaussian noise. Unlike the extension method, its privacy noise
scales with the released degree bound $\dmax$, but each query evaluates
only the sampled edge attributes.

For optimization, we normalize the answers by a public target sample size $m$
and allocate a single privacy budget across preprocessing and all queries (if the edge count $|E(G)|$ is not public, a private estimate can be computed with sensitivity $\dbound$ via Corollary~\ref{cor:Lip-ext-sum}; for simplicity we state accuracy and runtime when $|E(G)|=m$).
The following specialization records the resulting accuracy guarantee.

\begin{cor}[A normalized query oracle for optimization]
\label{cor:sco-vector-oracle}
    Fix a public target sample size $m\in\N$ and $\eps,\delta\in(0,1/2)$.
    There is an $(\eps,\delta)$-node-DP normalized oracle obtained from
    \Cref{thm:vector-sum-oracle}. The query budget $1\leq T\leq m^2$ may
    be chosen using the released $\dmax$. Set
    \[
        p=\min\left\{1,\ \max\left\{\frac1m,
            \frac{\eps}{\dmax\sqrt{T\log(1/\delta)}}\right\}\right\},
        \qquad
        L_T=\log(1/\delta)\log(\dmax T/\delta).
    \]
    On a $\dbound$-bounded graph with exactly $m$ edges, each answer
    $\widehat g_t$ is unbiased for
    $g_t=\frac1m\sum_{e\in E(G)}q_t(x(e))$ and satisfies
    \[
        \ex{}{\|\widehat g_t-g_t\|_2^2}
        \leq\frac{1-p}{mp}
            +O\left(\frac{\dmax^2dT L_T}{m^2\eps^2}\right).
    \]
    Both statements hold conditionally on $\dmax$, the preceding
    transcript, and $q_t$.
\end{cor}

\subsection{Applications to stochastic convex optimization}
\label{sec:SCO-application}

Let $\cW\subseteq\R^d$ be a nonempty compact convex set of Euclidean
diameter at most $1$. Each edge attribute $x$ specifies a convex,
$1$-Lipschitz loss $\ell(\cdot;x)$ on $\cW$.
We normalize the losses so that $\ell(w_0;x)=0$ for every $x$, at a fixed
reference point $w_0\in\cW$.
Assume access to a subgradient
$g(w;x)\in\partial\ell(w;x)$ with $\|g(w;x)\|_2\leq1$, computable in
$O(d)$ time, and to Euclidean projection $\Pi_{\cW}$.
For a distribution $P$ on attributes, define the population objective and
its optimum by
\[
    \cL_P(w)=\ex{X\sim P}{\ell(w;X)},
    \qquad \cL_P^*=\min_{w\in\cW}\cL_P(w).
\]
We fix a public target sample size $m\in\N$. Accuracy is measured on any
fixed $\dbound$-bounded topology with exactly $m$ edges, whose attributes
are drawn independently from $P$; expectations include both these draws
and the algorithm's randomness. Privacy holds for arbitrary graphs and
arbitrary loss attributes, without the degree, size, or i.i.d.\ assumptions.
The algorithms always normalize by the public target $m$.

Both methods use averaged projected subgradient descent, initialized at a
fixed $w_1\in\cW$:
\[
    w_{t+1}=\Pi_{\cW}(w_t-\eta_t\widehat g_t),
    \qquad \overline w=\frac1T\sum_{t=1}^T w_t.
\]
All running times below are in addition to at most $m^2$ projections onto $\cW$.
Losses with Lipschitz constant $L$ can be handled by rescaling, multiplying
the excess-loss bounds by $L$.

\begin{prop}[SCO using the vector-sum extension]
\label[prop]{prop:sco-extension}
    For every $\mu>0$, there is a $\mu$-node-GDP algorithm with expected
    excess population loss
    \[
        \ex{}{\cL_P(\overline w)-\cL_P^*}
        =O\left(\min\left\{1,\frac1{\sqrt m}
            +\frac{\dbound\sqrt d}{m\mu}\right\}\right).
    \]
    On $m$-edge inputs, the algorithm uses at most $m^3$ subgradient
    evaluations and runs in time
    \[
        O\left(m^2\left[
            m\,\kashintime(d)+dm\log m+d^2\right]\right).
    \]
\end{prop}

This method recomputes the vector-sum extension at each gradient step.
It achieves the accuracy of a promised degree bound $\dbound$, with no
additional error from preprocessing. Setting $\mu=\Theta(\eps/\sqrt{\log(1/\delta)})$ yields an $(\eps,\delta)$-node-DP algorithm with excess population loss $O\left(\min\left\{1,\frac{1}{\sqrt{m}}+\frac{\dbound\sqrt{d\log(1/\delta)}}{m\eps}\right\}\right)$.

\begin{prop}[SCO using a preprocessed graph]
\label[prop]{prop:sco-projection}
    For every $\eps,\delta\in(0,1/2)$, there is an
    $(\eps,\delta)$-node-DP algorithm that, conditional on its released
    degree bound $\dmax$, satisfies
    \[
        \ex{}{\cL_P(\overline w)-\cL_P^*\mid\dmax}
        =\widetilde O\left(\min\left\{1,\frac1{\sqrt m}
            +\frac{\dmax\sqrt d}{m\eps}\right\}\right),
    \]
    where $\widetilde O$ suppresses logarithmic factors in $m$, $\dmax$, and $1/\delta$.
    If $\eps=\Omega(1/n)$, the released bound satisfies
    \[
        \dmax\leq\dbound+O\left(\lambda_n\sqrt{\dbound}+\lambda_n^2\right),
    \]
    where $\lambda_n$ is as in \Cref{thm:private-size-stable-proj}, evaluated
    at $(\eps,\delta)$.
    In particular, if $\delta=n^{-\Theta(1)}$, then
    \[
        \dmax\leq\dbound+o(\dbound)
            +\frac{\operatorname{polylog}(n)}{\eps^2}.
    \]
    On $m$-edge inputs, the algorithm uses $O(m^2)$ expected subgradient
    evaluations. If $\eps=\Omega(1/n)$ and $\delta=n^{-\Theta(1)}$,
    its expected running time is $O(dm^2)$.
\end{prop}

This method preprocesses the topology once and uses
Corollary~\ref{cor:sco-vector-oracle} to estimate each gradient from a fresh sample.
The larger degree bound $\dmax$ increases the privacy noise, while
subsampling reduces the total number of subgradient evaluations to $O(m^2)$.

\printbibliography[heading=bibintoc]

\appendix

\section{Degree Clipping Versus Abstention: Comparison to JSW}
\label{app:jsw-brittleness}

In this appendix, we work out the example discussed in \Cref{sec:intro-proj-results}, showing how our randomized projection and the test-and-project framework of Jain, Smith, and Wagaman (JSW)~\cite{JainSW24} behave when a small dense subgraph violates the target degree cutoff $\dbound$. Because our projection acts as degree clipping rather than a global test, we can set $\dbound$ to match the typical degree of the graph without losing stability or abstaining, and still estimate population quantities accurately from the surviving edges.

\paragraph{The graph and the estimation problem.}
For an integer $r \geq 3$, let $n = r^2$ and consider the graph
\[
    G = K_r \ \sqcup\ sK_2,
    \qquad s = \frac{n-r}{2},
\]
consisting of a disjoint union of an $r$-clique ($r = \sqrt{n}$) and a matching of $s$ isolated edges on the remaining $n - \sqrt{n}$ vertices. Both the clique and the matching contain exactly $s$ edges, so $G$ has $m = 2s = n - \sqrt{n} = \Theta(n)$ edges in total. Only $\sqrt{n}$ vertices have degree greater than $1$, even though those vertices are incident to half of the edges in $G$.

Suppose each edge $e \in E(G)$ carries an i.i.d.\ attribute $X_e \sim P$ drawn from an unknown distribution $P$ supported on the unit Euclidean ball $B_2^d \subseteq \R^d$, and our goal is to estimate the population mean $\theta = \ex{X \sim P}{X}$ while guaranteeing $(\eps,\delta)$-node-DP on all graphs (as in \Cref{sec:summing-vectors,sec:SCO-application}). For simplicity, we treat $s$ as a public target sample size for normalization; the graph topology and i.i.d.\ assumption on $X_e$ are used only for the utility analysis. Privacy holds unconditionally on arbitrary input graphs and edge attributes.

\paragraph{Our projection at cutoff $\dbound = 1$.}
Suppose we run $\sadnsp$ (\Cref{thm:private-size-stable-proj}) with cutoff $\dbound = 1$, splitting the $(\eps,\delta)$ privacy budget evenly between the projection and the downstream mean estimate. Let $(H, \dmax)$ be the output of the projection. By the edge-preservation guarantee (\Cref{eq:proj-edge-preservation}), every omitted edge must have at least one endpoint of degree at least $\dbound = 1$ in $H$; consequently, all $s$ edges of the matching $sK_2$ are retained in $H$ with probability $1$, and
\[
    s \leq |E(H)| \leq s + \frac{r\dmax}{2}.
\]
For constant $\eps \in (0,1/2)$ and $\delta = n^{-\Theta(1)}$, \Cref{thm:private-size-stable-proj} guarantees $\dmax = O(\log^3 n)$ with probability $1$, so $H$ retains $|E(H)| = \frac{n}{2} + O(\sqrt{n}\log^3 n) = (\frac{1}{2} + o(1))n$ edges while clipping degrees in the clique to $O(\log^3 n)$.

To estimate $\theta$, we take the first $s$ edges of $H$ in lexicographic order (or all edges of $H$ if $|E(H)| < s$), denote this set by $S(H)$, and release the noisy average
\[
    \widehat{\theta} = \frac{1}{s}\sum_{e \in S(H)} X_e + Z,
    \qquad
    Z \mid \dmax \sim \mathcal{N}\bparen{0,\, \frac{C\dmax^2\log(1/\delta)}{s^2\eps^2} I_d},
\]
along with $\dmax$. On any $\dmax$-bounded graph $H$, removing one vertex deletes at most $\dmax$ edges from $S(H)$ and allows at most $\dmax$ subsequent edges of $H$ to take their place in the prefix of length $s$, so the sum $\sum_{e \in S(H)} X_e$ has $\ell_2$ node sensitivity at most $2\dmax$. By our composition lemma (\Cref{lem:projection-plus-dp}), releasing $(\dmax, \widehat{\theta})$ is $(\eps,\delta)$-node-DP on all graphs for a suitable constant $C > 0$.

On our example graph $G = K_r \sqcup sK_2$, we always have $|S(H)| = s$. Because both the projection and the choice of $S(H)$ depend only on the graph topology (and internal randomness independent of $\{X_e\}$), the attributes $\{X_e\}_{e \in S(H)}$ are still i.i.d.\ draws from $P$. Their sample mean therefore has root mean squared error at most $1/\sqrt{s}$ around $\theta$, and combining this with the Gaussian noise gives
\[
    \bparen{\ex{}{\|\widehat{\theta} - \theta\|_2^2}}^{1/2}
    \leq \frac{1}{\sqrt{s}} + O\bparen{\frac{\sqrt{d\log(1/\delta)}\,\sqrt{\ex{}{\dmax^2}}}{s\eps}}
    = \widetilde{O}\bparen{\frac{1}{\sqrt{n}} + \frac{\sqrt{d}}{n}}.
\]
Note that dropping most of the clique edges does not bias the estimate of the population mean $\theta$: the $s$ surviving edges already provide $\Theta(n)$ fresh i.i.d.\ samples.

\paragraph{Behavior of JSW on the same instance.}
In the framework of \citet{JainSW24}, a graph is called $(\dbound,\ell)$-bounded if at most $\ell$ vertices have degree exceeding $\dbound$~\cite[Definition~3.2]{JainSW24}. We use $\dbound$ for the projection cutoff throughout. JSW's privacy analysis uses a private safety test before releasing the result of a downstream algorithm; outside the tested class, the test outputs $\bot$ with high probability~\cite[Claim~4.6]{JainSW24}. On $G = K_r \sqcup sK_2$, every vertex in the clique has degree $r - 1 = \sqrt{n} - 1$, while every vertex in the matching has degree $1$. Thus, for any $\dbound \geq 1$,
\[
    \abs{\{v \in V(G) : \deg_G(v) > \dbound\}}
    = \begin{cases}
        r, & 1 \leq \dbound < r-1, \\
        0, & \dbound \geq r-1.
    \end{cases}
\]
Consequently, $G$ is $(\dbound,\ell)$-bounded only when
\begin{equation}\label{eq:jsw-clique-matching-budget}
    \dbound + \ell \geq r - 1 = \Omega(\sqrt{n}).
\end{equation}
If $\dbound + \ell \ll \sqrt{n}$ (for instance, if the analyst sets $\dbound = 1$ and $\ell = O(1)$ to target the matching), the JSW safety test rejects $G$ with high probability and outputs $\bot$. For reliable acceptance, the input must also be far in node distance from graphs outside the $(\dbound,\ell)$-bounded class. We ignore this additional slack below, giving JSW the more favorable comparison.\footnote{JSW's published wrapper enlarges the degree cutoff to provide this slack~\cite[Algorithm~5]{JainSW24}.}

\paragraph{JSW followed by an edge-private estimator.}
For node-neighboring inputs in the safe class, JSW's BBDS-based projection has output edge distance at most $\dbound+\ell$~\cite[Theorem~3.3(2a)]{JainSW24}. This is an upper bound, essentially tight in the worst case, rather than a lower bound on the local sensitivity at our particular graph. Calibrating a downstream edge-private algorithm using this bound and group privacy gives budgets $\eps_{\mathrm{edge}}=\Theta(\eps/(\dbound+\ell))$ and $\delta_{\mathrm{edge}}=\Theta(\delta/(\dbound+\ell))$, after reserving a constant fraction of the privacy budget for the safety test~\cite[Theorem~4.1(1)]{JainSW24}.

Consider the same statistic $\frac {1}{s}\sum_{e\in S(H)}X_e$ as above. An edge insertion or deletion changes at most two terms of the prefix, so its $\ell_2$ edge sensitivity is at most $2/s$. The standard Gaussian mechanism calibrated using the preceding budgets therefore adds noise with root mean squared $\ell_2$ magnitude
\[
    \Theta\paren{\frac{(\dbound+\ell)\sqrt{d\log((\dbound+\ell)/\delta)}}{s\eps}}
    = \widetilde{\Omega}\paren{\frac{\sqrt{d}}{\sqrt{n}}},
\]
where the last bound uses \Cref{eq:jsw-clique-matching-budget}, $s=\Theta(n)$, and constant $\eps$. JSW also analyze a DLL-based greedy projection, whose stated node-to-edge bound is $\dbound+2\ell\sqrt{\min\{\dbound,\ell\}}$; calibrating to that bound gives no smaller order of noise in this example~\cite[Theorem~3.3(2b)]{JainSW24}.

\paragraph{JSW followed by a node-private estimator.}
Both JSW projections also have a node-to-node distance bound of $2\ell+1$ on the safe class~\cite[Theorem~3.3(3)]{JainSW24}. Thus, their framework can instead use a downstream algorithm that is node-private on $\dbound$-bounded graphs, with budgets $\eps_{\mathrm{node}}=\Theta(\eps/(2\ell+1))$ and $\delta_{\mathrm{node}}=\Theta(\delta/(2\ell+1))$~\cite[Theorem~4.1(2)]{JainSW24}. This route avoids the factor $\dbound+\ell$ in the privacy-budget reduction, but the Gaussian estimator itself has node sensitivity at most $2\dbound/s$. Calibrating to this sensitivity and these budgets gives root mean squared privacy noise
\[
    \Theta\bparen{\frac{\dbound(2\ell+1)\sqrt{d\log((2\ell+1)/\delta)}}{s\eps}}.
\]
Since $\dbound\geq1$ and $\ell\geq0$, we have
\[
    \dbound(2\ell+1)\geq\dbound+\ell\geq r-1.
\]
Consequently, this Gaussian calibration also has noise magnitude $\widetilde{\Omega}(\sqrt{d}/\sqrt{n})$ whenever the graph belongs to the safe class. A small node-distance bound alone therefore does not resolve the comparison: increasing the cutoff reduces the number of high-degree exceptions but increases the downstream node sensitivity.

\paragraph{Three parameter choices.}
Setting $\dbound=1$ retains the matching and clips the clique, but requires $\ell\geq r$, inflating the noise under either calibration. Raising $\dbound$ to at least $r-1$ allows fewer exceptions, but the downstream node sensitivity grows with $\dbound$, so either calibration still pays a factor $\Omega(r)$. Keeping both $\dbound$ and $\ell$ small instead causes the test to reject with high probability.

Thus, both standard Gaussian applications of JSW's stability guarantees either abstain or use a privacy-noise scale of at least $\widetilde{\Omega}(\sqrt{d}/\sqrt{n})$, compared to $\widetilde{O}(\sqrt{d}/n)$ under our projection. This is a comparison of the specified calibrations to class-wide stability and sensitivity bounds, not an instance-specific sensitivity lower bound or an information-theoretic lower bound against arbitrary node-private estimators. In particular, when $d=n$, our estimator achieves vanishing error $\widetilde{O}(n^{-1/2})$, whereas these JSW-based Gaussian calibrations have nonvanishing privacy noise whenever they release an estimate.
\space

\section{A Quasilinear-Time Implementation of the Projection}
\label{app:fast-stable-proj}

In this appendix we sketch a variant of \Cref{alg:stable-proj} that avoids scanning all $\binom{n}{2}$ unordered pairs.
The key observation is that, for a fixed vertex $u$, the detector $t(u)$ is updated once for each other vertex $v\neq u$, in increasing order of the label of $v$.
Thus, if we process only the actual edges of $G$, then between two consecutive incident edges of $u$ the detector sees a block of updates with value $0$.
Since these updates do not change the true counter, they can be simulated in aggregate.

For $u\in [n]$ and $v\in [n]\setminus\{u\}$, let
\[
    \pi_u(v)=
    \begin{cases}
        v & \text{if } v<u,\\
        v-1 & \text{if } v>u.
    \end{cases}
\]
This is the position at which the pair $\{u,v\}$ is processed in the update stream of detector $t(u)$.

The following detector extends \Cref{alg:svt-detector} with a subroutine $\texttt{skip-to}(r)$ that advances directly to the $r$-th local time step.
Here $\mathrm{Geom}(p)$ denotes the geometric distribution on $\{1,2,\ldots\}$ with success probability $p$.

\begin{algorithm}
    \caption{Fast-forwardable SVT Detector}
    \label{alg:ff-svt-detector}
    \KwIn{Threshold $\tau$, privacy parameter $\teps>0$}
    \BlankLine
    \AlgoProc{\texttt{FF-SVT-initialize}($\tau,\teps$)}{
        \Initialize $\counter\gets 0$, $\flag\gets\bot$\;
        $s\gets 1$ \tcc{Next local time step to be processed}
        $\hat\tau\gets\tau+\eta$, where $\eta\sim\Lap(2/\teps)$\;
    }
    \BlankLine
    \AlgoProc{\texttt{skip-to}($r$)}{
        \If{$\flag=\bot$ \textbf{and} $s<r$}{
            $p_0\gets\Pr[\xi\sim\Lap(4/\teps)]{\counter+\xi>\hat\tau}$\;
            $J\sim\mathrm{Geom}(p_0)$\;
            \lIf{$J\le r-s$}{$\flag\gets\top$}
        }
        $s\gets r$\;
    }
    \BlankLine
    \AlgoProc{\texttt{process}($a$), where $a\in\{0,1\}$}{
        \If{$\flag=\bot$}{
            $\counter\gets\counter+a$\;
            $\hat c\gets\counter+\xi$, where $\xi\sim\Lap(4/\teps)$\;
            \lIf{$\hat c>\hat\tau$}{$\flag\gets\top$}
        }
        $s\gets s+1$\;
    }
    \BlankLine
    \AlgoProc{\flagproc()}{
        \Return $\flag$\;
    }
\end{algorithm}

We can now process only the actual edges of the graph.
The algorithm below assumes that the edges of $G$ are supplied in the same lexicographic order used by \Cref{alg:stable-proj}; otherwise, one may first sort the edges by this order.

\begin{algorithm}
    \caption{Fast-forwarded Distributionally Node-stable Projection}
    \label{alg:stable-proj-fast}
    \KwIn{$G=(V,E)$ with vertex set $V=[n]$, integer parameters $0<\dbound<\dmax$, and $\teps>0$}
    \KwOut{A subgraph of $G$}
    \tcc{After every \texttt{skip-to} or \texttt{process} call, if a detector raises its flag when its counter is below $\dbound$, or reaches $\dmax$ without raising its flag, return any maximal $\dmax$-bounded subgraph of $G$.}
    \BlankLine
    \Initialize $\tau \gets (\dbound + \dmax)/2$\;
    \For{$u \in V$}{
        $t(u) \gets$ \texttt{FF-SVT-initialize}($\tau,\teps$)\;
    }
    \BlankLine
    List the edges of $G$ as $e_1,\ldots,e_m$ in lexicographic order on unordered pairs\;
    \Initialize $H \gets (V,\emptyset)$\;
    \For{$j=1,\ldots,m$}{
        Let $e_j=\{u,v\}$ with $u<v$\;
        $r_u \gets \pi_u(v)$ and $r_v \gets \pi_v(u)$\;
        $t(u)\text{\rm .skip-to}(r_u)$\;
        $t(v)\text{\rm .skip-to}(r_v)$\;
        \If{$t(u)\text{\rm .}\flagproc() = t(v)\text{\rm .}\flagproc() = \perp$}{
            Add $(u,v)$ to $H$\;
            $a\gets 1$\;
        }
        \Else{
            $a\gets 0$\;
        }
        $t(u)\text{\rm .process}(a)$\;
        $t(v)\text{\rm .process}(a)$\;
    }
    \For{$u \in V$}{
        $t(u)\text{\rm .skip-to}(n)$ \tcc{Simulate trailing zero updates after the last incident edge}
    }
    \BlankLine
    \Return $H$\;
\end{algorithm}

The following lemma records both the correctness and the running time of the fast-forwarded implementation.

\begin{lem}\label{lem:stable-proj-fast}
    Let $G=(V,E)$ be an $n$-vertex graph with $m$ edges.
    \begin{enumerate}
        \item Fix integers $0<\dbound<\dmax$ and $\teps>0$, and suppose the edges of $G$ are listed in the same lexicographic order used by \Cref{alg:stable-proj} (and that both algorithms use the same rule to select a maximal $\dmax$-bounded subgraph in the fallback step). Then the output distribution of \Cref{alg:stable-proj-fast} on input $G$ is identical to the output distribution of \Cref{alg:stable-proj} on the same input.
        \item If the edges of $G$ are given in lexicographic order, then \Cref{alg:stable-proj-fast} runs in time $O(n+m)$. In general, \Cref{alg:stable-proj-fast} can be implemented in time $O(n+m\log m)$ (which is $O(m\log m)$ when $n=O(m)$).
    \end{enumerate}
\end{lem}

\begin{proof}
    For part 1, we compare the two algorithms by examining the local update stream seen by each detector $t(u)$.
    In \Cref{alg:stable-proj}, detector $t(u)$ receives exactly one update for each $v\neq u$, in increasing order of the label of $v$.
    If the current state of $t(u)$ is $(\counter,\flag,\hat\tau)$ with $\flag=\bot$, then every skipped non-edge contributes an update of value $0$, so during such a block the counter remains equal to $\counter$ throughout.

    Consider a block of $\ell$ consecutive zero updates to some detector $t(u)$ while $\flag=\bot$ and $\counter=c$.
    In \Cref{alg:svt-detector}, at each step in this block the detector draws an independent $\xi\sim \Lap(4/\teps)$ and raises its flag exactly when $c+\xi>\hat\tau$.
    Thus each zero update causes the flag to raise with the same probability
    \[
        p_0=\Pr[\xi\sim \Lap(4/\teps)]{c+\xi>\hat\tau},
    \]
    independently across the $\ell$ steps.
    Therefore the first step in the block at which the flag raises has distribution $\mathrm{Geom}(p_0)$, truncated to the set $\{1,\ldots,\ell\}$, and the detector remains unflagged throughout the block precisely when this geometric random variable exceeds $\ell$.
    This is exactly the distribution simulated by $\texttt{skip-to}$ in \Cref{alg:ff-svt-detector}.
    In particular, after simulating a block of zero updates, the joint distribution of the detector state $(\counter,\flag)$ is the same in the two algorithms.

    Couple the executions by deferring the simulation of zero updates
    until a detector is next queried. Immediately before an actual edge
    $\{u,v\}$ is processed, advance its two endpoint detectors through
    their intervening zero blocks. The preceding argument couples these
    blocks with the full scan, so the endpoint states agree and both
    executions make the same edge decision. Their single-step updates
    can then use the same noises. Induction over the actual edges gives
    the same tentative graph in both executions, unless the fast algorithm
    has already returned the fallback.

    The fast algorithm checks for fallback after every detector operation.
    During a skipped zero block the counter is constant, so a flag raised
    below $\dbound$ is detected when the block is simulated. A counter can
    first reach $\dmax$ only in a single-step update, where failure to flag
    is detected immediately. The terminal calls to $\texttt{skip-to}(n)$
    cover any remaining zero updates. Thus the fast algorithm returns the
    fallback exactly when the full scan's historical condition holds.
    Using the same fallback rule gives identical output distributions.

    For part 2, initializing the $n$ detectors and running the final $\texttt{skip-to}(n)$ pass together take $O(n)$ time.
    Once the edges are listed in lexicographic order, the main loop of \Cref{alg:stable-proj-fast} iterates once per edge.
    In each iteration, the algorithm computes $\pi_u(v)$ and $\pi_v(u)$, calls $\texttt{skip-to}$ and $\texttt{process}$ on the two endpoints, checks their flags, and possibly inserts one edge into $H$.
    Each of these operations takes $O(1)$ time, so the total cost of the loop is $O(m)$.
    Each fallback check takes $O(1)$ time. If fallback is triggered, a maximal $\dmax$-bounded subgraph of $G$ can be constructed greedily in $O(n+m)$ time by scanning the edge list once.
    Hence the overall running time is $O(n+m)$ when the input edges are already sorted.

    If the edge list is not already in lexicographic order, we may first sort the $m$ edges, which takes $O(m\log m)$ time.
    Adding this preprocessing cost gives total running time
    $O(n+m\log m)$, which is $O(m\log m)$ when $n=O(m)$.
\end{proof}

\section{Proof of the Private-Size Projection Wrapper}
\label{app:private-size-projection}

\begin{proof}[Proof of \Cref{thm:private-size-stable-proj}]
    We describe the wrapper and verify the guarantees.  Split the stability budget into two constant-factor pieces, say $(\eps_s,\delta_s)=(\eps/2,\delta/2)$ for estimating the size and $(\eps_p,\delta_p)=(\eps/2,\delta/2)$ for the projection.

    First, we need a private upper estimate for the number of vertices.  Let
    \[
        B=\left\lceil \frac{1}{\eps_s}\log\paren{\frac{4}{\delta_s}}\right\rceil.
    \]
    Let $Z$ be a discrete Laplace random variable with parameter $\eps_s$, conditioned on the event $|Z|\leq B$; that is, $\Pr[Z=z]\propto e^{-\eps_s |z|}$ for integers $z\in[-B,B]$.
    On input $G=(V,E)$, set
    \[
        \widehat n=\max\{1,|V|+B+Z\}.
    \]
    Then deterministically
    \[
        \max\{1,|V|\}\leq \widehat n\leq |V|+2B
        =
        |V|+O\paren{\frac{\log(1/\delta)}{\eps}}.
    \]
    Before clamping, the estimate $|V|+B+Z$ is $(\eps_s,\delta_s)$-indistinguishable under node-neighboring changes: for neighboring inputs whose sizes differ by one, the probabilities of common outputs differ by at most a factor $e^{\eps_s}$, and the only non-overlapping endpoint masses have probability at most $e^{-\eps_s B}\leq \delta_s/4$ each. Clamping to at least $1$ preserves this guarantee by post-processing and ensures that the parameters below are defined even for the empty graph.

    Conditioned on a released value $\widehat n=N$, the wrapper pads the input with isolated dummy vertices so that the projection is run on a graph with vertex set $[N]$.
    Specifically, let $\pi\colon V(G)\hookrightarrow [N]$ be a uniformly random injection, and let $\overline{G}$ be the graph on $[N]$ in which $\{\pi(u),\pi(v)\}$ is an edge for each $\{u,v\}\in E(G)$ and the remaining $N-|V(G)|$ vertices in $[N]\setminus \pi(V(G))$ are isolated.

    We now choose the parameters for the base projection.  Let
    \[
        L_N=\log(9N/\delta_p),
        \qquad
        L_\delta=\ln(3/\delta_p),
        \qquad
        \lambda_N=\frac{L_N\sqrt{L_\delta}}{\eps_p}.
    \]
    For sufficiently large and small absolute constants $C_0,c_0>0$, set
    \[
        \dmaxN
        =
        \dbound+\left\lceil C_0\paren{\lambda_N\sqrt{\dbound}+\lambda_N^2}\right\rceil,
        \qquad
        \teps_N=\frac{c_0\eps_p}{\sqrt{\dmaxN L_\delta}}.
    \]
    Define $\dmax(G)=\dmaxN$, and let $\cP(G)$ be the graph on $V(G)$ obtained by taking the induced subgraph of $\cP_{\dbound,\dmaxN,\teps_N}(\overline{G})$ on $\pi(V(G))$ and applying $\pi^{-1}$ to restore the original vertex labels.  The wrapper outputs the pair $(\cP(G),\dmax(G))$.

    We check that these parameters satisfy the hypotheses and conclusion of \Cref{lem:stable-proj}.  The resulting privacy parameter in \Cref{lem:stable-proj} is at most
    \[
        c_0\eps_p\sqrt{2}
        +
        \dmaxN\teps_N\frac{e^{\teps_N}-1}{e^{\teps_N}+1}
        \leq
        c_0\eps_p\sqrt{2}+c_0^2\eps_p,
    \]
    where the last inequality uses $\teps_N\leq 1$ and $L_\delta\geq 1$, which hold after decreasing $c_0$ if necessary.  Thus this term is at most $\eps_p\leq 1$ for small enough $c_0$.
    The gap condition in \Cref{lem:stable-proj} asks for
    \[
        \dmaxN-\dbound
        \geq
        \frac{2c_1L_N}{\teps_N}
        =
        \frac{2c_1}{c_0}\lambda_N\sqrt{\dmaxN}.
    \]
    Since
    \[
        \lambda_N\sqrt{\dmaxN}
        \leq
        O\paren{\lambda_N\sqrt{\dbound}+\lambda_N^2}
    \]
    for our choice of $\dmaxN$, this holds by taking $C_0$ large enough as a function of $c_0$ and $c_1$.
    Therefore \Cref{lem:stable-proj} implies that, for each fixed released value $N$, the padded projection is $(\eps_p,\delta_p)$-distributionally node-stable on $N$-node graphs.

    Now consider node-neighboring graphs $G,G'$ where $G'$ is obtained from $G$ by adding $v^*$.  Condition on any common released size $N$ for which both padded inputs are defined.  We couple the random injections in the two executions by drawing a uniformly random injection $\pi\colon V(G')\hookrightarrow [N]$ and using its restriction $\pi|_{V(G)}$ on input $G$.  Under any fixed realization of $\pi$, every vertex $u\in V(G)$ receives the same label $\pi(u)\in [N]$ in both $\overline{G}$ and $\overline{G'}$, while $w^*=\pi(v^*)$ is an isolated dummy vertex in $\overline{G}$ and the differing vertex in $\overline{G'}$.  Thus $\overline{G}$ and $\overline{G'}$ are fixed-vertex node-neighbors on $[N]$ differing at $w^*$, so \Cref{lem:stable-proj} gives
    \[
        \cP_{\dbound,\dmaxN,\teps_N}(\overline{G})_{-w^*}
        \approx_{\eps_p,\delta_p}
        \cP_{\dbound,\dmaxN,\teps_N}(\overline{G'})_{-w^*}.
    \]
    For each fixed $\pi$, deleting the remaining dummy vertices in $[N]\setminus \pi(V(G'))$, applying $\pi^{-1}$ to restore the vertex labels in $V(G)$, and keeping the released value $\dmaxN$ are deterministic post-processing operations, so the same bound holds for each $\pi$ and hence after averaging over $\pi$.  Combining this with the $(\eps_s,\delta_s)$-indistinguishability of the released size estimate by the standard adaptive composition theorem gives the displayed $(\eps,\delta)$-indistinguishability of
    \[
        \paren{\cP(G),\dmax(G)}
        \quad\text{and}\quad
        \paren{\cP(G')_{-v^*},\dmax(G')}.
    \]

    The projection and identity properties follow immediately from \Cref{lem:stable-proj}: the padded graph $\overline{G}$ contains no edges incident to dummy vertices, and if $G$ is $\dbound$-bounded then so is $\overline{G}$.  The graph output is therefore a subgraph of $G$, equals $G$ on $\cG_\dbound$, and has maximum degree at most the released $\dmaxN$.  These properties hold for every realization, so conditioning on the released degree bound shows that the first component is a $(\dbound,\dmax)$-projection.
    Removing the isolated dummy vertices changes no degrees of original vertices, so \Cref{eq:proj-edge-preservation} also holds for the wrapper's output, including after conditioning on the released degree bound.

    Finally, because $\widehat n\leq n+O(\log(1/\delta)/\eps)$ deterministically, the displayed bound on $\dmax(G)$ follows by substituting this upper bound for $N$ in the definition of $\lambda_N$ and absorbing absolute constant factors into the big-$O$ notation.
    The fast-forwarded implementation initializes $\widehat n$ detectors and
    processes the sorted edge list in $O(m)$ time. Including sorting and
    using $n=O(m)$ for the original input gives the running time
    \[
        O\paren{\widehat n+m\log \widehat n}
        =
        O\paren{\frac{\log(1/\delta)}{\eps}+m\log\paren{n+\frac{\log(1/\delta)}{\eps}}}.
    \]
    If $\eps\geq c/n$ for a fixed $c>0$ and $\delta\leq1/2$, then
    \[
        n+\frac{\log(1/\delta)}{\eps}
        \leq n\bigl(1+c^{-1}\log(1/\delta)\bigr),
        \qquad
        \log\paren{\frac{n+\log(1/\delta)/\eps}{\delta}}
        =O\bigl(\log(n/\delta)\bigr).
    \]
    This gives the simpler choice of $\lambda_n$. The same inequalities,
    together with $n=O(m)$, give the simplified runtime
    $O(m\log(n/\delta))$. When $\delta=n^{-\Theta(1)}$, this is
    $O(m\log n)=O(m\log m)$.
\end{proof}

\section{Proof of the Projection Composition Lemma}
\label{app:projection-plus-dp}

\begin{proof}[Proof of \Cref{lem:projection-plus-dp}]
    Fix node-neighboring graphs $G,G'$, where $G'$ is obtained from $G$ by adding $v^*$ (and its adjacent edges). We can now consider an algorithm which has $G$ and $G'$ hardwired, and takes as input a single bit $b\in\{0,1\}$ that selects $G$ or $G'$.  We analyze an augmented two-stage algorithm on this two-element input domain.  

    In the first stage, run the projection $\sadnspdetail{\dbound}{\peps}{\pdelta}$ on the selected input, writing its output as $(H,\dmax)$ on $G$ and $(H',\dmax')$ on $G'$, and release
    \[
        Z(b)=
        \begin{cases}
            (H,\dmax), & b=0,\\
            (H'_{-v^*},\dmax'), & b=1.
        \end{cases}
    \]
    By \Cref{thm:private-size-stable-proj}, the process $Z()$ is $(\peps,\pdelta)$-differentially private on the input space $\{0,1\}$.

    For the second stage, fix a first-stage output $z=(h,\ell)$.  On input $b=0$, output $\cA_\ell(h)$.  On input $b=1$, draw the full projection $H'$ from its conditional distribution given $(H'_{-v^*},\dmax')=(h,\ell)$ and output $\cA_\ell(H')$. (In the case that the conditioning event has probability zero, define the second-stage output to be $\cA_\ell(h)$.) 
    This conditional sampling preserves the joint distribution of the truncated projection and the full projection from the original execution. The joint output of the first and second stages is thus distributed as either
    $$
        (\dmax, H, A_{\dmax}(H)) \ \text{for}\ b=0 
        \quad \text{and} \quad
        (\dmax', H'_{-v^*}, A_{\dmax'}(H'))\ \text{for}\ b=1 
        \, . 
    $$
    The second stage is also differentially private, since the every graph that can be sampled when $b=1$ is an $\ell$-bounded node-neighbor of $h$. 
    
    Basic adaptive composition (e.g., \cite[Theorem~3.16 and Appendix~B]{DworkR14}) now gives $(\peps+\aeps,\pdelta+\adelta)$-differential privacy for the transcript $(Z,Y)$, where $Y$ is the second-stage answer.  The actual output of our algorithm $\cB$, which includes only only the degree bound and $Y$ (not the projected graph), is a post-processing of $(Z,Y)$.  Since the neighboring pair was arbitrary, this proves the first claim.

    For the second claim, if $G\in\cG_\dbound$, then $H=G$ with probability $1$.  Consequently, conditioned on $\dmax=\widehat d$, the second output has exactly the distribution of $\cA_{\widehat d}(G)$.
\end{proof}

\section{Proofs from \Cref{sec:applications}}
\label{app:applications-proofs}

In this appendix, we give the proofs for the vector-summation results in \Cref{sec:summing-vectors} (\Cref{lem:vector-sum-ext,thm:vector-sum-gdp,thm:vector-sum-oracle} and Corollary~\ref{cor:sco-vector-oracle}) and succinct proof sketches for the stochastic convex optimization results in \Cref{sec:SCO-application} (\Cref{prop:sco-extension,prop:sco-projection}).

\subsection{Proof of \Cref{lem:vector-sum-ext}}
\label{app:proof-vector-sum-ext}

\begin{proof}[Proof of \Cref{lem:vector-sum-ext}]
    Fix the public Kashin frame $U\in\R^{d\times k}$ with $k=O(d)$ and $UU^\top=I_d$, together with the deterministic encoder $a\colon B_2^d\to [-C_0/\sqrt{k},C_0/\sqrt{k}]^k$ satisfying $Ua(z)=z$ and $\|a(z)\|_\infty\leq C_0/\sqrt{k}$ for all $z\in B_2^d$.
    Let $M=C_0/\sqrt{k}$. Given an input graph $G=(V,E,x)\in\cGvec{d}$, for each edge $e\in E$ and each coordinate $j\in[k]$, decompose the $j$-th coordinate of $a(x(e))$ into scaled non-negative parts in $[0,1]$:
    \[
        a_j^+(e)=\frac{\max\{a(x(e))_j,0\}}{M}\in[0,1],
        \qquad
        a_j^-(e)=\frac{\max\{-a(x(e))_j,0\}}{M}\in[0,1],
    \]
    so that $a(x(e))_j = M\bigl(a_j^+(e)-a_j^-(e)\bigr)$.
    Let $G_j^+=(V,E,a_j^+)$ and $G_j^-=(V,E,a_j^-)$ denote the graph $G$ equipped with scalar edge weights $a_j^+$ and $a_j^-$, respectively. Using the deterministic $\dbound$-Lipschitz scalar extension $\tfsum$ from Corollary~\ref{cor:Lip-ext-sum}, define the vector $\widetilde{a}(G)\in\R^k$ coordinate-wise by
    \[
        \widetilde{a}(G)_j = M\bigl(\tfsum(G_j^+) - \tfsum(G_j^-)\bigr)
        \qquad\text{for } j\in[k],
    \]
    and set $F_{\dbound}(G) = U\widetilde{a}(G)$.

    If $G\in\cGvecbound{d}{\dbound}$, then $G_j^+$ and $G_j^-$ are $\dbound$-bounded for every $j\in[k]$. By Corollary~\ref{cor:Lip-ext-sum}, $\tfsum(G_j^\pm)=\fsum(G_j^\pm)$, so $\widetilde{a}(G)_j = \sum_{e\in E} M(a_j^+(e)-a_j^-(e)) = \sum_{e\in E} a(x(e))_j$. By linearity of $U$ and the reconstruction property $Ua(z)=z$,
    \[
        F_{\dbound}(G) = U\sum_{e\in E} a(x(e)) = \sum_{e\in E} Ua(x(e)) = \vsum(G),
    \]
    establishing item~1.

    For item~2, consider any pair of node-neighboring graphs $G,G'\in\cGvec{d}$. For each $j\in[k]$, $G_j^+$ and $(G')_j^+$ are node-neighbors with edge weights in $[0,1]$, and likewise for $G_j^-$ and $(G')_j^-$. Since $\tfsum$ is $\dbound$-Lipschitz by Corollary~\ref{cor:Lip-ext-sum}, the triangle inequality gives
    \[
        \abs{\widetilde{a}(G)_j - \widetilde{a}(G')_j}
        \leq M\abs{\tfsum(G_j^+)-\tfsum((G')_j^+)} + M\abs{\tfsum(G_j^-)-\tfsum((G')_j^-)}
        \leq 2M\dbound = \frac{2C_0\dbound}{\sqrt{k}}.
    \]
    Summing the squared differences over all $k$ coordinates yields $\|\widetilde{a}(G)-\widetilde{a}(G')\|_2 \leq \sqrt{k}\cdot(2C_0\dbound/\sqrt{k}) = 2C_0\dbound$. Since $UU^\top=I_d$, $U$ has spectral norm $\|U\|_{\mathrm{op}}=1$, and therefore
    \[
        \|F_{\dbound}(G)-F_{\dbound}(G')\|_2
        = \|U(\widetilde{a}(G)-\widetilde{a}(G'))\|_2
        \leq \|\widetilde{a}(G)-\widetilde{a}(G')\|_2
        \leq 2C_0\dbound =: C\dbound.
    \]

    Finally, encoding $m$ edge vectors takes $O(m\,\kashintime(d))$ time, evaluating $\tfsum$ across the $2k=O(d)$ coordinate graphs takes $O(dm\log m)$ time by Corollary~\ref{cor:Lip-ext-sum}, and multiplying $\widetilde{a}(G)\in\R^k$ by $U\in\R^{d\times k}$ takes $O(dk)=O(d^2)$ time.
\end{proof}

\subsection{Proof of \Cref{thm:vector-sum-gdp}}
\label{app:proof-vector-sum-gdp}

\begin{proof}[Proof of \Cref{thm:vector-sum-gdp}]
    By \Cref{lem:vector-sum-ext}(2), the map $F_{\dbound}$ has global $\ell_2$ node sensitivity at most $C\dbound$ on $\cGvec{d}$. By the Gaussian mechanism characterization of Gaussian differential privacy~\cite{DongRS19}, adding spherical Gaussian noise $Z\sim\mathcal{N}(0,\sigma^2 I_d)$ with $\sigma=C\dbound/\mu$ satisfies $\mu$-node-GDP on all of $\cGvec{d}$.
    For any $\dbound$-bounded input $G\in\cGvecbound{d}{\dbound}$, \Cref{lem:vector-sum-ext}(1) ensures $F_{\dbound}(G)=\vsum(G)$, so $A(G)=\vsum(G)+Z$ and
    \[
        \bparen{\ex{}{\|A(G)-\vsum(G)\|_2^2}}^{1/2}
        = \bparen{\ex{}{\|Z\|_2^2}}^{1/2}
        = \sigma\sqrt{d}
        = \frac{C\dbound\sqrt{d}}{\mu}.\qedhere
    \]
\end{proof}

\subsection{Proof of \Cref{thm:vector-sum-oracle}}
\label{app:proof-vector-sum-oracle}

\begin{proof}[Proof of \Cref{thm:vector-sum-oracle}]
    \emph{Preprocessing.}
    Given $G=(V,E,x)$, run $\sadnspdetail{\dbound}{\peps}{\pdelta}$ (\Cref{thm:private-size-stable-proj}) on the underlying unweighted graph $(V,E)$ to obtain a subgraph $H$ and a released degree bound $\dmax$, and store pointers to the edges of $E(H)$ in an array. By \Cref{thm:private-size-stable-proj}, when $\peps=\Omega(1/n)$ and $\pdelta\in(0,1/2)$, this preprocessing runs in $O(m\log(n/\pdelta))$ time (which is $O(m\log m)$ when $\pdelta=n^{-\Theta(1)}$) and outputs $\dmax\leq \dbound + O(\lambda_n\sqrt{\dbound}+\lambda_n^2)$. Applying Young's inequality $\lambda_n\sqrt{\dbound}\leq \eta\dbound + \lambda_n^2/(4\eta)$ with $\eta=1/\log n$ gives $\dmax\leq \dbound+o(\dbound)+\operatorname{polylog}(n)/\peps^2$ when $\pdelta=n^{-\Theta(1)}$.

    \mypar{Adaptive query mechanism on $\cG_\ell$.}
    For each possible released degree bound $\ell\in\N$, define a $T$-round adaptive query mechanism $\cA_\ell$ on $\ell$-bounded graphs $H\in\cG_\ell$ as follows. Given a sampling rate $p\in[\seps/\ell,1]$, in each round $t\in[T]$, upon receiving a query $q_t\colon\cX\to B_2^d$, form a random mini-batch $B_t\subseteq E(H)$ by including each edge $e\in E(H)$ independently with probability $p$, and output
    \[
        \hatvsumt{t} = \frac{1}{p}\sum_{e\in B_t} q_t(x(e)) + Z_t,
        \qquad
        Z_t\sim\mathcal{N}(0,\sigma_\ell^2 I_d),
        \qquad
        \sigma_\ell = \frac{C_1\ell\sqrt{\log(2\ell/\sdelta)}}{\seps},
    \]
    for a sufficiently large universal constant $C_1>0$.

    \mypar{Privacy on arbitrary graphs (item 1).}
    Fix $\ell\in\N$ and consider a single round $t$ of $\cA_\ell$ on $H\in\cG_\ell$. Under \emph{edge} adjacency on $H$ (adding or removing a single edge), the rescaled output $p\hatvsumt{t} = \sum_{e\in B_t} q_t(x(e)) + pZ_t$ is a Poisson-subsampled sum of vectors in $B_2^d$ (which has $\ell_2$ sensitivity $1$ on the subsample) perturbed by Gaussian noise of standard deviation $p\sigma_\ell$. Let $\eps_0 = \seps/(2\ell)$ and $\delta_0 = \sdelta/(2\ell)$. Set $\bar\eps=\eps_0/(2p)\leq1/4$, using $p\geq\seps/\ell$. For large enough $C_1$, the Gaussian mechanism before subsampling is $(\bar\eps,\delta_0)$-edge-DP. Poisson amplification~\cite{BalleBG18} then gives privacy parameters
    \[
        \log\bigl(1+p(e^{\bar\eps}-1)\bigr)\leq2p\bar\eps=\eps_0,
        \qquad p\delta_0\leq\delta_0.
    \]
    Thus the sampled query is $(\seps/(2\ell),\sdelta/(2\ell))$-edge-DP.
    Because $H\in\cG_\ell$ has maximum degree at most $\ell$, adding or removing a vertex in $\cG_\ell$ changes at most $\ell$ edges. By group privacy of size $\ell$ (using $\seps\leq 1/2$ so that $e^{\seps/2}\ell\cdot(\sdelta/(2\ell))\leq \sdelta$), each round of $\cA_\ell$ is $(\seps,\sdelta)$-node-DP on $\cG_\ell$.
    By the advanced composition theorem for adaptive mechanisms~\cite{DworkRV10}, for any $\delta_c\in(0,1)$, the $T$-round transcript of $\cA_\ell$ is
    \[
        \bparen{O\bparen{\seps\sqrt{T\log(1/\delta_c)}+T\seps^2},\ T\sdelta+\delta_c}\text{-node-DP on }\cG_\ell.
    \]
    Composing the preprocessing $\sadnspdetail{\dbound}{\peps}{\pdelta}$ with the family $\{\cA_\ell\}_{\ell\in\N}$ via \Cref{lem:projection-plus-dp}(1) yields item~1.

    \mypar{Accuracy and query time (items 2 and 3).}
    If $G\in\cGvecbound{d}{\dbound}$, then \Cref{lem:projection-plus-dp}(2) guarantees that $H=G$ with probability $1$. Conditioned on $\dmax$, the preceding transcript, and $q_t$, the indicators $I_{e,t}=\mathbf{1}\{e\in B_t\}\sim\mathrm{Bernoulli}(p)$ are independent across $e\in E(G)$ and independent of $Z_t$. Hence $\ex{}{\hatvsumt{t}}=\vsumt{t}(G)$ and
    \[
        \ex{}{\|\hatvsumt{t}-\vsumt{t}(G)\|_2^2}
        = \sum_{e\in E(G)} \var{}{\frac{I_{e,t}}{p}}\|q_t(x(e))\|_2^2 + \ex{}{\|Z_t\|_2^2}
        \leq m\frac{1-p}{p} + d\sigma_{\dmax}^2,
    \]
    which gives item~2. Finally, because the edges of $H$ are stored in an array, we can sample $B_t$ in $O(1+|B_t|)$ time by drawing geometric gaps $\mathrm{Geom}(p)$ between consecutive sampled indices, and evaluating $q_t(x(e))$ for $e\in B_t$ takes $O(d|B_t|)$ time. Taking expectations gives $O(d(1+p|E(H)|))$, proving item~3.
\end{proof}

\subsection{Proof of Corollary~\ref{cor:sco-vector-oracle}}
\label{app:proof-sco-vector-oracle}

\begin{proof}[Proof of Corollary~\ref{cor:sco-vector-oracle}]
    Allocate $(\peps,\pdelta)=(\eps/2,\delta/3)$ to preprocessing.
    For each released value $\ell$ of $\dmax$, fix the chosen budget
    $T=T(\ell)$ and set
    \[
        \delta_c=\delta/3,\qquad
        \sdelta=\frac{\delta}{3T},\qquad
        \seps=\frac{c\eps}{\sqrt{T\log(1/\delta)}}
    \]
    for a sufficiently small universal constant $c>0$.
    The stated sampling rate satisfies $p\geq\seps/\ell$.
    Advanced composition gives $(\eps/2,2\delta/3)$-node-DP for the
    query transcript on $\cG_\ell$, uniformly in $\ell$.
    Applying \Cref{lem:projection-plus-dp} therefore gives
    $(\eps,\delta)$-node-DP even when $T$ depends on the released bound.
    Finally, divide the answers by the public target $m$.
    On a $\dbound$-bounded input $G$ with $|E(G)|=m$, dividing the unbiasedness and variance bounds of \Cref{thm:vector-sum-oracle}(2) by $m$ and $m^2$, respectively, yields $\ex{}{\widehat{g}_t}=g_t$ and
    \[
        \ex{}{\|\widehat{g}_t-g_t\|_2^2}
        \leq \frac{1-p}{mp} + O\bparen{\frac{\dmax^2 d\log(\dmax T/\delta)}{m^2\seps^2}}
        = \frac{1-p}{mp} + O\bparen{\frac{\dmax^2 d T L_T}{m^2\eps^2}}.
    \]
    Note also for use in \Cref{prop:sco-projection} below that $\widehat{g}_t = u_t + \xi_t$, where $u_t = \frac{1}{mp}\sum_{e\in E(G)} I_{e,t} q_t(x(e))$ is the subsampled empirical average and $\xi_t = Z_t/m$ is the Gaussian noise; since $mp\geq 1$ and $\|g_t\|_2\leq 1$, we have $\ex{}{\|u_t-g_t\|_2^2}\leq \frac{1-p}{mp}\leq 1$ and hence $\ex{}{\|u_t\|_2^2} = \|g_t\|_2^2 + \ex{}{\|u_t-g_t\|_2^2}\leq 2$.
\end{proof}

\subsection{Proof sketches for \Cref{prop:sco-extension,prop:sco-projection}}
\label{app:proof-sco-extension}
\label{app:proof-sco-projection}

Both \Cref{prop:sco-extension} and \Cref{prop:sco-projection} follow by combining our noisy vector-sum estimators with the standard optimization and uniform-stability analysis for projected subgradient descent on non-smooth convex losses due to Bassily, Feldman, Guzm\'an, and Talwar~\cite{BassilyFGT20}.

\begin{proof}[Proof sketch for \Cref{prop:sco-extension}]
    Let $a = \frac{\dbound\sqrt{d}}{m\mu}$ and $\alpha = \min\set{1,\ m^{-1/2}+a}$. If $\alpha=1$, outputting any fixed $w_1\in\cW$ achieves excess loss at most $1$ (since $\cW$ has unit diameter and $\ell(\cdot;x)$ is $1$-Lipschitz) and is $0$-node-GDP. Otherwise, set
    \[
        T = \lceil \alpha^{-4}\rceil \leq m^2,
        \qquad
        \eta = \alpha^3.
    \]
    In each step $t\in[T]$, let $G_t=(V,E,q_t)$ be the input graph with edge vectors $q_t(e) = g(w_t; x(e))\in B_2^d$, and compute the noisy subgradient via \Cref{thm:vector-sum-gdp} with parameter $\mu_0 = \mu/\sqrt{T}$:
    \[
        \widehat{g}_t = \frac{1}{m}\bparen{F_{\dbound}(G_t) + Z_t},
        \qquad
        Z_t\sim\mathcal{N}\bparen{0,\ \frac{C^2\dbound^2 T}{\mu^2}I_d}.
    \]
    By adaptive composition of $T$ mechanisms that are each $(\mu/\sqrt{T})$-node-GDP~\cite{DongRS19}, the $T$-step algorithm is $\mu$-node-GDP on all graphs, and its runtime is $T$ times the runtime of \Cref{lem:vector-sum-ext}.

    Now suppose $G$ is a fixed $\dbound$-bounded graph with $|E(G)|=m$ and edge attributes $\{x(e)\}_{e\in E(G)}$ drawn i.i.d.\ from $P$. Then the empirical objective $\widehat{\cL}_G(w) = \frac{1}{m}\sum_{e\in E(G)} \ell(w; x(e))$ is a standard $m$-sample i.i.d.\ empirical risk, and by \Cref{lem:vector-sum-ext}(1) the update at step $t$ is $w_{t+1} = \Pi_{\cW}(w_t - \eta(g_t + \xi_t))$, where $g_t = \frac{1}{m}\sum_{e\in E(G)} g(w_t; x(e))\in\partial\widehat{\cL}_G(w_t)$ satisfies $\|g_t\|_2\leq 1$ and $\xi_t = Z_t/m$ satisfies $\ex{}{\|\xi_t\|_2^2} = \frac{C^2\dbound^2 d T}{m^2\mu^2} = C^2 T a^2$.
    By the standard regret bound for projected subgradient descent on a unit-diameter domain, the empirical optimization error of the average iterate $\overline{w}$ satisfies
    \[
        \ex{}{\widehat{\cL}_G(\overline{w}) - \min_{w\in\cW}\widehat{\cL}_G(w)}
        \leq \frac{1}{2\eta T} + \frac{\eta}{2}\bparen{1 + C^2 T a^2}.
    \]
    Furthermore, for two $m$-edge datasets differing on a single edge attribute $x(e)$, coupling the Gaussian noises $\xi_t$ across the two runs shows that projected subgradient descent with step size $\eta$ over $T$ steps has uniform argument stability $O\bigl(\eta\sqrt{T} + \frac{\eta T}{m}\bigr)$~\cite[Lemma~3.1]{BassilyFGT20}, which bounds the expected generalization error $\ex{}{\cL_P(\overline{w}) - \widehat{\cL}_G(\overline{w})}$. Summing the empirical optimization and generalization bounds gives
    \begin{equation}\label{eq:sco-combined-bound}
        \ex{}{\cL_P(\overline{w}) - \cL_P^*}
        = O\bparen{\frac{1}{\eta T} + \eta(1 + T a^2) + \eta\sqrt{T} + \frac{\eta T}{m}}.
    \end{equation}
    Substituting $T=\lceil\alpha^{-4}\rceil$ and $\eta=\alpha^3$ (and noting $\eta T a^2 \leq 2\alpha^{-1}a^2 \leq 2\alpha$ since $a\leq\alpha$) bounds each term in~\eqref{eq:sco-combined-bound} by $O(\alpha) = O(m^{-1/2}+a)$.
\end{proof}

\begin{proof}[Proof sketch for \Cref{prop:sco-projection}]
    Run the normalized oracle of Corollary~\ref{cor:sco-vector-oracle}
    on $G$ and release $\dmax$. Set
    \[
        L_* = \log(1/\delta)\log(\dmax m^2/\delta),
        \qquad
        a = \frac{\dmax\sqrt{d L_*}}{m\eps},
        \qquad
        \alpha = \min\set{1,\ m^{-1/2}+a}.
    \]
    As above, if $\alpha=1$ we may output $w_1$; otherwise set $T=\lceil\alpha^{-4}\rceil\leq m^2$ and $\eta=\alpha^3$, and in each step $t\in[T]$ query the oracle of Corollary~\ref{cor:sco-vector-oracle} with $q_t(x)=g(w_t;x)$. By Corollary~\ref{cor:sco-vector-oracle}, the entire procedure is $(\eps,\delta)$-node-DP on arbitrary graphs.

    Now fix a $\dbound$-bounded graph $G$ with $|E(G)|=m$ and i.i.d.\ edge attributes from $P$. Because the projection preprocessing reads only the topology $(V,E)$, conditioning on the released degree bound $\dmax$ leaves the edge attributes $\{x(e)\}_{e\in E(G)}$ i.i.d.\ from $P$, and $H=G$ by \Cref{lem:projection-plus-dp}(2). In each step $t$, the oracle returns $\widehat{g}_t = u_t + \xi_t$, where $\ex{}{u_t\mid w_t} = g_t\in\partial\widehat{\cL}_G(w_t)$, $\ex{}{\|u_t\|_2^2}\leq 2$ (as shown in the proof of Corollary~\ref{cor:sco-vector-oracle}), and $\ex{}{\|\xi_t\|_2^2} = O(T a^2)$.
    The projected-SGD regret bound applies as before. For stability, couple
    the Bernoulli samples and Gaussian noises on two datasets differing
    at edge $j$. Conditional on $\dmax$, the schedules are the same.
    Write $K_t=|B_t|$ and $L_t=K_t/(mp)$ for the Lipschitz constant of
    the sampled objective. The argument of~\cite[Lemma~3.1]{BassilyFGT20}
    gives, for the distance $\Delta_t$ between the two iterates,
    \[
        \Delta_{t+1}\leq
        \sqrt{\Delta_t^2+4\eta^2L_t^2}
        +\frac{2\eta I_{j,t}}{mp}.
    \]
    The common Gaussian noise cancels before projection; the last term
    accounts for the changed edge when it is sampled. Unrolling and using
    $\ex{}{L_t^2}=1+(1-p)/(mp)\leq2$ and $\ex{}{I_{j,t}}=p$ yields
    \[
        \ex{}{\Delta_{T+1}}
        \leq2\eta\sqrt{2T}+\frac{2\eta T}{m}.
    \]
    The averaged iterates satisfy the same bound, so the conditional
    excess population loss again obeys~\eqref{eq:sco-combined-bound}.
    Substituting the schedules gives $O(\alpha)$, as claimed.

    Finally, since $\eps<1/2$, $\dmax\geq1$, and
    $\log(1/\delta)>\log2$, the sampling rate satisfies
    $p\leq\max\{1/m,1/\sqrt T\}$. Thus
    $Tmp\leq T+m\sqrt T=O(m^2)$, giving $O(m^2)$ expected subgradient
    evaluations. If $\eps=\Omega(1/n)$ and $\delta=n^{-\Theta(1)}$,
    preprocessing takes $O(m\log m)$ time. Adding the expected query
    cost $O(dT(1+mp))$ gives $O(dm^2)$ total expected time.

\end{proof}

\end{document}